\documentstyle[11pt,epsf]{article}
 1
 1
 1

\def\lsim{\mathrel{\raise.3ex\hbox{$<$\kern-.75em\lower1ex\hbox{$\sim$}}}}
\def\gsim{\mathrel{\raise.3ex\hbox{$>$\kern-.75em\lower1ex\hbox{$\sim$}}}}
\catcode`\@=11
\def\slash{\mathpalette\make@slash}
\def\make@slash#1#2{\setbox\z@\hbox{$#1#2$}%
  \hbox to 0pt{\hss$#1/$\hss\kern-\wd0}\box0}
\catcode`\@=12 

\begin{document}
\noindent
\thispagestyle{empty}
\renewcommand{\thefootnote}{\fnsymbol{footnote}}
\begin{flushright}
{\bf CERN-TH/99-59}\\
{\bf DESY 99-047}\\
{\bf hep-ph/9904468}\\
{\bf April 1999}\\
\end{flushright}
\vspace{.0cm}
\begin{center}
  \begin{Large}\bf
Top Quark Pair Production Close to Threshold:\\[2mm]
\noindent 
Top Mass, Width and Momentum Distribution 
  \end{Large}
  \vspace{1.2cm}

\begin{large}
 A.~H.~Hoang$^{a}$ and
 T. Teubner$^{b}$ 
\end{large}

\vspace{1.2cm}
\begin{it}
${}^a$ Theory Division, CERN,\\
   CH-1211 Geneva 23, Switzerland\\[.5cm]
${}^b$  Deutsches Elektronen-Synchrotron DESY,\\
   D-22603 Hamburg, Germany
\end{it}

  \vspace{2.cm}
  {\bf Abstract}\\
\vspace{0.3cm}

\noindent
\begin{minipage}{14.0cm}
\begin{small}
The complete NNLO QCD corrections to the total cross section 
$\sigma(e^+e^-\to Z^*,\gamma^*\to t\bar t)$,
in the kinematic region close to the top-antitop threshold, are
calculated by solving the corresponding Schr\"odinger equations
exactly in momentum space in a consistent momentum cutoff
regularization scheme. The corrections coming from the same NNLO QCD
effects to the top quark three-momentum distribution 
$d\sigma/d |\vec k_t|$
are determined. We discuss the origin of the large NNLO corrections
to the peak position and the normalization of the total cross section
observed in previous works and propose a new top mass definition, the
$1S$ mass $M_{1S}$, which stabilizes the peak in the total cross
section. If the influence of beamstrahlung and initial state radiation
on the mass determination is small, a theoretical
uncertainty on the $1S$ top mass measurement of
200~MeV from the total cross section at the linear collider seems
possible. We discuss how well the $1S$ mass can be
related to the $\overline{\mbox{MS}}$ mass. We propose a consistent
way to implement the top quark width at NNLO by including electroweak
effects into the NRQCD matching coefficients, which can then become
complex. 
\\[3mm]
\end{small}
\end{minipage}
\end{center}
\setcounter{footnote}{0}
\renewcommand{\thefootnote}{\arabic{footnote}}
\vspace{1.2cm}
{\bf CERN-TH/99-59}\\
{\bf April 1999}
%
%
%
\newpage
\noindent

\section{Introduction}

It will be one of the primary goals of a future $e^+e^-$ linear
collider (LC) or $\mu^+\mu^-$ pair collider (FMC) to measure and
determine 
the properties of the top quark, whose existence has been confirmed at
the Tevatron ($M_t=173.8\pm 5$~GeV~\cite{Vancouver1}). Although the
top quark will also be object
of intense studies at the Run II at the Tevatron and at the LHC, the
measurements at a LC are important to fill the gaps left by the
measurements in the environment of the hadron colliders. One of the 
most dramatic improvements attainable at a LC can be expected in 
the determination of the top quark mass. At the LHC, where the mass 
is extracted from the peak in the top invariant mass spectrum of the 
$W$ and $b$ originating from the top decay, a final 
(systematics-dominated) top mass uncertainty at the level of 
2--3~GeV seems realistic.
More precision will be difficult owing to
unavoidable conceptual and practical problems and ambiguities in
disentangling the top quark invariant mass from numerous effects in
the environment of hadron colliders. 
At a LC the top quark mass can be determined from a
measurement of the line-shape of the total cross section  
$\sigma(e^+e^-\to Z^*,\gamma^*\to t\bar t)$ for centre-of-mass
energies around the threshold, $\sqrt{q^2}\approx 350$~GeV. The rise
of the cross section with increasing centre-of-mass energy is
directly correlated to the mass of the top quark. Because the total
cross section describes the rate of colour singlet top-antitop events,
it is theoretically and practically much better under control than the
top quark invariant mass distribution.
Because of the large top width
($\Gamma(t\to b\,W)= \frac{G_F}{\sqrt{2}}\frac{M_t^3}{8 \pi}\approx
1.5\,\mbox{GeV}$) the top-antitop pair cannot hadronize into toponium
resonances, and the cross section represents a smooth line-shape
showing only a moderate peak-like enhancement, which is the broad
remnant of the $1S$ resonance. At the same time the top width
effectively serves as an infrared cutoff~\cite{Fadin1} and as a natural
smearing mechanism~\cite{Poggio1}, which allows us to calculate the cross
section in the threshold region to high precision using perturbative
QCD. It is therefore possible to reliably relate the cross
section line-shape to the parameters of the Standard Model, most
notably the top quark mass and the strong coupling. LC simulation
studies have demonstrated that for 50--100~$fb^{-1}$ total integrated
luminosity an experimental uncertainty of order 100--200~MeV can be
expected in the top mass determination from a line-shape scan of the
total cross section~\cite{Orange1}. Evidently, at this level of
precision an adequate control over theoretical uncertainties has to be
achieved. In particular, a precise definition of the top mass
extracted from the experiment has to be given. 
  
For centre-of-mass energies close to the top-antitop threshold, the top
quarks are produced with non-relativistic velocities $v\ll
1$. Therefore the relevant physical scales, which govern the
top-antitop dynamics, the top mass $M_t$, the relative momentum $M_t
v$ and the top kinetic energy $M_t v^2$, are widely separated. Because
ratios of the three scales arise, the cross section close to threshold
cannot be calculated using the standard multi-loop expansion in the
strong coupling $\alpha_s$, but rather a double expansion in $\alpha_s$
and $v$. In the non-relativistic limit the most prominent indication of
this feature is known as the ``Coulomb singularity'', which originates
from the ratio $M_t/(M_t v)$. The Coulomb singularity is visible as a
singular $(\alpha_s/v)^n$
behaviour in the $n$-loop QCD correction to the amplitude $\gamma\to
t\bar t$ for $v\to 0$. The most economic and systematic way to tackle
this problem is to employ the concept of effective theories by using the
hierarchy $M_t\gg M_t v\gg M_t v^2 > \Gamma_t\gg\Lambda_{QCD}$ and  by
successively integrating out higher momentum effects. At leading order
(LO) and next-to-leading order (NLO) in the non-relativistic
expansion\footnote{
We will define what is meant by LO, NLO, NNLO, etc., in the
framework of the non-relativistic expansion at the beginning of
Sec.~\ref{sectionNRQCDstable}. 
},
the use of effective field theoretical methods seems not to be vital,
because, clearly, in the first approximation the top-antitop pair can
be described by a non-relativistic Schr\"odinger
equation~\cite{Fadin1,Fadin2} and
because, luckily, the relevant current operators do not have any
anomalous dimension at the one-loop level. Beyond NLO, however,
anomalous
dimensions arise when relativistic effects suppressed by $v^2$ are
included. This
makes the use of effective field theoretical methods mandatory. In
addition, the effective field theoretical approach allows for the
development of a power counting scheme, which allows for a systematic
identification of all effects contributing to a certain order of
approximation. Those power counting rules in fact confirm that at
next-to-next-to-leading order (NNLO) the top-antitop pair can be
completely described by a conventional
Schr\"odinger equation containing an instantaneous potential.
In general, the same conclusions cannot be drawn for non-relativistic
bottom-antibottom or charm-anticharm systems. 

A large number of theoretical studies at LO and 
NLO~\cite{Strassler1,Kwong1,Jezabek1,Sumino1,japaner} 
have been carried out in
the past in order to study the feasibility of the threshold scan and
other measurements at the top-antitop threshold.
Recently, first NNLO QCD calculations for the total 
vector-current-induced cross section $\sigma(e^+e^-\to \gamma^*\to
t\bar t)$ have
been performed in Refs.~\cite{Hoang3,Melnikov3,Yakovlev1}.
These analyses were
based on non-relativistic quantum chromodynamics
(NRQCD)~\cite{Caswell1,Bodwin1} and on the direct matching
procedure~\cite{Hoang1,Hoang2}. 
In this work we extend the calculations and also determine the   
NNLO QCD relativistic corrections to the top quark three-momentum
distribution. Interconnection effects caused by gluon exchange
among top-antitop decay and production processes are not considered in
this work. They are known to vanish at NLO for the total cross
section~\cite{Khoze1,Khoze2,Melnikov1}, but can lead to sizeable
corrections in the
momentum distribution~\cite{Harlander1,Sumino2}. We also include the
total cross section and
the three-momentum distribution induced by the axial-vector current.
Because the latter quantities are suppressed by $v^2$ with respect to
the vector-current-induced ones, we only determine them at leading
order in the non-relativistic expansion. It turns out that the size of
the axial-vector contributions is smaller than the theoretical
uncertainties contained in the dominant vector-current-induced cross
section. A discussion on the size of the axial-vector-current-induced
contributions can also be found in~\cite{Kuehn1}.
The three-momentum distributions presented in this work represent a
first step towards an exclusive treatment of
the top-antitop final state at NNLO close to threshold. 
We analyse in detail the origin of the large NNLO corrections to the
peak position and the normalization in the total cross section already
observed in Refs.~\cite{Hoang3,Melnikov3,Yakovlev1}, and show that the
instabilities in the 
peak position are a consequence of the use of the pole mass scheme.
We show that the pole mass parameter is irrelevant to the peak
position and define a new top quark mass, the $1S$ mass,  which is
more suitable to parametrize the total cross section. Whereas the
$1S$ mass leads to a considerable stabilization of the peak position
it does not affect the large corrections to the normalization. 
In this paper
we also propose a NNLO generalization of the energy replacement rule
``$E\to E+i \Gamma_t$'', by Fadin and Khoze, for the implementation of
the top quark
width by including electroweak corrections into the matching
conditions of NRQCD. In general this leads to
NRQCD short-distance coefficients that have an imaginary part. 

The program of this paper is as follows:
in Sec.~\ref{sectionNRQCDstable} we review the conceptual framework of
the effective theories NRQCD and potential (P) NRQCD as far as it is
relevant to the NNLO calculations carried out in this work. 
Section~\ref{sectionlippmannschwinger} contains a derivation of the
integral equations that have to be solved and
Sec.~\ref{sectionregularization} describes our cutoff regularization
scheme.
In Secs.~\ref{sectionNRQCDstable}, \ref{sectionlippmannschwinger} and
\ref{sectionregularization} the top quark width is neglected.
In Sec.~\ref{sectionwidth} we discuss the effects of the top quark
width from the point of view of (P)NRQCD. Some details about our
numerical methods to solve the integral equations are given in
Sec.~\ref{sectionnumerics}. A first analysis of the total cross section
and the three momentum distribution in the pole mass scheme is given in
Sec.~\ref{sectionpolescheme}. Section~\ref{sectionuncertainties}
concentrates on the origin and interpretation of the large NNLO
corrections in the pole mass scheme and introduces the $1S$
mass. The relation of the $1S$ mass to other mass definitions is
discussed. Section~\ref{sectionconclusion} contains our
conclusions. In 
Appendix~\ref{appendixshortdistance}, details of the NNLO matching
calculation are given in the framework of our regularization scheme.

\vspace{1.5cm}
\section{The Conceptual Framework - NRQCD and PNRQCD}
\label{sectionNRQCDstable}
In this section we review the effective field theories NRQCD
and PNRQCD, which form the conceptual framework in which
the NNLO corrections to the top-antitop cross section close to
threshold are calculated. By N$^k$LO ($k=0,1,2,\ldots$)
for the total cross section we mean a resummation of all terms
proportional to $\alpha_s^m v^n$, with $m+n=1,\ldots,k+1$, in
perturbation
theory in $\alpha_s$ supplemented by a subsequent expansion in the
top quark velocity, i.e. in the limit $\alpha_s\ll v\ll 1$.
Thus at the NNLO level all terms proportional to
$v\sum_{n=0}^\infty(\alpha_s/v)^n [ 1; \alpha_s, v; \alpha_s^2,
\alpha_s v, v^2 ]$ have to be resummed to all orders in conventional
perturbation theory in $\alpha_s$, where the dominant terms in the
non-relativistic limit are determined by a non-relativistic
Schr\"odinger equation with a Coulomb potential $V({\mbox{\boldmath
$r$}})=- C_F \alpha_s/r$, $C_F=4/3$. 
In this context one has to count the strong coupling $\alpha_s$ of
order $v$, as long as the renormalization scale is much larger than
the typical hadronization scale $\Lambda_{QCD}$. For simplicity we
postpone the effects of the top quark width until
Section~\ref{sectionwidth}.

NRQCD~\cite{Caswell1,Bodwin1} is an effective field theory of QCD
specifically
designed to handle non-relativistic heavy quark-antiquark
systems. NRQCD is based on the separation of the low momentum scales 
$M_t v$ and $M_t v^2$, which govern the non-relativistic quark-antiquark
dynamics, from the high momentum scale $M_t$, which is relevant for hard
effects involved in the quark-antiquark production process and
quark-antiquark and quark-gluon interactions. The NRQCD Lagrangian for
the top-antitop system is obtained from QCD by integrating out all
hard quark and gluon momenta of order $M_t$ or larger, and the
corresponding antiparticle poles of the small components. Treating
all quarks except the top as massless, the resulting
non-renormalizable Lagrangian reads
\begin{eqnarray}
\lefteqn{
{\cal{L}}_{\mbox{\tiny NRQCD}} \, = \, 
- \frac{1}{2} \,\mbox{Tr} \, G^{\mu\nu} G_{\mu\nu} 
+ \sum_{q=u,d,s,c,b} \bar q \, i \slash{D} \, q
}\nonumber\\& &
+\, \psi^\dagger\,\bigg[\,
i D_t 
+ c_2\,\frac{{\mbox{\boldmath $D$}}^2}{2\,M_t} 
+ c_4\,\frac{{\mbox{\boldmath $D$}}^4}{8\,M_t^3}
+ \ldots 
\nonumber\\[2mm] & &
\hspace{8mm}
+  \frac{c_F\,g_s}{2\,M_t}\,{\mbox{\boldmath $\sigma$}}\cdot
    {\mbox{\boldmath $B$}}
+ \, \frac{c_D\,g_s}{8\,M_t^2}\,(\,{\mbox{\boldmath $D$}}\cdot 
  {\mbox{\boldmath $E$}}-{\mbox{\boldmath $E$}}\cdot 
  {\mbox{\boldmath $D$}}\,)
+ \frac{c_S\,g_s}{8\,M_t^2}\,i\,{\mbox{\boldmath $\sigma$}}\,
  (\,{\mbox{\boldmath $D$}}\times 
  {\mbox{\boldmath $E$}}-{\mbox{\boldmath $E$}}\times 
  {\mbox{\boldmath $D$}}\,)
+\ldots
 \,\bigg]\,\psi 
\nonumber\\[2mm] & &
+\, \chi^\dagger\,\bigg[\,
i D_t 
- c_2\,\frac{{\mbox{\boldmath $D$}}^2}{2\,M_t} 
- c_4\,\frac{{\mbox{\boldmath $D$}}^4}{8\,M_t^3}
+ \ldots 
\nonumber\\[2mm] & &
\hspace{8mm}
-  \frac{c_F\,g_s}{2\,M_t}\,{\mbox{\boldmath $\sigma$}}\cdot
    {\mbox{\boldmath $B$}}
+ \, \frac{c_D\,g_s}{8\,M_t^2}\,(\,{\mbox{\boldmath $D$}}\cdot 
  {\mbox{\boldmath $E$}}-{\mbox{\boldmath $E$}}\cdot 
  {\mbox{\boldmath $D$}}\,)
+ \frac{c_S\,g_s}{8\,M_t^2}\,i\,{\mbox{\boldmath $\sigma$}}\,
  (\,{\mbox{\boldmath $D$}}\times 
  {\mbox{\boldmath $E$}}-{\mbox{\boldmath $E$}}\times 
  {\mbox{\boldmath $D$}}\,)
+\ldots
 \,\bigg]\,\chi 
\,,
\label{NRQCDLagrangian}
\end{eqnarray}
where only those terms are displayed explicitly which are relevant to
the NNLO calculations in this work. The gluonic and light quark
degrees of freedom are described by the
conventional relativistic Lagrangian, where $G^{\mu\nu}$ is the gluon
strength field tensor, $q$ the Dirac spinor of a massless quark.
The non-relativistic top and antitop
quark are described by the Pauli spinors $\psi$ and $\chi$,
respectively. For convenience all colour indices are suppressed and
summations over colour indices are understood. 
$D_t$ and {\boldmath $D$} are the time and space components of the
gauge covariant derivative $D_\mu$, and $E^i = G^{0 i}$ and $B^i =
\frac{1}{2}\epsilon^{i j k} G^{j k}$ the electric and magnetic
components of the gluon field strength tensor.
The short-distance coefficients $c_2,c_4,c_F,c_D,c_S$ are normalized to one
at the Born level. The subscripts $F$, $D$ and $S$ stand for Fermi,
Darwin and spin-orbit.
We emphasize that the mass parameter $M_t$ used for the formulation of
NRQCD is the top quark pole mass. Although it is known that
this choice can lead to a bad behaviour of the perturbative
coefficients at large orders, the pole mass is still the most
convenient mass
parameter to be used at this stage, because the formulation of NRQCD is
particularly simple in this scheme. 
 
In addition to integrating out hard quark and gluon momenta and the
small components in the QCD
Lagrangian one also has to do the same in
the vector ($j_\mu^v=\bar t\gamma_\mu t$) and
the axial-vector currents ($j_\mu^a=\bar t\gamma_\mu\gamma_5 t$),
which produce and annihilate the top-antitop pair close to
the threshold with centre-of-mass energy $\sqrt{q^2}$. This means that we have
to expand the respective QCD currents
in terms of NRQCD currents carrying the proper quantum numbers. In
momentum space representation the expansion of the
QCD vector current in terms of ${}^3\!S_1$ NRQCD currents reads ($k=1,2,3$)   
\begin{eqnarray}
\tilde j_k^v(q) & = & c^v_1\,\Big({\tilde \psi}^\dagger \sigma_k 
\tilde \chi\Big)(q) -
\frac{c^v_2}{6 M_t^2}\,\Big({\tilde \psi}^\dagger \sigma_k
(\mbox{$-\frac{i}{2}$} 
\stackrel{\leftrightarrow}{\mbox{\boldmath $D$}})^2
 \tilde \chi\Big)(q) + \ldots
\,,
\label{vectorcurrentexpansion1}
\\
\tilde j_k^v(-q) & = & c^v_1\,\Big({\tilde \chi}^\dagger \sigma_k 
\tilde \psi\Big)(-q) -
\frac{c^v_2}{6 M_t^2}\,\Big({\tilde \chi}^\dagger \sigma_k
(\mbox{$-\frac{i}{2}$} 
\stackrel{\leftrightarrow}{\mbox{\boldmath $D$}})^2
 \tilde \psi\Big)(-q) + \ldots
\,,
\label{vectorcurrentexpansion2}
\end{eqnarray}
and the expansion of the QCD axial-vector current in terms of
${}^3\!P_1$ NRQCD currents     
\begin{eqnarray}
\tilde j_k^a(q) & = & \frac{c^a_1}{M_t}\,\Big({\tilde \psi}^\dagger 
(\mbox{$-\frac{i}{2}$} 
\stackrel{\leftrightarrow}{\mbox{\boldmath $D$}}\!\!
       \times{\mbox{\boldmath $\sigma$}})_k 
\tilde \chi\Big)(q) + \ldots
\,,
\label{axialvectorcurrentexpansion1}
\\
\tilde j_k^a(-q) & = &  \frac{c^a_1}{M_t}\,\Big({\tilde \chi}^\dagger 
(\mbox{$-\frac{i}{2}$} 
\stackrel{\leftrightarrow}{\mbox{\boldmath $D$}}\!\!
       \times{\mbox{\boldmath $\sigma$}})_k 
\tilde \psi\Big)(-q) + \ldots
\,,
\label{axialvectorcurrentexpansion2}
\end{eqnarray}
where the constants $c^v_{1,2}$ and $c^a_1$ are the short-distance
coefficients normalized to one at the Born level. 
The time components of the currents do not contribute because the
trace over the massless lepton fields that describe the
$e^+e^- $ annihilation process is proportional to
$(\delta^{ij}-e^i e^j)$, $(e^1,e^2,e^3)$ being the unit-vector
pointing into the centre-of-mass electron direction. In addition, the
zero component of the vector current vanishes. The dominant NRQCD
current in the expansion of the QCD vector current has dimension
three. Thus for a NNLO description of the cross section we have to expand
the QCD vector current in NRQCD currents up to dimension five. 
The QCD axial-vector current only needs to be expanded up to dimension
four. For the NNLO calculation of the cross section only the
${\cal{O}}(\alpha_s^2)$ short-distance corrections to the coefficient
$c^v_1$ have to be calculated.

To formulate the total $t\bar t$ production cross sections in $e^+e^-$
annihilation in the non-relativistic region at NNLO
in NRQCD, we first define the vector and axial-vector-current-induced
cross sections using the optical theorem and starting from their
corresponding expressions in full QCD:
\begin{eqnarray}
R^v(q^2) & = &
\frac{4\,\pi}{q^2}\,\mbox{Im}\,\bigg[\,
-i\,\int\,d^4x\,e^{i\,q.x}\,
  \langle\, 0\,| T\,j^v_i(x) \, j^{v\,i}(0)\, |\, 0\,\rangle\,\bigg]
\nonumber\\[2mm] & \equiv &
\frac{4\,\pi}{q^2}\,\mbox{Im}\,[\,-i\,
\langle\, 0\,| T\, \tilde j^v_i(q) \,
 \tilde j^{v\,i}(-q)\, |\, 0\,\rangle\,]
\,,
\label{vectorcrosssectioncovariant}
\\[3mm]
R^a(q^2) & = &
\frac{4\,\pi}{q^2}\,\mbox{Im}\,\bigg[\,
-i\,\int\,d^4x\,e^{i\,q.x}\,
  \langle\, 0\,| T\,j^a_i(x) \, j^{a\,i}(0)\, |\, 0\,\rangle\,\bigg]
\nonumber\\[2mm] & \equiv &
\frac{4\,\pi}{q^2}\,\mbox{Im}\,[\,-i\,
\langle\, 0\,| T\, \tilde j^a_i(q) \,
 \tilde j^{a\,i}(-q)\, |\, 0\,\rangle]
\,.
\label{axialvectorcrosssectioncovariant}
\end{eqnarray}
In terms of $R^v$ and $R^a$ the total cross section 
$\sigma_{tot}^{\gamma,Z}(e^+e^-\to \gamma^*, Z^*\to t\bar t)$ reads
\begin{eqnarray}
\sigma_{tot}^{\gamma,Z}(q^2) & = &
\sigma_{pt}\, \bigg[\,
Q_t^2 - 
2\,\frac{q^2}{q^2-M_Z^2}\,v_e\,v_t\,Q_t +
\bigg(\frac{q^2}{q^2-M_Z^2}\bigg)^2\,
\Big[ v_e^2+a_e^2 \Big]\,v_t^2
\,\bigg]\,R^v(q^2)
\nonumber \\[2mm] & &
+\,\sigma_{pt}\,\bigg(\frac{q^2}{q^2-M_Z^2}\bigg)^2\,
\Big[ v_e^2 + a_e^2 \Big]\,a_t^2\,R^a(q^2)
\,,
\label{totalcrossfullQCD}
\end{eqnarray}
where
\begin{eqnarray}
\sigma_{pt} & = & \frac{4\,\pi\,\alpha^2}{3\,q^2}
\,,
\\[3mm]
v_f & = & \frac{T_3^f - 2\,Q_f\,\sin^2\theta_W}{2\, \sin\theta_W
  \,\cos\theta_W}
\,,
\\[2mm]
a_f & = & \frac{T_3^f}{2\, \sin\theta_W\, \cos\theta_W}
\,.
\end{eqnarray}
Here,
$\alpha$ is the fine structure constant, $Q_t=2/3$ the electric
charge of the top quark, $\theta_W$ the Weinberg angle, and $T_3^f$
refers to the third component of the weak isospin;
$Q_t^2 R^v$ is equal to the
total normalized photon-induced cross section, which is usually
referred to as the $R$-ratio.
To determine $R^v$ and $R^a$ at NNLO in NRQCD we
insert the expansions in 
Eqs.~(\ref{vectorcurrentexpansion1})--(\ref{axialvectorcurrentexpansion2})
into Eqs.~(\ref{vectorcrosssectioncovariant}) and
(\ref{axialvectorcrosssectioncovariant}).
This leads to the expressions
\begin{eqnarray}
R_{\mbox{\tiny NNLO}}^{v,\mbox{\tiny thr}}(q^2) & = &
\frac{4\,\pi}{q^2}\,C^v\,
\mbox{Im}\Big[\,
{\cal{A}}^v(q^2)
\,\Big]
+ \ldots
\,,
\label{vectorcrosssectionexpanded}
\\[3mm]
R_{\mbox{\tiny NNLO}}^{a,\mbox{\tiny thr}}(q^2) & = &
\frac{4\,\pi}{q^2}\,C^a\,
\mbox{Im}\Big[\,
{\cal{A}}^a(q^2)
\,\Big]
+ \ldots
\,,
\label{axialvectorcrosssectionexpanded}
\end{eqnarray}
where
\begin{eqnarray}
{\cal{A}}^v & = & i\,\Big\langle \, 0 \, \Big| 
\, \Big({\tilde\psi}^\dagger \vec\sigma \, \tilde \chi
+ \frac{1}{6 M_t^2} {\tilde\psi}^\dagger  \vec\sigma \,
(\mbox{$-\frac{i}{2}$} 
\stackrel{\leftrightarrow}{\mbox{\boldmath $D$}})^2
\tilde \chi
\Big)\,
\, \Big({\tilde\chi}^\dagger \vec\sigma \, \tilde \psi
+ \frac{1}{6 M_t^2}{\tilde\chi}^\dagger \vec\sigma \,
(\mbox{$-\frac{i}{2}$} 
\stackrel{\leftrightarrow}{\mbox{\boldmath $D$}})^2
\tilde \psi
\Big)\,
\Big| \, 0 \, \Big\rangle
\,,
\label{correlatorV}
\\[2mm]
{\cal{A}}^a & = & i\,\Big\langle \, 0 \, \Big| 
\, \Big({\tilde\psi}^\dagger
(\mbox{$-\frac{i}{2}$} 
\stackrel{\leftrightarrow}{\mbox{\boldmath $D$}}\!\!
       \times{\mbox{\boldmath $\sigma$}}) \, \tilde \chi \Big)\,
\, \Big({\tilde\chi}^\dagger
(\mbox{$-\frac{i}{2}$} 
\stackrel{\leftrightarrow}{\mbox{\boldmath $D$}}\!\!
       \times{\mbox{\boldmath $\sigma$}}) \, \tilde \psi \Big)\,
\Big| \, 0 \, \Big\rangle
\,,
\label{correlatorA}
\end{eqnarray}
and 
\begin{eqnarray}
C^v & = & (c_1^v)^2
\,,
\label{Cvdef}
\\[2mm]
C^a & = & 1
\,.
\label{Cadef}
\end{eqnarray}
The expressions for $R^v$ and $R^a$ at NNLO in the non-relativistic
expansion
in Eqs.~(\ref{vectorcrosssectionexpanded}) and
(\ref{axialvectorcrosssectionexpanded}) represent an application of
the factorization formalism proposed by Bodwin, Braaten and
Lepage~\cite{Bodwin1}. The cross sections are written as a sum of
absorptive
parts of non-relativistic current correlators, each of which is
multiplied by a short-distance coefficient. The vector correlator
${\cal{A}}^v$ describes the top-antitop system produced in an S-wave
spin triplet state and the axial-vector correlator ${\cal{A}}^a$
describes the system in a corresponding P-wave triplet state. 
The axial vector-current-induced cross section is suppressed by $v^2$
with respect to the vector-current-induced cross section. We note
that the
correlators are defined within a proper regularization scheme.
For convenience we have not expanded ${\cal{A}}^v$ into a sum of a
dimension six and a dimension eight current correlator.
The short distance coefficients $C^v$ and $C^a$ encode the effects
of quark and gluon momenta of order the top mass or larger in
top-antitop production and annihilation vertex diagrams. As the
non-relativistic correlators they are 
regularization-scheme-dependent. In principle the non-relativistic
correlators ${\cal{A}}^v$
and ${\cal{A}}^a$ could be calculated from the Feynman rules derived
from the NRQCD Lagrangian~(\ref{NRQCDLagrangian}). Such a task,
however, would be quite cumbersome, because an infinite number of
diagrams would still have to be resummed. Clearly, we would like to
derive an
equation of motion for the off-shell top quark four point Green
function, which in the non-relativistic limit reduces to the well known
Schr\"odinger equation, which automatically carries out the resummation
of the relevant diagrams. A formal and systematic way to achieve that
is to also integrate out top quark and gluon modes carrying momenta
of the order the inverse Bohr radius $\sim M_t v$. The resulting
effective field theory is called ``potential NRQCD''
(PNRQCD)~\cite{Pineda1}. The basic ingredient to construct PNRQCD is to
identify the relevant physical momentum regions in the description of
heavy quark-antiquark systems in the framework of NRQCD. Those
momentum regions have been
identified in Ref.~\cite{Beneke1} by constructing an asymptotic
expansion of
Feynman diagrams describing heavy quark-antiquark production or
annihilation close to threshold in terms of the heavy quark
centre-of-mass velocity. Because NRQCD is not Lorentz-covariant, the
time and the spatial components of the momenta have to be treated
independently. There are momentum regions where time and spatial
components are of a different order in the velocity counting. The
relevant momentum regions are ``soft'' ($k^0\sim
M_t v$, $\vec k\sim M_t v$), ``potential'' ($k^0\sim M_t v^2$, $\vec
k\sim M_t v$) and ``ultrasoft'' ($k^0\sim M_t v^2$, $\vec k\sim M_t
v^2$). It can be shown that heavy quarks and gluons can have soft and
potential momenta, but only gluons can have ultrasoft momenta. 
A momentum region with $k^0\sim M_t v$, $\vec k\sim M_t v^2$
does not exist. It is in principle not excluded that there are 
momentum regions scaling like $M_t v^n$ with $n > 2$, but even if such
regions existed they would be irrelevant to top quark production
because they would represent modes below the hadronization scale. 
To construct PNRQCD one integrates out ``soft'' heavy
quarks and gluons ($k^0\sim M_t v$, $\vec k\sim M_t v$) and
``potential'' gluons ($k^0\sim M_t v^2$, $\vec k\sim M_t
v$), supplemented by an expansion in momentum components of
order $M_t v^2$. Heavy quarks carrying potential momenta
and gluons with ultrasoft momenta are kept as dynamical
fields. This leads to spatially non-local four (heavy) quark
operators which represent a coupling of a quark-antiquark pair
separated by distances of order of the Bohr radius $\sim 1/M_t v$. For a
quark-antiquark pair in a colour singlet state this
non-local interaction is nothing else than the instantaneous potential
of a quark-antiquark separated by a distance of order the inverse Bohr
radius. Generically the PNRQCD Lagrangian has the form
\begin{eqnarray}
{\cal{L}}_{\mbox{\tiny PNRQCD}} & = &
\tilde{\cal{ L}}_{\mbox{\tiny NRQCD}} \, + \,
\int d^3 \mbox{\boldmath $r$} \Big(\psi^\dagger \psi\Big)
  (\mbox{\boldmath $r$})\,V(\mbox{\boldmath $r$})\,
\Big(\chi^\dagger \chi\Big)(0)
\,,
\label{PNRQCDLagrangian}
\end{eqnarray}
where the tilde above ${\cal{ L}}_{\mbox{\tiny NRQCD}}$ on the RHS of
Eq.~(\ref{PNRQCDLagrangian}) indicates that the corresponding
operators only describe potential quark and ultrasoft gluonic
degrees of freedom. In addition, an expansion in momentum components
$\sim M_t v^2$ is understood. $V$ is the heavy quark-antiquark
potential and is given below. Using the velocity counting rules for
potential heavy quarks mentioned above we see that the LO contribution
to $V$, the Coulomb potential $-C_F\alpha_s/|\mbox{\boldmath $r$}|$,
counts as $v^2$, i.e. it is of the same order as the kinetic energy.
Thus, as is well known, in the non-relativistic limit, the Coulombic
interaction between the heavy 
quark pair has to be treated exactly rather than perturbatively.
We note that the expansion of the heavy quark currents in
terms of NRQCD currents (Eqs.~(\ref{vectorcurrentexpansion1}
)--(\ref{axialvectorcurrentexpansion2}))
is in general also affected by going from
NRQCD to PNRQCD, because also in the NRQCD currents soft and
potential gluonic degrees of freedom have to be integrated
out. However, at NNLO this does not affect the results displayed in 
Eqs.~(\ref{vectorcurrentexpansion1}) and
(\ref{axialvectorcurrentexpansion2}). The only (and fortunate)
practical consequence is that we can neglect the gluonic contribution
in the covariant derivatives.
Collecting all terms from the PNRQCD Lagrangian which contribute at NNLO,
i.e. count as $v^2$, $v^3$ or $v^4$, one can derive the following
equation of motion in momentum space representation for the Green
function of the time-independent 
Schr\"odinger equation, valid at NNLO in the non-relativistic expansion:
\begin{equation}
\bigg[\,
 \frac{\mbox{\boldmath $k$}^2}{M_t} - 
\frac{\mbox{\boldmath $k$}^4}{4M_t^3}
\,-\,\bigg(\,
\frac{p_0^2}{M_t} - \frac{p_0^4}{4 M_t^3}
\,\bigg)
\,\bigg]\,
\tilde G(\mbox{\boldmath $k$},\mbox{\boldmath $k$}^\prime;q^2) \,+\, 
\int\frac{d^3 \mbox{\boldmath $p$}^\prime}{(2\,\pi)^3}\,
\tilde V(\mbox{\boldmath $k$},\mbox{\boldmath $p$}^\prime)\,
\tilde G(\mbox{\boldmath $p$}^\prime,\mbox{\boldmath $k$}^\prime;q^2)
\, = \, 
(2\,\pi)^3\,\delta^{(3)}(\mbox{\boldmath $k$}-\mbox{\boldmath $k$}^\prime) 
\,,
\label{NNLOSchroedinger}
\end{equation}
where
\begin{eqnarray}
\tilde V(\mbox{\boldmath $k$},\mbox{\boldmath $k$}^\prime) & = & 
\tilde V_c(\mbox{\boldmath $k$}-\mbox{\boldmath $k$}^\prime) + 
\tilde V_{\mbox{\tiny BF}}(\mbox{\boldmath $k$},
  \mbox{\boldmath $k$}^\prime) + 
\tilde V_{\mbox{\tiny NA}}(\mbox{\boldmath $k$}-
  \mbox{\boldmath $k$}^\prime)
\,,
\label{NNLOpotential}
\end{eqnarray}
and
\begin{eqnarray}
p_0^2 & = & \frac{q^2}{4} - M_t^2
\,.
\end{eqnarray}
The parameter
$p_0$ is equal to the centre-of-mass three-momentum of the top quarks.
We have chosen this rather unusual representation for the energy
parameter since it greatly simplifies, because of its symmetric form
in Eq.~(\ref{NNLOSchroedinger}), the analytic matching calculations we
carry out in App.~\ref{appendixshortdistance}. The same trick has
already been used in~\cite{Hoang2} and later in
\cite{Hoang3,Melnikov3,Yakovlev1}. It is a non-trivial
fact that the PNRQCD operators, which describe
interactions of the heavy quarks with ultrasoft gluons, do not
contribute at NNLO, and that a conventional two-body Schr\"odinger
equation with an instantaneous potential is fully capable of resumming all
terms $\propto\alpha_s^m v^n$ that belong to NNLO. We come back to
this issue at the end of this section.
 
The individual potentials in momentum space representation read
($a_s\equiv\alpha_s(\mu)$, $C_A=3$, $C_F=4/3$,
$T=1/2$, 
$\mbox{\boldmath $Q$}\equiv\mbox{\boldmath $k$}-\mbox{\boldmath
  $k$}^\prime$): 
\begin{eqnarray}
\tilde V_c(\mbox{\boldmath $k$}) & = & 
-\,\frac{4\,\pi\,C_F\,a_s}{\mbox{\boldmath $k$}^2}\,
\bigg\{\, 1 +
\Big(\frac{a_s}{4\,\pi}\Big)\,\Big[\,
-\beta_0\,\ln\Big(\frac{\mbox{\boldmath $k$}^2}{\mu^2}\Big) + a_1
\,\Big]
\nonumber\\[2mm] & & \quad
 + \Big(\frac{a_s}{4\,\pi}\Big)^2\,\Big[\,
\beta_0^2\,\ln^2\Big(\frac{\mbox{\boldmath $k$}^2}{\mu^2}\Big)  
- \Big(2\,\beta_0\,a_1 +
\beta_1\Big)\,\ln\Big(\frac{\mbox{\boldmath $k$}^2}{\mu^2}\Big) 
+ a_2
\,\Big]
\,\bigg\}
\,,
\label{NNLOCoulomb}
\\[3mm]
\tilde V_{\mbox{\tiny BF}}(\mbox{\boldmath $k$},
  \mbox{\boldmath $k$}^\prime) & = &
\frac{\pi\,C_F\,a_s}{M_t^2} +
\frac{4\,\pi\,C_F\,a_s}{M_t^2}\,
\bigg[\,
\mbox{\boldmath $S_t$}\mbox{\boldmath $S_{\bar t}$} - 
\frac{(\mbox{\boldmath $Q$\boldmath $S_t$})
 (\mbox{\boldmath $Q$}\mbox{\boldmath $S_{\bar t}$})}
  {\mbox{\boldmath $Q$}^2}
\,\bigg]
\nonumber \\[2mm] & &
-\,\frac{\pi\,C_F\,a_s}{M_t^2}\,
\bigg[\,
\frac{(\mbox{\boldmath $k$}+\mbox{\boldmath $k$}^\prime)^2}
   {\mbox{\boldmath $Q$}^2} -
\frac{(\mbox{\boldmath $k$}^2-{\mbox{\boldmath $k$}^\prime}^2)^2}
   {\mbox{\boldmath $Q$}^4}
\,\bigg]
\nonumber\\[2mm] & &
+\,6\,i\,\frac{\pi\,C_F\,a_s}{M_t^2}\,
   \frac{(\mbox{\boldmath $S_t$}+\mbox{\boldmath $S_{\bar t}$})
   (\mbox{\boldmath $k$}\times\mbox{\boldmath $k$}^\prime)}{\mbox{\boldmath $Q$}^2}
\label{NNLOBF}
\,,\\[3mm]
\tilde V_{\mbox{\tiny NA}}(\mbox{\boldmath $k$}) & = &
-\,\frac{\pi^2\,C_A\,C_F\,a_s^2}{M_t\,|\mbox{\boldmath $k$}|} 
\,,
\label{NNLONA}
\end{eqnarray}
where
$\mbox{\boldmath $S_t$}$ and $\mbox{\boldmath $S_{\bar t}$}$
are the top and antitop spin operators and ($n_l=5$)
\begin{eqnarray}
\beta_0 & = & \frac{11}{3}\,C_A - \frac{4}{3}\,T\,n_l
\,,
\nonumber\\[2mm]
\beta_1 & = & \frac{34}{3}\,C_A^2 
-\frac{20}{3}C_A\,T\,n_l
- 4\,C_F\,T\,n_l
\,,
\nonumber\\[2mm]
a_1 & = &  \frac{31}{9}\,C_A - \frac{20}{9}\,T\,n_l
\,,
\nonumber\\[2mm]
a_2 & = & 
\bigg(\,\frac{4343}{162}+4\,\pi^2-\frac{\pi^4}{4}
 +\frac{22}{3}\,\zeta_3\,\bigg)\,C_A^2 
-\bigg(\,\frac{1798}{81}+\frac{56}{3}\,\zeta_3\,\bigg)\,C_A\,T\,n_l
\nonumber\\[2mm] & &
-\bigg(\,\frac{55}{3}-16\,\zeta_3\,\bigg)\,C_F\,T\,n_l 
+\bigg(\,\frac{20}{9}\,T\,n_l\,\bigg)^2
\,.
\end{eqnarray}
The constants $\beta_0$ and $\beta_1$ are the one- and two-loop
coefficients of the QCD beta function and $\gamma=0.577216\ldots\,$ 
is the
Euler constant; $V_c$ is the Coulomb (static) potential. Its
${\cal{O}}(\alpha_s)$ and ${\cal{O}}(\alpha_s^2)$ corrections, which
come from loops carrying soft momenta, have
been determined in~\cite{Fischler1,Billoire1} and
\cite{Schroeder1,Peter1},
respectively.\footnote{
The constant $a_2$ was first calculated in Ref.~\cite{Peter1}.
In Ref.~\cite{Schroeder1} an error in the coefficient of the term
$\propto \pi^2 C_A^2$ was corrected. 
} 
$V_{\mbox{\tiny BF}}$ is the Breit-Fermi potential known from
positronium. It describes the Darwin and the spin-orbit interactions
mediated by longitudinal gluons and the hyperfine 
interactions mediated by transverse gluons in the potential momentum
region. Then, $V_{\mbox{\tiny NA}}$ is a purely non-Abelian potential
generated (in Coulomb gauge) through a non-analytic term in a soft
momentum one-loop vertex correction to the Coulomb potential involving
the triple gluon vertex~\cite{Gupta1,Gupta2} (see also
Ref.~\cite{Kummer1}).

We have already noted that it is a non-trivial fact that ultrasoft
gluons do not contribute at NNLO, which means that a common
two-body Schr\"odinger equation, i.e. a wave equation containing an
instantaneous interaction potential, is indeed capable of resumming all
the terms that we count as NNLO.
Although this is a well accepted fact for positronium and the
hydrogen atom in QED~\cite{Bethe1}, it is not at all trivial to
understand, even in the QED
case, if one has to rely on arguments that are not in the framework of
(P)NRQCD. In QED the corrections caused by ultrasoft photons are known 
as retardation effects. The Lamb shift in hydrogen is the most famous
example. 
Based on the identification of the various momentum regions for
quarks and gluons mentioned above, however, one has transparent
power counting rules at hand; these show that the non-instantaneous
exchange of gluons among the top quarks does not lead to any effects
at NNLO. For the validity of the argument it is important that the
scale $M_t v^2$ is much larger than the typical 
hadronization scale $\Lambda_{QCD}$.\footnote{
Even for energies very close to threshold, the scale $M_t v^2$ cannot
become smaller than $\Gamma_t$ since the dominant effect of the top
width is to effectively shift the energy into the positive complex
plane by an amount $\Gamma_t$ (see Section~\ref{sectionwidth}).
}
To see that retardation effects are suppressed by at least three powers
of $v$ with respect to the non-relativistic limit (LO), we recall that
the only source of non-instantaneous interactions in PNRQCD are
the ultrasoft gluons, for which all momentum components are of order
$M_t v^2$. In addition, only transversely polarized gluons need to be
considered as ultrasoft, since we can work in the Coulomb gauge
where the time component of the longitudinal gluons vanishes. Thus, an
exchange of an ultrasoft gluon among the heavy quark-antiquark pair is
already suppressed by $v^2$ with respect to LO from the coupling of
transverse gluons to the heavy quarks. To see that an additional power
of $v$ arises from the loop integration over the ultrasoft gluon
momentum, we compare the $v$-counting of the integration measure and
the gluon propagator for ultrasoft and potential momenta. In the
ultrasoft case, the product of the integration measure $d^4 k$ and the
gluon propagator $1/k^2$ counts as $v^8\times v^{-4}=v^4$, whereas in the
potential case the result reads $v^5\times v^{-2}=v^3$. Because the
potential momenta contribute at LO we find that the ultrasoft gluons
can indeed only lead to effects beyond NNLO. Even if the gluon self
coupling is taken into account, this conclusion remains true, because
it also leads to additional powers of $v$ if ultrasoft momenta are
involved. We note that in arbitrary gauge the conclusions are true
only after all gauge cancellations have been taken into account. 
The relation $M_t v^2 \gg \Lambda_{QCD}$ is needed for the
above argumentation: otherwise, the coupling of ultrasoft gluons
to the heavy quarks or among themselves, $\alpha_s(M_t v^2)$, could be
of order $1$. Therefore our conclusion that retardation effects
do not contribute at NNLO would not be valid for the bottom
quark\footnote{
For the case of bottom-antibottom quark sum rules the conclusions
can, however, still be correct if the effective smearing range is
chosen larger than $\Lambda_{QCD}$.~\cite{Hoang5}
}
or even the charm quark case.

\vspace{1.5cm}
\section{Lippmann-Schwinger Equation}
\label{sectionlippmannschwinger}
The non-relativistic current correlators in the NRQCD factorization
formulae for the total top-antitop cross section close to threshold,
Eqs.~(\ref{vectorcrosssectionexpanded}) and
(\ref{axialvectorcrosssectionexpanded}), are directly related to the
Green function $\tilde G({\mbox{\boldmath $k$}},{\mbox{\boldmath
$k$}^\prime};q^2)$ of the Schr\"odinger
equation~(\ref{NNLOSchroedinger}), which describes off-shell elastic
scattering of a top-antitop pair with centre-of-mass three momentum
$\pm{\mbox{\boldmath $k$}}$ into a top-antitop pair with three
momentum $\pm{\mbox{\boldmath $k$}^\prime}$. We emphasize that the
Green function does not describe the scattering of on-shell top quarks
because the common three-dimensional formulation in the form of the
Schr\"odinger equation~(\ref{NNLOSchroedinger}) already contains an implicit
integration over the zero-components of the momenta in the heavy quark
propagators. Because the heavy quark potential is energy-independent
this integration is trivial by residues. In the
first part of this section we give the relations of the current
correlators to the Green function in the three-dimensional
formulation.
In the second part, we present the generalization to four dimensions 
for those results that are essential for a proper treatment
of the $W^+W^-b\bar b$ phase space once the top quark width is taken
into account. For this section we still assume that the top quark is
stable.

Taking into account the
partial wave decomposition of the Green function of the Schr\"odinger
equation~(\ref{NNLOSchroedinger})
\begin{eqnarray}
\tilde G({\mbox{\boldmath $k$}},{\mbox{\boldmath $k$}^\prime}) & = &
\sum_{l=0}^\infty 
\tilde G^l({\mbox{\boldmath $k$}},{\mbox{\boldmath $k$}^\prime})
\,,
\label{Greenfunctiondecomposition}
\end{eqnarray}
where $l$ is the total angular momentum quantum number, we find the
following relation between the NRQCD non-relativistic current
correlators~(\ref{correlatorV}) and (\ref{correlatorA}) and the $S$
and $P$ wave contributions to $\tilde G$:
\begin{eqnarray}
{\cal{A}}^v(q^2) & = & N_c\,\mbox{Tr}
\int\frac{d^3 \mbox{\boldmath $k$}}{(2\pi)^3}
\int\frac{d^3 \mbox{\boldmath $k$}^\prime}{(2\pi)^3}\,
\mbox{\boldmath $\sigma$}\, 
\bigg(1+\frac{\mbox{\boldmath $k$}^2}{6 M_t^2}\bigg)
\tilde G^0({\mbox{\boldmath $k$}},{\mbox{\boldmath $k$}^\prime})\,
\bigg(1+\frac{{\mbox{\boldmath $k$}^\prime}^2}{6 M_t^2}\bigg)\,
\mbox{\boldmath $\sigma$}
\nonumber\\[2mm] & = & 
6\,N_c\,
\int\frac{d^3 \mbox{\boldmath $k$}}{(2\pi)^3}
\int\frac{d^3 \mbox{\boldmath $k$}^\prime}{(2\pi)^3}\, 
\bigg(1+\frac{\mbox{\boldmath $k$}^2}{6 M_t^2}\bigg)
\tilde G^0({\mbox{\boldmath $k$}},{\mbox{\boldmath $k$}^\prime})\,
\bigg(1+\frac{{\mbox{\boldmath $k$}^\prime}^2}{6 M_t^2}\bigg)\,
\,,
\nonumber
\label{correlatorGreenV}
\\[4mm]
{\cal{A}}^a(q^2) & = & N_c\,\mbox{Tr}
\int\frac{d^3 \mbox{\boldmath $k$}}{(2\pi)^3}
\int\frac{d^3 \mbox{\boldmath $k$}^\prime}{(2\pi)^3}\,
\frac{\mbox{\boldmath $k$}\times \mbox{\boldmath $\sigma$}}{M_t}
\tilde G^1({\mbox{\boldmath $k$}},{\mbox{\boldmath $k$}^\prime})
\frac{\mbox{\boldmath $k$}^\prime\times \mbox{\boldmath $\sigma$}}{M_t}
\nonumber\\[2mm] & = &  4\,N_c\,
\int\frac{d^3 \mbox{\boldmath $k$}}{(2\pi)^3}
\int\frac{d^3 \mbox{\boldmath $k$}^\prime}{(2\pi)^3}\,
\frac{\mbox{\boldmath $k$}\mbox{\boldmath $k$}^\prime}{M_t^2}\,
\tilde G^1({\mbox{\boldmath $k$}},{\mbox{\boldmath $k$}^\prime})
\,,
\label{correlatorGreenA}
\end{eqnarray}
where a proper UV regularization is understood. 
For simplicity we have
dropped the energy argument of the Green function $\tilde G$. 
In this work we use the Lippmann-Schwinger equation, the Fourier
transform of Eq.~(\ref{NNLOSchroedinger}) with respect to
$\mbox{\boldmath $k$}^\prime$,
\begin{eqnarray}
\bigg[\,
 \frac{\mbox{\boldmath $k$}^2}{M_t} - 
\frac{\mbox{\boldmath $k$}^4}{4M_t^3}
\,-\,\bigg(\,
\frac{p_0^2}{M_t} - \frac{p_0^4}{4 M_t^3}
\,\bigg)
\,\bigg]\,
G(\mbox{\boldmath $k$},\mbox{\boldmath $x$}) 
\, = \, 
\exp(i\,\mbox{\boldmath $k$}\mbox{\boldmath $x$})
\,-\, 
\int\frac{d^3 \mbox{\boldmath $p$}^\prime}{(2\,\pi)^3}\,
\tilde V(\mbox{\boldmath $k$},\mbox{\boldmath $p$}^\prime)\,
G(\mbox{\boldmath $p$}^\prime,\mbox{\boldmath $x$})
\,,
\label{LippmannSchwinger}
\end{eqnarray}
to derive integral equations for
$\tilde G^0$ and $\tilde G^1$, which are then solved numerically. 
Using the partial
wave decomposition of $\exp(i\,\mbox{\boldmath $k$}\mbox{\boldmath
  $x$})$ 
($k\equiv|\mbox{\boldmath $k$}|$, $x\equiv|\mbox{\boldmath $x$}|$):
\begin{equation}
\exp(i \,\mbox{\boldmath $k$}\mbox{\boldmath $x$}) \, = \,
\exp(i\, k\,x\,\cos\theta) \, = \,
\sum_{l=0}^\infty i^l\,(2\,l+1)\,j_l(k\,x)\,P_l(\cos\theta)
\,,
\end{equation}
where $j_l$ are the spherical Bessel functions
$j_l(x)=(-x)^l(\frac{1}{x} \frac{d}{d x})^l [\frac{\sin x}{x}]$ and
$P_l$ the Legendre polynomials, one arrives at the following equations
for $G^0$ and $G^1$:
\begin{eqnarray}
\bigg[\,
 \frac{\mbox{\boldmath $k$}^2}{M_t} - 
\frac{\mbox{\boldmath $k$}^4}{4M_t^3}
\,-\,\bigg(\,
\frac{p_0^2}{M_t} - \frac{p_0^4}{4 M_t^3}
\,\bigg)
\,\bigg]\,
G^0(\mbox{\boldmath $k$},\mbox{\boldmath $x$})  & = &
\frac{\sin(k \,x)}{k \,x} \, - \,
\int\frac{d^3 \mbox{\boldmath $p$}^\prime}{(2\pi)^3}\,
\tilde V(\mbox{\boldmath $k$},\mbox{\boldmath $p$}^\prime)\,
G^0(\mbox{\boldmath $p$}^\prime,\mbox{\boldmath $x$})
\,,
\label{inteqG0}
\\[4mm]
\bigg[\,
 \frac{\mbox{\boldmath $k$}^2}{M_t}
\,-\,
\frac{p_0^2}{M_t} 
\,\bigg]\,
G^1(\mbox{\boldmath $k$},\mbox{\boldmath $x$}) & = &
3\, i\,\frac{\mbox{\boldmath $k$}\mbox{\boldmath $x$}}{k^2 \,x^2}\,
\bigg(\,\frac{\sin(k \,x)}{k \,x}-\cos(k \,x)\,\bigg) 
\nonumber
\\[2mm] & & - \,
\int\frac{d^3 \mbox{\boldmath $p$}^\prime}{(2\pi)^3}\,
\tilde V(\mbox{\boldmath $k$},\mbox{\boldmath $p$}^\prime)\,
G^1(\mbox{\boldmath $p$}^\prime,\mbox{\boldmath $x$})
\,.
\label{inteqG1}
\end{eqnarray}
Because in Eq.~(\ref{inteqG0}) only S-wave states are considered, one
can, instead of the complicated form of $\tilde V_{\mbox{\tiny BF}}$,
use the angular average with respect to the
angle between $\mbox{\boldmath $p^\prime$}$ and $\mbox{\boldmath $k$}$ 
of the $1/\mbox{\boldmath $Q$}^4$ term on the RHS of
Eq.~(\ref{NNLOBF}). Evaluating also the spin matrices for the
S-wave state, the Breit-Fermi potential simplifies to
\begin{eqnarray}
\tilde V^s_{\mbox{\tiny BF}}(\mbox{\boldmath $k$},
\mbox{\boldmath $k$}^\prime) & = &
\frac{1}{4\pi}\,\int d\Omega\,
\tilde V_{\mbox{\tiny BF}}(\mbox{\boldmath $k$},
\mbox{\boldmath $k$}^\prime)  
\nonumber
\\[2mm] & = & 
\frac{11}{3}\,\frac{\pi\,C_F\,a_s}{M_t^2}\,
-2\,\frac{\pi\,C_F\,a_s}{M_t^2}\,
\frac{\mbox{\boldmath $k$}^2+{\mbox{\boldmath $k$}^\prime}^2}
   {\mbox{\boldmath $Q$}^2} 
\,.
\end{eqnarray}
In Eq.~(\ref{inteqG1}), on the other hand,
we will just use the LO Coulomb potential because the axial-vector
contribution to the total cross section is already suppressed by
$v^2$. For the same reason we do not include any kinematic
relativistic corrections in Eq.~(\ref{inteqG1}).
Defining the S-wave and the P-wave vertex Green function as
\begin{eqnarray}
\label{Svertex}
S(\mbox{\boldmath $k$}) & = &
\int\frac{d^3 \mbox{\boldmath $p$}^\prime}{(2\,\pi)^3}\,
\tilde G^0(\mbox{\boldmath $k$},\mbox{\boldmath $p$}^\prime)\,
\bigg(\,1+\frac{{\mbox{\boldmath $p$}^\prime}^2}{6 M_t^2}\,\bigg)
\,,
\\[3mm]
\label{Pvertex}
P(\mbox{\boldmath $k$}) & = &
\int\frac{d^3 \mbox{\boldmath $p$}^\prime}{(2\,\pi)^3}\,
\frac{\mbox{\boldmath $k$}\mbox{\boldmath $p$}^\prime}
{\mbox{\boldmath $k$}^2}\,
\tilde G^1(\mbox{\boldmath $k$},\mbox{\boldmath $p$}^\prime)
\,,
\end{eqnarray}
we finally arrive at the integral equations for 
$S(\mbox{\boldmath $k$})$ and $P(\mbox{\boldmath $k$})$,
\begin{eqnarray}
S(\mbox{\boldmath $k$}) & = & 
G^f(\mbox{\boldmath $k$})\,
\bigg(\,1+\frac{{\mbox{\boldmath $k$}^2}}{6 M_t^2}\,\bigg)
\,-\, 
G^f(\mbox{\boldmath $k$})\,
\int\frac{d^3 \mbox{\boldmath $p$}^\prime}{(2\pi)^3}\,
\tilde V(\mbox{\boldmath $k$},\mbox{\boldmath $p$}^\prime)\,
S(\mbox{\boldmath $p$}^\prime)
\,,
\label{Sintegral}
\\[3mm]
P(\mbox{\boldmath $k$}) & = &  G^f(\mbox{\boldmath $k$})
\,-\,
G^f(\mbox{\boldmath $k$})\,
\int\frac{d^3 \mbox{\boldmath $p$}^\prime}{(2\pi)^3}\,
\frac{\mbox{\boldmath $k$} \mbox{\boldmath $p$}^\prime}
  {\mbox{\boldmath $k$}^2}\,
\tilde V_c^{\tiny \mbox{LO}}(\mbox{\boldmath $k$},
      \mbox{\boldmath $p$}^\prime)\,
P(\mbox{\boldmath $p$}^\prime)
\,,
\label{Pintegral}
\end{eqnarray}
where 
\begin{eqnarray}
G^f(\mbox{\boldmath $k$}) & = &
\frac{M_t}{\mbox{\boldmath $k$}^2-p_0^2-i\epsilon}\,\bigg[\,
1 + \frac{\mbox{\boldmath $k$}^2+p_0^2}{4\,M_t^2}
\,\bigg]
\label{Gfree}
\end{eqnarray}
is the free vertex function.  We note that
$(2\pi)^3\,\delta^{(3)}(\mbox{\boldmath $k$}-
\mbox{\boldmath $k$}^\prime)\,G^f(\mbox{\boldmath $k$})$ is the
Green function of Eq.~(\ref{NNLOSchroedinger}) for $\tilde V=0$.

The vertex functions $S(\mbox{\boldmath $k$})$ and $P(\mbox{\boldmath
$k$})$ only depend on the spatial momentum $\mbox{\boldmath
$k$}$. As mentioned at the beginning of this section, their dependence
on the time component $k^0$ has been eliminated by a trivial
integration by residues. For a proper integration over the phase space
of the top-antitop decay products, which is carried out in
Sec.~\ref{sectionwidth},
we also need the full dependence of the vertex functions on 
$k^0$. It can be recovered by comparing $G^f(\mbox{\boldmath $k$})$ to
the product of a free PNRQCD top and antitop propagator at NNLO
carrying momenta 
$k_{t}=(k^0+(\frac{p_0^2}{2M_t}-\frac{p_0^4}{8 M_t^3}),
\mbox{\boldmath $k$})$ and 
$k_{\bar t}=(k^0-(\frac{p_0^2}{2M_t}-\frac{p_0^4}{8 M_t^3}),
\mbox{\boldmath $k$})$, respectively\footnote{
This choice corresponds to a situation in top-antitop ladder diagrams
where half of the centre-of-mass energy is flowing through the top and
half through the antitop line, see Fig.~\ref{figroutingladder}.
}:
\begin{eqnarray}
G^f(k^0,\mbox{\boldmath $k$}) & \equiv &
\frac{i}{k^0+(\frac{p_0^2}{2M_t}-\frac{p_0^4}{8 M_t^3}) - 
  (\frac{\mbox{\boldmath $k$}^2}{2M_t}-
  \frac{\mbox{\boldmath $k$}^4}{8 M_t^3}) + i\epsilon }\,
\frac{i}{k^0-(\frac{p_0^2}{2M_t}-\frac{p_0^4}{8 M_t^3}) +
  (\frac{\mbox{\boldmath $k$}^2}{2M_t}-
  \frac{\mbox{\boldmath $k$}^4}{8 M_t^3}) - i\epsilon }
\nonumber
\\[2mm] & = &
\frac{-1}{{k^0}^2-
   \Big(\frac{p_0^2}{2M_t}-\frac{\mbox{\boldmath $k$}^2}{2M_t}
   +i\epsilon\Big)^2} + 
\frac{
\Big(\frac{p_0^2}{2M_t}-\frac{\mbox{\boldmath $k$}^2}{2M_t}\Big)^2
\Big(\frac{p_0^2}{2M_t^2}+\frac{\mbox{\boldmath $k$}^2}{2M_t^2}\Big)
}{\Big[{k^0}^2-
   \Big(\frac{p_0^2}{2M_t}-\frac{\mbox{\boldmath $k$}^2}{2M_t}
   +i\epsilon\Big)^2\Big]^2}
\,.
\label{Gfreecov}
\end{eqnarray}
We emphasize that an expansion of the NNLO
relativistic effects is understood.
The relation between $G^f(\mbox{\boldmath $k$})$ and 
$G^f(k^0,\mbox{\boldmath $k$})$ reads
\begin{eqnarray}
G^f(\mbox{\boldmath $k$}) & = &
-i\,\int\limits_{-\infty}^{+\infty}\frac{d k^0}{2\pi}\,
G^f(k^0,\mbox{\boldmath $k$})
\,.
\label{GthreetoGfourrelation}
\end{eqnarray}
Recalling that the potentials as well as the production and
annihilation vertex corrections do not depend on the zero-components
of the momenta, the integral equations for the generalized vertex
functions read
\begin{eqnarray}
S(k^0,\mbox{\boldmath $k$}) & = & 
G^f(k^0,\mbox{\boldmath $k$})\,
\bigg(\,1+\frac{{\mbox{\boldmath $k$}^2}}{6 M_t^2}\,\bigg)
\,+\,i\, 
G^f(k^0,\mbox{\boldmath $k$})\,
\int\frac{d^3 \mbox{\boldmath $p$}^\prime}{(2\pi)^3}\,
\int\limits_{-\infty}^{+\infty}\frac{d {p^\prime}^0}{2\pi}\,
\tilde V(\mbox{\boldmath $k$},\mbox{\boldmath $p$}^\prime)\,
S({p^\prime}^0,\mbox{\boldmath $p$}^\prime)
\,,
\label{Sintegralcov}
\\[3mm]
P(k^0,\mbox{\boldmath $k$}) & = &  
G^f(k^0,\mbox{\boldmath $k$})
\,+\,i\,
G^f(k^0,\mbox{\boldmath $k$})\,
\int\frac{d^3 \mbox{\boldmath $p$}^\prime}{(2\pi)^3}\,
\int\limits_{-\infty}^{+\infty}\frac{d {p^\prime}^0}{2\pi}\,
\frac{\mbox{\boldmath $k$} \mbox{\boldmath $p$}^\prime}{k^2}\,
\tilde V_c^{\tiny \mbox{LO}}(\mbox{\boldmath $k$},
      \mbox{\boldmath $p$}^\prime)\,
P({p^\prime}^0,\mbox{\boldmath $p$}^\prime)
\,.
\label{Pintegralcov}
\end{eqnarray}
The three- and four-dimensional versions of the vertex functions $S$
and $P$ are related by an equation similar to
Eq.~(\ref{GthreetoGfourrelation}). We note that, by construction, the
relation
\begin{eqnarray}
\frac{X(k^0,\mbox{\boldmath $k$})}{G^f(k^0,\mbox{\boldmath $k$})}
 & = &
\frac{X(\mbox{\boldmath $k$})}{G^f(\mbox{\boldmath $k$})}
\,,
\qquad
(X\, = \, S, P)
\,,
\label{amputatedrelation}
\end{eqnarray}
holds for the amputated vertex functions.

It is straightforward to formulate the optical theorem, which relates
the imaginary part of the correlators ${\cal{A}}^v$ and ${\cal{A}}^a$
to explicit phase space integrals over the modulus squared of the
vertex functions $S$ and $P$. The relations read
\begin{eqnarray}
\mbox{Im}\Big[\,{\cal{A}}^v(q^2)\,\Big]
& = & 12\,\pi^2\,N_c\,
\int\frac{d^3 \mbox{\boldmath $k$}}{(2\pi)^3}\,
\int\limits_{-\infty}^{+\infty}\frac{d k^0}{2\pi}\,
\bigg|\,
\frac{S(k^0,\mbox{\boldmath $k$})}{G^f(k^0,\mbox{\boldmath $k$})}
\,\bigg|^2\,
\delta\bigg(\frac{\mbox{\boldmath $k$}^2}{2 M_t}-
   \frac{\mbox{\boldmath $k$}^4}{8 M_t^3}-
\Big(\frac{p_0^2}{2 M_t}-\frac{p_0^4}{8 M_t^3}\Big) + k^0
\bigg)
\nonumber\\ & &
\hspace{2cm}
\times\,
\delta\bigg(\frac{\mbox{\boldmath $k$}^2}{2 M_t}-
   \frac{\mbox{\boldmath $k$}^4}{8 M_t^3}-
\Big(\frac{p_0^2}{2 M_t}-\frac{p_0^4}{8 M_t^3}\Big) - k^0
\bigg)
\nonumber\\[2mm]  & = & 
6\,\pi\,N_c\,
\int\frac{d^3 \mbox{\boldmath $k$}}{(2\pi)^3}\,
\bigg|\,
\frac{S(\mbox{\boldmath $k$})}{G^f(\mbox{\boldmath $k$})}
\,\bigg|^2\,
\delta\bigg(\frac{\mbox{\boldmath $k$}^2}{M_t}-
   \frac{\mbox{\boldmath $k$}^4}{4 M_t^3}-
\Big(\frac{p_0^2}{M_t}-\frac{p_0^4}{4 M_t^3}\Big)
\bigg)
\,,
\label{Sopticalstable}
\\[4mm]
\mbox{Im}\Big[\,{\cal{A}}^a(q^2)\,\Big]
& = & 8\,\pi^2\,N_c\,
\int\frac{d^3 \mbox{\boldmath $k$}}{(2\pi)^3}\,
\int\limits_{-\infty}^{+\infty}\frac{d k^0}{2\pi}\,
\frac{\mbox{\boldmath $k$}^2}{M_t^2}\,
\bigg|\,
\frac{P(k^0,\mbox{\boldmath $k$})}{G^f(k^0,\mbox{\boldmath $k$})}
\,\bigg|^2\,
\delta\bigg(\frac{\mbox{\boldmath $k$}^2}{2 M_t}
-
\frac{p_0^2}{2 M_t}
+ k^0
\bigg)
\nonumber\\ & &
\hspace{2cm}
\times\,
\delta\bigg(\frac{\mbox{\boldmath $k$}^2}{2 M_t}
-
\frac{p_0^2}{2 M_t}
- k^0
\bigg)
\nonumber\\[2mm]  & = & 
4\,\pi\,N_c\,
\int\frac{d^3 \mbox{\boldmath $k$}}{(2\pi)^3}\,
\frac{\mbox{\boldmath $k$}^2}{M_t^2}\,
\bigg|\,
\frac{P(\mbox{\boldmath $k$})}{G^f(\mbox{\boldmath $k$})}
\,\bigg|^2\,
\delta\bigg(\frac{\mbox{\boldmath $k$}^2}{M_t}
-
\frac{p_0^2}{M_t}
\bigg)
\,.
\label{Popticalstable}
\end{eqnarray}

\vspace{1.5cm}
\section{Regularization Scheme and Short-Distance Coefficients}
\label{sectionregularization}
All equations derived previously have to be considered in the
framework of a proper UV regularization scheme. In fact, UV
linear and logarithmic divergences arise in Eq.~(\ref{Sintegral}) from
the NNLO non-Coulombic potentials and from the kinetic energy and
vertex corrections. However, we emphasize that even in the case when
no UV divergences arise,
all integrals have to be consistently regularized, because only in a
consistent regularization scheme can the short-distance coefficients 
be defined properly. In principle, the regularization scheme ``of
choice'' would be an analytic scheme like $\overline{\mbox{MS}}$ as it
is usually used in modern perturbative QCD calculations. The
preference for the $\overline{\mbox{MS}}$ scheme arises from the fact
that it naturally preserves gauge invariance, Ward identities and,
particularly important in the framework of effective field theories,
power counting rules. Unfortunately an analytic solution of
Eqs.~(\ref{NNLOSchroedinger}), (\ref{Sintegral}) and (\ref{Pintegral})
is not even known for four dimensions. Thus, the only sensible way to
use the $\overline{\mbox{MS}}$ scheme is to start from the
known Coulomb solution of the non-relativistic Schr\"odinger equation
and include NLO and NNLO corrections via time-independent perturbation
theory, and then explicitly construct the spectral representation of
the Green function at the NNLO level. While this program
might still be feasible for the determination of the total cross
section, it is rather cumbersome for the calculations of distributions
at NNLO. 

\begin{figure}[t!] 
\begin{center}
\leavevmode
\epsfxsize=6cm
\leavevmode
\epsffile[240 420 440 460]{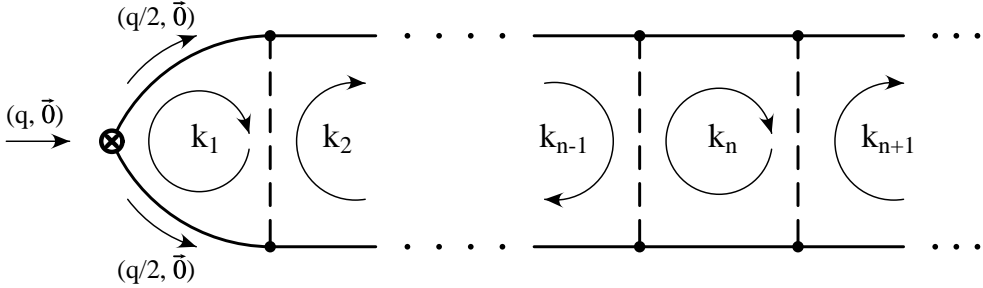}
%
%
\vskip  2.8cm
 \caption{\label{figroutingladder}
Routing convention for loop momenta in ladder diagrams.
}
 \end{center}
\end{figure}

\begin{figure}[t!] 
\begin{center}
\leavevmode
\epsfxsize=2.5cm
\leavevmode
\epsffile[240 420 440 460]{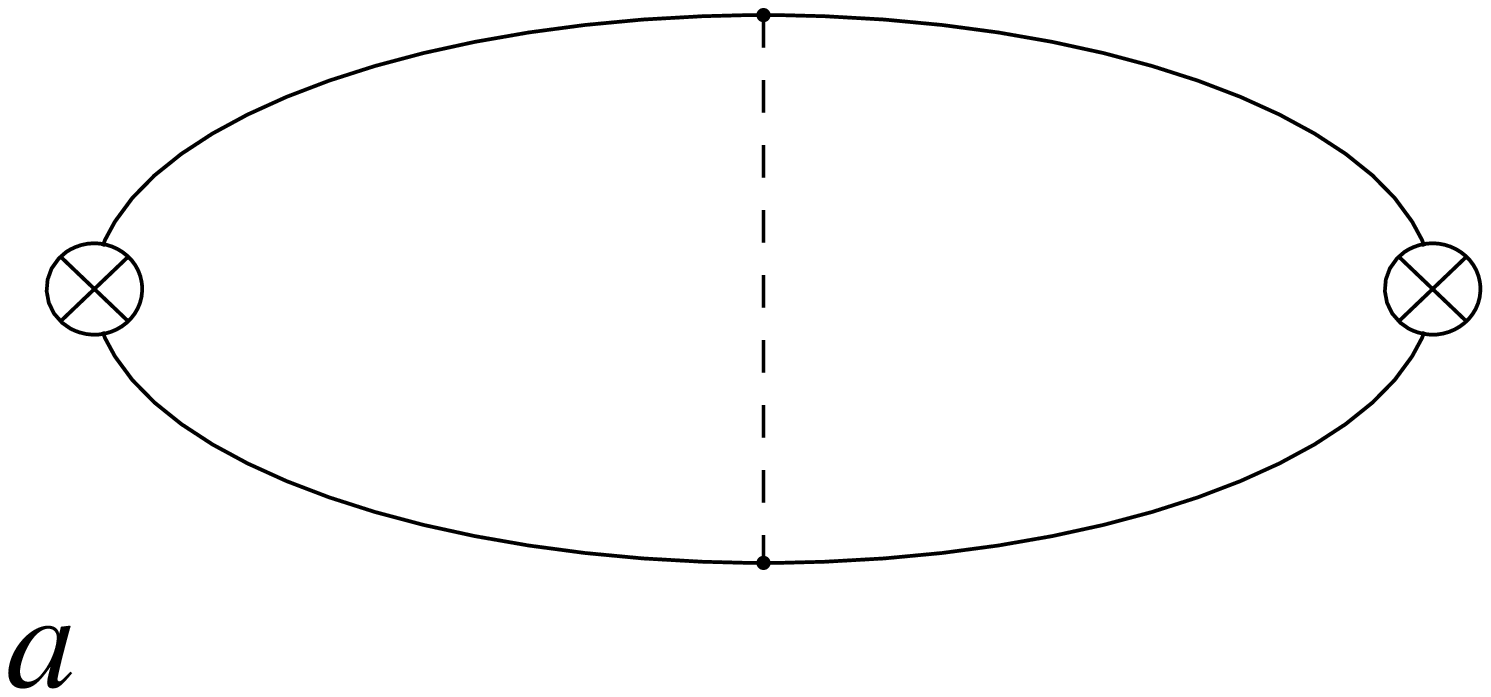}
\hspace{4cm}
\leavevmode
\epsfxsize=2.5cm
\leavevmode
\epsffile[240 420 440 460]{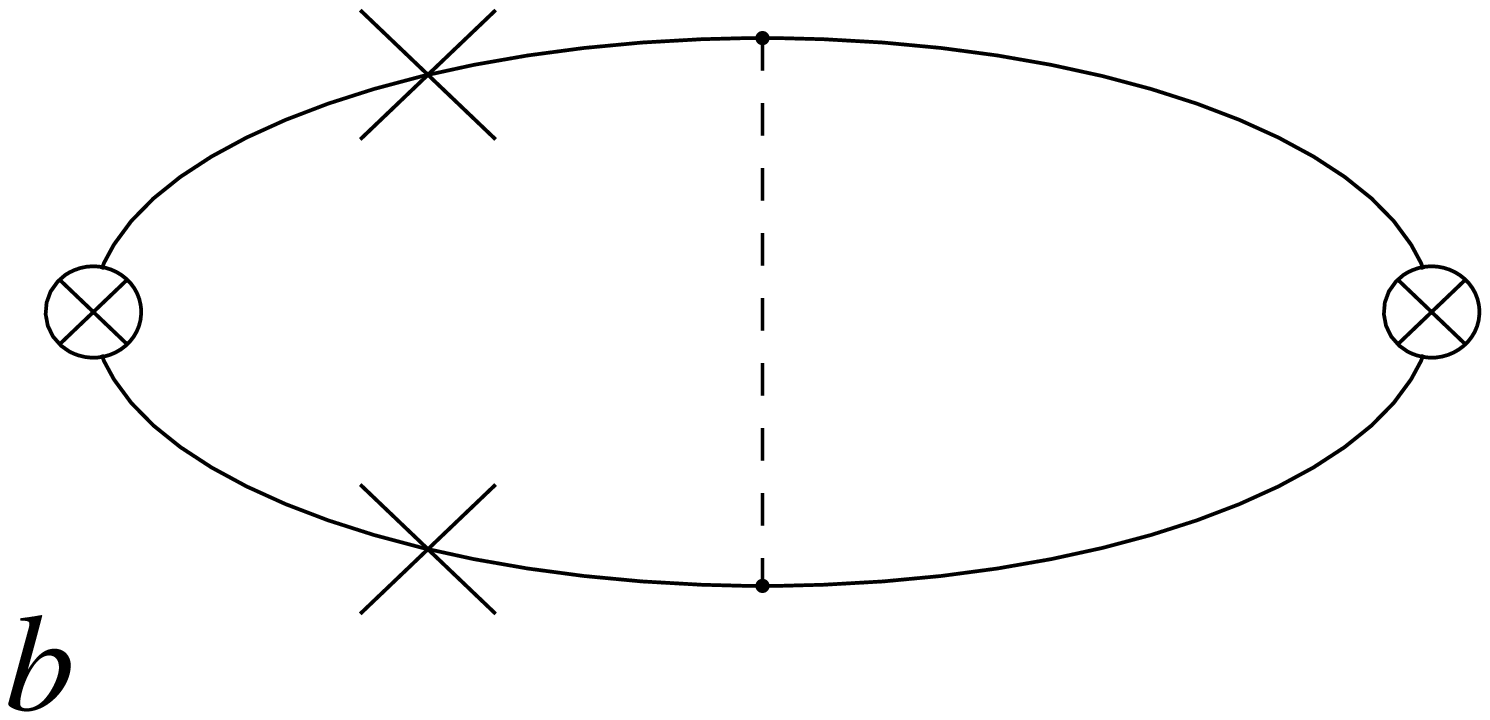}
%
%
\vskip  2.2cm
 \caption{\label{figcorrelatorCoulomb}
The two-loop NRQCD vector current correlator with the exchange of one
Coulomb gluon without relativistic corrections (a) and with the
kinetic energy corrections (b). 
}
 \end{center}
\end{figure}

In this work we use a momentum cutoff regularization scheme by simply
excluding momenta that have a spatial component larger than the
cutoff $\Lambda$. This is in fact the most natural regularization
scheme for a numerical solution of Eqs.~(\ref{Sintegral}) and
(\ref{Pintegral}). However, a cutoff scheme contains a number of
subtleties, which shall be briefly discussed in the following. As
indicated before, a cutoff scheme leads to violations of gauge
invariance and Ward identities in the (P)NRQCD calculation. These effects,
however, are generated at the cutoff and are therefore cancelled by
corresponding terms with a different sign in the short-distance
coefficients. Thus the cross section, which contains the proper
combination of non-relativistic correlators and short-distance
coefficients, is gauge-invariant and satisfies all Ward identities up
to terms beyond the order at which the matching calculation has been
carried out. Another subtlety of a cutoff prescription is that it is
only well defined if a specific routing convention for the momenta in
loops is adopted. For the calculations of the top-antitop cross
section at NNLO, it is straightforward and easy to find such a routing
convention, because only ladder-type diagrams are involved. It is
natural to choose the routing used in the integral
equations~(\ref{Sintegral}) and (\ref{Pintegral}), which we have, for
clarity, depicted graphically in Fig~\ref{figroutingladder}.
It is important to use exactly the same routing for the (P)NRQCD diagrams
calculated in the matching procedure to obtain consistent results for
the NNLO short-distance coefficients. Finally, it has to be mentioned
that a cutoff scheme inevitably leads to power counting
breaking effects. This means in our case that a term in
the Schr\"odinger equation~(\ref{NNLOSchroedinger}), which is NNLO
according to the velocity counting, can in principle
lead to lower order contributions in the non-relativistic
current correlator. Like the terms that violate gauge invariance and
Ward identities, the power counting breaking terms are also generated
at the cutoff; they are therefore cancelled
in the combination of the correlators and the corresponding
short-distance coefficients. However, all this happens only if the cutoff
is chosen of the order of $M_t$. To illustrate this issue let us first
consider the LO two-loop NRQCD vector current correlator, which
contains the exchange of a Coulomb gluon as shown in
Fig.~\ref{figcorrelatorCoulomb}a.
It is straightforward to calculate
the absorptive part of the diagram as an expansion in $p_0$ (see
App.~\ref{appendixshortdistance}):
\begin{eqnarray}
I_1^{(1)}  & = & 
\frac{C_F\,\alpha_s\,M_t^2}{4\, \pi^2}\,\bigg[\,
\frac{\pi^2}{2}-\frac{4\,p_0}{\Lambda} + 
{\cal{O}}(p_0^3)
\,\bigg]
\,.
\label{2loopCoulomb}
\end{eqnarray}
The first term in the brackets on the RHS of Eq.~(\ref{2loopCoulomb})
is the well
known Coulomb singularity, which leads to a finite cross section at
order $\alpha_s$. The second term is cutoff-dependent and leads to a
short-distance correction $\propto
\frac{\alpha_s}{\pi}\frac{M_t}{\Lambda}$. (In
App.~\ref{appendixshortdistance} the reader can convince her/himself
that the scale of $\alpha_s$ in this term is indeed of order $M_t$.)
Obviously, to avoid a breakdown in the separation of long- and
short-distance contributions\footnote{
In this context ``long-distance'' effects are not understood as
``non-perturbative'' effects, which come from scales of order
$\Lambda_{QCD}$ , but rather as effects governed by scales of order
$M_t v$ or $M_t v^2$.
},
$\Lambda$ has to be chosen of order $M_t$. (Choosing $\Lambda$ of order
$M_t v$ would be absurd anyway, because this would cut off a large part
of the dynamics. In this respect the cutoff $\Lambda$ acts in a way
completely different from the
``cutoff'' scale in the  $\overline{\mbox{MS}}$ scheme, which
would naturally be chosen of order $M_t v$.)
Let us now consider the kinetic energy corrections in the same
two-loop NRQCD diagram, as shown in
Fig.~\ref{figcorrelatorCoulomb}b. Using the result obtained in
App.~\ref{appendixshortdistance} the expression for the diagram
including combinatorial factors reads
\begin{eqnarray}
I_3^{(1)} & = & 
\frac{C_F\,\alpha_s\,M_t^2}{4\, \pi^2}\,\bigg[\, 
\frac{\Lambda\,p_0}{M_t^2} +
\frac{p_0^2\,\pi^2}{2\,M_t^2} +  
{\cal{O}}( p_0^3 )
\,\bigg]
\,.
\label{2loopCoulombkinetic}
\end{eqnarray}
As expected from power counting arguments, the kinetic energy
correction leads to a contribution suppressed by $p_0^2/M_t^2$ with
respect to the pure Coulomb exchange diagram in
Eq.~(\ref{2loopCoulomb}). However, there is also a term proportional
to $p_0/M_t$ because $\Lambda$ is of order $M_t$. This is an example
for a power counting breaking term. In the short-distance coefficient
$C^v$ this term leads to a NNLO contribution
$\propto\alpha_s\frac{\Lambda}{M_t}$. 
Because this term only
arises if NNLO relativistic effects are taken into account, we have to
count it as NNLO. 
Similar terms are caused by the
Breit-Fermi potential $V_{\mbox{\tiny BF}}$, the non-Abelian potential
$V_{\mbox{\tiny NA}}$ and the dimension-5 NRQCD vector current. We
emphasize again that, as is the case for the terms that violate gauge
invariance and Ward identities, all power counting breaking terms are
automatically cancelled in the combination with the short-distance
coefficients up to terms beyond the order at which one carries out the
matching procedure. In our case, where the matching is carried out at
order $\alpha_s^2$, power counting breaking terms of order $\alpha_s^3$
remain uncancelled, but they are beyond NNLO accuracy. 

Taking into account the issues discussed above, it is straightforward
to determine the matching coefficient $C^v$. Details of
this calculation are given in App.~\ref{appendixshortdistance}. 
The result reads ($a_s\equiv\alpha_s(\mu)$)
\begin{eqnarray}
C^v & = & 1 + 
\bigg\{\,
  \frac{4\,C_F\,a_s}{\pi}\,
\bigg[\,
-1 + \frac{M_t}{\Lambda}
\,\bigg]
\,\bigg\}^{\mbox{\tiny NLO}}
\nonumber
\\[3mm] & &
+ \,\bigg\{\,
\frac{4\,C_F\,a_s\,\Lambda}{3\,\pi\,M_t}
 +
\frac{C_F^2\,a_s^2}{\pi^2}\,\bigg[\,
\frac{\beta_0}{C_F}\,\bigg(\,
- \frac{\Lambda^2 + 12\,M_t^2}{6\,\Lambda\,M_t} 
+ \frac{\Lambda^2 - 12\,M_t^2}
     {6\,\Lambda\,M_t}\,\ln\Big(\frac{\Lambda}{\mu}\Big) 
+ 2\,\ln\Big(\frac{M_t}{\mu}\Big)
\,\bigg)
\nonumber
\\[2mm] & &
- \frac{a_1}{C_F}\,\frac{\Lambda^2 - 12\,M_t^2}{12\,\Lambda\,M_t} 
+ \pi^2\,\bigg(\,\frac{2}{3}+\frac{C_A}{C_F}\,\bigg)\,
    \ln\Big(\frac{2\,M_t}{\Lambda}\,\Big)
+ \frac{\pi^2\,\kappa}{C_F^2} 
\nonumber
\\[2mm] & &
+ \frac{16\,\Lambda^2}{9\,M_t^2} 
- \frac{16\,\Lambda}{3\,M_t} 
- \frac{16\,M_t}{\Lambda} 
+ \frac{M_t^2\,(20 + \pi^2)}{2\,\Lambda^2}
- \frac{53\,\pi^2}{24} - \frac{C_A\,\pi^2}{C_F} + \frac{25}{6} + 
\frac{7}{3}\zeta_3
\,\bigg]
\,\bigg\}^{\mbox{\tiny NNLO}}
\,.
\label{CVshortdistance}
\end{eqnarray}
As explained in App.~\ref{appendixshortdistance} we have displayed the
NLO and NNLO contributions to $C^v$ separately. 
Appendix~\ref{appendixshortdistance} also contains a discussion on
the convergence of the perturbative series in $C^v$ in our
cutoff scheme compared to the  $\overline{\mbox{MS}}$
scheme~\cite{Melnikov4,Beneke2}.

\vspace{1.5cm}
\section{Top Quark Width}
\label{sectionwidth}
Up to now we have treated the top quark as a stable particle. For top
quark production close to threshold this is not appropriate because
the kinetic energy of the top quark $\sim M_t v^2\sim M_t
\alpha_s^2\sim$~2--3~GeV
is of the same order as its decay width. It has been shown by Fadin and
Khoze~\cite{Fadin1} that in the non-relativistic limit the top width
can be consistently implemented by calculating the cross section for
stable
top quarks supplemented by the replacement $E\to E+i \Gamma_t$, where
$E$ is the centre-of-mass energy measured with respect to the
two-particle threshold. For the total cross section, this prescription
remains valid even at NLO because, in this case, order $\alpha_s$ QCD
radiative corrections in form of a gluon that connects top quark
production and decay vanish~\cite{Melnikov1}. A consistent
implementation of the top width at NNLO has been missing so far. 

In the framework of NRQCD, and if one is not interested in any
differential information on the top decay products, the top width can
be understood as a modification of the NRQCD matching conditions
caused by electroweak corrections. Because the particles involved in
these corrections can be lighter than the top quark (i.e. if the top
quark can decay weakly) they can lead to non-zero imaginary parts in
the matching conditions and, likewise, in the short-distance
coefficients of the NRQCD Lagrangian and the NRQCD currents. This is a
well known concept in quantum mechanics of inelastic processes where
particle decay and absorption processes are represented by potentials
and couplings carrying complex coefficients, if one is not
interested in the details of the decay and absorption process. In this
context the effects of the top quark decay are only a small (but
nevertheless the
most important) part of a whole array of electroweak corrections
relevant to top-antitop quark pair production close to threshold. In
this work we only consider the effects from the on-shell top decay
into a $W$ boson and a bottom quark, assuming that the $W$ and the $b$
are themselves stable. Thus the results presented here are, by
definition, gauge-invariant. The consistent treatment of all
electroweak effects, including a proper handling of the off-shell decays
of the top quark, interconnection effects and of
gauge-invariance-violating contributions,
is beyond the scope of this paper. Such a treatment is carried out
in~\cite{BenekeHoang}.

Before we determine the modifications of the NRQCD
Lagrangian, Eq.~(\ref{NRQCDLagrangian}), through the top decay, some
remarks about the velocity counting of the top quark width are in
order. Comparing the numerical size of the top width
\begin{equation}
\Gamma_t \, \approx \,  \frac{G_F}{\sqrt{2}}\frac{M_t^3}{8 \pi}
\sim \, 1.5~\mbox{GeV}
\end{equation}
with the binding energy of a fictitious toponium $1S$ bound state
in the pole mass scheme ($\alpha_s\sim 0.15$) 
\begin{equation}
E_{1S} \approx \frac{M_t (C_F \alpha_s)^2}{4} \, \sim \,
1.75~\mbox{GeV}
\,,
\end{equation}
we find that we have to count $\Gamma_t/M_t$ as order $v^2$, i.e.
\begin{equation}
\frac{\Gamma_t}{M_t} \, \sim \,
\alpha_s^2 \, \sim \, v^2
\,.
\label{gammavcounting}
\end{equation}
Recalling the velocity counting of the operators of the NRQCD
Lagrangian, this means that at NNLO and to first order in
$\Gamma_t/M_t$ we have to determine the coefficients of the NRQCD
operators $\psi^\dagger i \Gamma_t \psi$,  
$\psi^\dagger i\frac{\Gamma_t}{M_t} D_t \psi$, 
$\psi^\dagger i \Gamma_t \frac{\mbox{\boldmath $D$}^2}{M_t^2} \psi$ and
those operators, where the top quark Pauli spinors are replaced by the
antitop ones. As we show later in this section there are also
contributions proportional to $i\Gamma_t$ to the 
photon and Z boson wave function renormalization constants,
which have to be included to account for
a proper treatment of the top quark decay phase space. We note that
the velocity counting~(\ref{gammavcounting}) implies that
$g\sim\alpha_s\sim v$, where $g$ is the SU(2) gauge coupling. Thus in
a complete calculation of all electroweak effects one 
also has to determine ${\cal{O}}(g^4)$ electroweak contributions and
${\cal{O}}(\alpha_s^2)$ QCD corrections to the top decay width. In the
framework of the Standard Model those corrections are
known~\cite{Jezabek2,Czarnecki1,Denner1}. For our purposes, however,
it is sufficient to
consider the top quark width as an independent parameter, which we
treat only to first order and which is not subject to higher-order
corrections. 

The matching coefficients of the bilinear top spinor
operators given above can be obtained by sandwiching the absorptive
part of the top quark self energy $\Sigma_t$ (in the full electroweak
theory) between 
top quark Dirac spinors and expanding the result around the complex
pole position. Keeping only the terms proportional to $\Gamma_t$, the
result for the bilinear top Pauli spinor terms reads
($k=(k^0,\mbox{\boldmath $k$})$)
\begin{eqnarray}
\bar u(\mbox{\boldmath $k$})\,\bigg[\,
\mbox{Im} \Sigma_t(k)
\,\bigg]\, u(\mbox{\boldmath $k$}) & \Rightarrow &
\bar u(\mbox{\boldmath $k$})\,i \frac{\Gamma_t}{2}\, 
u(\mbox{\boldmath $k$})
\nonumber\\ & = &
\tilde \psi^\dagger\,\bigg[\,
i\,\frac{\Gamma_t}{2}\,\frac{M_t}{E_k}
\,\bigg]\,\tilde\psi \, = \, 
\tilde \psi^\dagger\,\bigg[\,
i\,\frac{\Gamma_t}{2}\,\bigg(\,
1 - \frac{\mbox{\boldmath $k$}^2}{2\,M_t^2}
\,\bigg)\,\bigg]\,\tilde\psi
\,.
\end{eqnarray} 
The
corresponding result for the bilinear antitop Pauli spinor terms reads
\begin{eqnarray}
\bar v(-\mbox{\boldmath $k$})\,\bigg[\,
\mbox{Im} \Sigma_t(k)
\,\bigg]\, v(-\mbox{\boldmath $k$}) & \Rightarrow &
\bar v(-\mbox{\boldmath $k$})\,i \frac{\Gamma_t}{2}\, 
v(-\mbox{\boldmath $k$})
\nonumber\\ & = &
-\,\tilde \chi^\dagger\,\bigg[\,
i\,\frac{\Gamma_t}{2}\,\frac{M_t}{E_k}
\,\bigg]\,\tilde\chi \, = \, 
-\,\tilde \chi^\dagger\,\bigg[\,
i\,\frac{\Gamma_t}{2}\,\bigg(\,
1 - \frac{\mbox{\boldmath $k$}^2}{2\,M_t^2}
\,\bigg)\,\bigg]
\,\tilde\chi
\,,
\end{eqnarray} 
where $E_k\equiv(\mbox{\boldmath $k$}^2+M_t^2)^{1/2}$ and
\begin{equation}
u(\mbox{\boldmath $k$}) \, = \, 
\sqrt{\frac{E_k+M_t}{2 E_k}}\, 
\left(\begin{array}{cc} \tilde\psi \\ 
    \frac{\mbox{\boldmath $\sigma$}\mbox{\boldmath $k$}}{E_k+M_t}\,
     \tilde\psi
      \end{array}
\right)\,,
\hspace{1cm}
v(-\mbox{\boldmath $k$}) \, = \, 
\sqrt{\frac{E_k+M_t}{2 E_k}}\, 
\left(\begin{array}{cc}
    -\frac{\mbox{\boldmath $\sigma$}\mbox{\boldmath $k$}}{E_k+M_t}\,
     \tilde\chi\\
     \tilde\chi 
      \end{array}
\right)\,,
\end{equation}
using the usual non-relativistic normalization for Dirac spinors. 
Thus in the presence of top decay we have to modify the NRQCD
Lagrangian by adding the terms
\begin{eqnarray}
\delta {\cal{L}}_{\mbox{\tiny NRQCD}} & = &
\psi^\dagger\,i\,\frac{\Gamma_t}{2}\,\bigg[\,1 + 
 \frac{\mbox{\boldmath $D$}^2}{2\,M_t^2}
\,\bigg]\,\psi - 
\chi^\dagger\,i\,\frac{\Gamma_t}{2}\,\bigg[\,1 + 
 \frac{\mbox{\boldmath $D$}^2}{2\,M_t^2}
\,\bigg]\,\chi
\,.
\end{eqnarray}
Physically, the dimension-5 operators multiplying the top width
correspond to the time dilatation correction.
This leads to the following modified versions of the Schr\"odinger
equation~(\ref{NNLOSchroedinger}) in momentum space representation
\begin{eqnarray}
\lefteqn{
\bigg[\,
 \frac{\mbox{\boldmath $k$}^2}{M_t} - 
\frac{\mbox{\boldmath $k$}^4}{4M_t^3}
\,-\,\bigg(\,
\frac{p_0^2}{M_t} - \frac{p_0^4}{4 M_t^3}
\,\bigg)
\,-\,i\,\Gamma_t\,\bigg(\,
1 - \frac{\mbox{\boldmath $k$}^2}{2\,M_t^2}
\,\bigg)\,
\,\bigg]\,
\tilde G(\mbox{\boldmath $k$},\mbox{\boldmath $k$}^\prime;q^2) 
}
\nonumber
\\[3mm] & &
 + 
\int\frac{d^3 \mbox{\boldmath $p$}^\prime}{(2\,\pi)^3}\,
\tilde V(\mbox{\boldmath $k$},\mbox{\boldmath $p$}^\prime)\,
\tilde G(\mbox{\boldmath $p$}^\prime,\mbox{\boldmath $k$}^\prime;q^2)
\, = \,
(2\,\pi)^3\,\delta^{(3)}(\mbox{\boldmath $k$}-\mbox{\boldmath
  $k$}^\prime) 
\,,
\label{NNLOSchroedingergamma}
\end{eqnarray}
and the free vertex functions $G^f(\mbox{\boldmath $k$})$ and 
$G^f(k^0,\mbox{\boldmath $k$})$ in the integral
equations~(\ref{Sintegral}), (\ref{Pintegral}), (\ref{Sintegralcov})
and (\ref{Pintegralcov}):
\begin{eqnarray}
G^f(\mbox{\boldmath $k$}) & = &
\frac{M_t}{\mbox{\boldmath $k$}^2-p_0^2-\frac{\Gamma_t^2}{4}-
         i\,M_t\,\Gamma_t}\,\bigg[\,
1 + \frac{\mbox{\boldmath $k$}^2+p_0^2}{4\,M_t^2}-
i\,\frac{\Gamma_t}{4\,M_t}
\,\bigg]
\,,
\label{Gfreegamma}
\\
G^f(k^0,\mbox{\boldmath $k$}) & = &
\frac{-1}{{k^0}^2-
   \Big(\frac{1}{2M_t}(p_0^2+\frac{\Gamma_t^2}{4})-
        \frac{\mbox{\boldmath $k$}^2}{2M_t}
        +i\frac{\Gamma_t}{2}\Big)^2} + 
\frac{
\Big(\frac{p_0^2}{2M_t}-\frac{\mbox{\boldmath $k$}^2}{2M_t}
            +i\frac{\Gamma_t}{2}\Big)^2
\Big(\frac{p_0^2}{2M_t^2}+\frac{\mbox{\boldmath $k$}^2}{2M_t^2}
            -i\frac{\Gamma_t}{2M_t}\Big)
}{\Big[{k^0}^2-
   \Big(\frac{1}{2M_t}(p_0^2+\frac{\Gamma_t^2}{4})-
        \frac{\mbox{\boldmath $k$}^2}{2M_t}
        +i\frac{\Gamma_t}{2}\Big)^2
\Big]^2}
\,.
\nonumber\\
\label{Gfreecovgamma}
\end{eqnarray}
We note that it is not coercive to keep the term
$-\frac{\Gamma_t^2}{4}$ in the LO propagator, because it is of NNLO
according to the power counting. We have adopted this convention
because this choice leads to a
simplification of the analytic form for the rest of the NNLO
corrections in Eqs.~(\ref{Gfreegamma}) and (\ref{Gfreecovgamma}).
From the modified version of the Schr\"odinger
equation~(\ref{NNLOSchroedingergamma}) we can immediately see that the
Fadin-Khoze replacement rule is valid at LO and NLO in the non-relativistic
expansion, but it is inappropriate at NNLO. In fact it is not possible at
all to consistently implement the top quark width at NNLO simply by
shifting the centre-of-mass energy in a calculation for
stable quarks.
The resulting form of the optical theorem
relations reads
\begin{eqnarray}
\mbox{Im}\Big[\,{\cal{A}}^v(q^2)\,\Big]
& = & 3\,N_c\,
\int\frac{d^3 \mbox{\boldmath $k$}}{(2\pi)^3}\,
\int\limits_{-\infty}^{+\infty}\frac{d k^0}{2\pi}\,
|S(k^0,\mbox{\boldmath $k$})|^2\,
\Gamma_t^2\,\bigg(1 - \frac{\mbox{\boldmath $k$}^2}{M_t^2}
\,\bigg)
\nonumber\\[2mm]  & = & 
6\,N_c\,
\int\frac{d^3 \mbox{\boldmath $k$}}{(2\pi)^3}\,
|S(\mbox{\boldmath $k$})|^2\,
\Gamma_t\,\bigg(1 - \frac{\mbox{\boldmath $k$}^2}{2\,M_t^2}
\,\bigg)
\,,
\label{Sopticalunstable}
\\[4mm]
\mbox{Im}\Big[\,{\cal{A}}^a(q^2)\,\Big]
& = & 2\,N_c\,
\int\frac{d^3 \mbox{\boldmath $k$}}{(2\pi)^3}\,
\int\limits_{-\infty}^{+\infty}\frac{d k^0}{2\pi}\,
\frac{\mbox{\boldmath $k$}^2}{M_t^2}\,
|P(k^0,\mbox{\boldmath $k$})|^2\,\Gamma_t^2
\nonumber\\[2mm]  & = & 
4\,N_c\,
\int\frac{d^3 \mbox{\boldmath $k$}}{(2\pi)^3}\,
\frac{\mbox{\boldmath $k$}^2}{M_t^2}\,
|P(\mbox{\boldmath $k$})|^2\,\Gamma_t
\,.
\label{Popticalunstable}
\end{eqnarray}
The first equality in
Eqs.~(\ref{Sopticalunstable}) and (\ref{Popticalunstable}) is an
explicit integration over the phase space of the top decay products
for off-shell top decay keeping only the term proportional to the top
and antitop width for the top decay sub phase space.
We emphasize again that in Eqs.~(\ref{Sopticalunstable}) and
(\ref{Popticalunstable}) no physical phase space boundaries for the
integration over the top-antitop four momentum are implemented.
All integrations are defined in the
framework of the regularization scheme. 

The reader might ask whether the inclusion of the top quark width
into the top quark propagators
leads to non-trivial modifications of the NRQCD short-distance
coefficient $C^v$.\footnote{
One could equally well ask whether there is a need to introduce the
operators $\frac{\Gamma_t}{M_t} \psi^\dagger \mbox{\boldmath
  $\sigma$}\chi$ and  $\frac{\Gamma_t}{M_t} \chi^\dagger \mbox{\boldmath
  $\sigma$}\psi$ in the non-relativistic expansion of the currents,
which produce and annihilate the top-antitop pair.
At this point we prefer the language used in the text.
} 
Because $\frac{\Gamma_t}{M_t}\sim
v^2\sim\alpha_s^2$, it is not
inconceivable that there could be terms $\propto\frac{\Gamma_t}{M_t}$
in $C^v$ at NNLO. To show that this is not the case we recall that
$C^v$ is the modulus squared of the short-distance coefficient $c_1^v$
of the non-relativistic current 
$\psi^\dagger\mbox{\boldmath $\sigma$}\chi$ in
Eq.~(\ref{vectorcurrentexpansion1}). As mentioned in
App.~\ref{appendixshortdistance},
the short-distance coefficient $c_1^v$
is the sum of amputated vertex diagrams in the full theory where the
three momenta in the loop integrations are larger than the cutoff
$\Lambda\sim M_t$ for $\sqrt{s}=2 M_t$. Thus, there is no integration
over the top-antitop pole located at three momenta $\sim
p_0\sim M_t v$, and we can conclude that the short-distance
coefficient does not contain non-analytic terms involving the top
quark width. Therefore $c_1^v$ 
can be expanded in $i\Gamma_t$. For the same reason, the
coefficients of an expansion in $i\Gamma_t$ do not contain a
top-antitop cut and are real numbers. 
Because $C^v$ is the modulus squared of $c_1^v$, the first
non-vanishing term of an expansion of $C^v$ is proportional to
$(\Gamma_t/M_t)^2\sim\alpha_s^4$, which is indeed beyond
NNLO. However, we
emphasize that, in a complete calculation of all electroweak effects,
$C^v$ will receive corrections $\propto g^2, {g^\prime}^2\sim G_F
M_W^2$, $g^\prime$ being the U(1) gauge coupling, which are formally
of order $\Gamma_t/M_t$. These corrections do not come from the width
contained in the top and antitop propagators but for instance from
electroweak corrections to the vertices $\gamma, Z\to t\bar
t$~\cite{Guth1}. These corrections exist even in the case when the top
quark is treated as a stable particle.

If the top quark width is included in the NRQCD framework there is
only one additional source of a linear dependence on $\Gamma_t$ which
comes from the phase space integration in
Eqs.~(\ref{Sopticalunstable}) and (\ref{Popticalunstable}). In
contrast to the case of a stable top quark, where the phase space is
restricted to those top quark four momenta allowed by the centre-of-mass
energy, the physical boundaries of the phase space integration
$\int\frac{d^4 k}{(2\pi)^4}$ in the case of
unstable top quarks
are determined by the allowed invariant masses of the top quark decay
products. Assuming that the bottom quark is massless and that bottom
quark and $W$ boson are stable, and taking into account our routing
convention (see Fig.~\ref{figroutingladder}), the boundaries of the
physical four-dimensional phase space
integrations read ($k^2\equiv|\mbox{\boldmath $k$}|^2$)
\begin{equation}
\int\limits_{\tiny\mbox{phase space}}\frac{d^4 k}{(2\pi)^4} \, = \,
\int\limits_0^{\sqrt{\frac{q^2}{4}-M_W^2}}\,
\frac{d^3 \mbox{\boldmath $k$}}{(2\pi)^3}
\int\limits_{-(\frac{\sqrt{q^2}}{2}-\sqrt{k^2+M_W^2})}^{
  (\frac{\sqrt{q^2}}{2}-\sqrt{k^2+M_W^2})}\,\frac{d k^0}{2\pi}
\,.
\label{phasespacephysical}
\end{equation}
It is obvious that the physical
limits of integration are not equivalent to the actual limits of
integration on the RHS of Eqs.~(\ref{Sopticalunstable}) and
(\ref{Popticalunstable}) as defined through our cutoff regularization
scheme:
\begin{equation}
\int\limits_{\tiny\mbox{cutoff scheme}}\frac{d^4 k}{(2\pi)^4} \, = \,
\int\limits_0^{|\mbox{\boldmath $k$}|<\Lambda}\,
\frac{d^3 \mbox{\boldmath $k$}}{(2\pi)^3}\,\,
\int\limits_{-\infty}^{+\infty}\,\frac{d k^0}{2\pi}
\,.
\label{phasespaceactual}
\end{equation}
For the total cross section on the LHS of
Eqs.~(\ref{Sopticalunstable}) and (\ref{Popticalunstable}), the
difference between using the phase space
integrations~(\ref{phasespaceactual}) or
(\ref{phasespacephysical}) can be expanded in $p_0^2$ and $\Gamma_t$,
and
can be accounted for by introducing additional photon and Z boson
wave function renormalization constants into the NRQCD Lagrangian,
which are proportional to $i\Gamma_t$.
The calculation of these counter-terms is straightforward. 
This leads to the following form of the
optical theorem relations for the vector- and 
axial-vector-current-induced correlators,
\begin{eqnarray}
\mbox{Im}\,\bigg[\,
{\cal{A}}^v + 
i \frac{3\,N_c\,M_t\,\Gamma_t}{2\,\pi^2}\,\Big[\,
2\frac{M_t}{\Lambda}-(1+\sqrt{3})
\,\Big]
\,\bigg] 
& = &
3\,N_c\,
\int\limits_{\tiny\mbox{phase space}}\frac{d^4 k}{(2\pi)^4}\,
|S(k^0,\mbox{\boldmath $k$})|^2\,
\Gamma_t^2\,\bigg(1 - \frac{\mbox{\boldmath $k$}^2}{M_t^2}
\,\bigg)
\,,
\label{Sopticalunstableproper}
\\[5mm]
\mbox{Im}\,\bigg[\,
{\cal{A}}^a + 
i \frac{N_c\,M_t\,\Gamma_t}{\pi^2}\,\Big[\,
-2\frac{\Lambda}{M_t}+2\,(\sqrt{3}-1)
\,\Big]
\,\bigg] & = &
2\,N_c\,
\int\limits_{\tiny\mbox{phase space}}\frac{d^4 k}{(2\pi)^4}\,
\frac{\mbox{\boldmath $k$}^2}{M_t^2}\,
|P(k^0,\mbox{\boldmath $k$})|^2\,\Gamma_t^2
\,,
\label{Popticalunstableproper}
\end{eqnarray}
where we have displayed the counter-terms for the phase space, each
at the Born level, to first order in $\Gamma_t$ and first order in the
non-relativistic expansion. For simplicity we have set the bottom quark
and the W boson masses to zero. As far as the velocity
counting of the width is concerned, we in principle should have
included also the ${\cal{O}}(\alpha_s)$ contributions for the 
counter-term in Eq.~(\ref{Sopticalunstableproper}), because we
formally have to count the vector-current-induced total cross section
as order $v$. However, the counter-term contributions are only at the
per cent level and not (yet) important phenomenologically, compared
with the
much larger uncertainties in the normalization of the total cross
section from QCD, which are discussed in
Sec.~\ref{subsectionnormalization}. We also note that the phase space
counter-terms can be calculated entirely within the
non-relativistic effective theory, and that there is no need to match
to the full theory. This is because within the physical phase space
boundaries the non-relativistic expansion of the phase space
integration should be convergent.\footnote{
This statement is a conjecture. In this context
``convergence'' for the phase space counter-terms is not
meant to be associated with an expansion in $\Gamma_t$ or $p_0$, but to
the convergence in the coefficient multiplying the term linear in
$\Gamma_t$. All higher-order terms in the non-relativistic expansion
under the phase space integral can contribute to the term linear in
$\Gamma_t$. 
} We also note that the calculation of
the phase space  counter-term for the S-P-wave interference,
contributing e.g. in the top quark angular distribution, is in complete
analogy to the calculation of the counter-terms in
Eqs.~(\ref{Sopticalunstableproper}) and (\ref{Popticalunstableproper}).
We emphasize that the non-relativistic current correlators
${\cal{A}}^v$, ${\cal{A}}^a$ and the vertex functions $S$ and $P$ in
relations~(\ref{Sopticalunstableproper}) and
(\ref{Popticalunstableproper}) are still calculated in our cutoff
scheme presented in Sec.~\ref{sectionregularization}. 

The expressions
for the vector-current- and axial-vector-current-induced total cross
sections valid at NNLO in the non-relativistic
expansion, and properly including all effects of the top width at the
Born level and leading order in the non-relativistic expansion, read
\begin{eqnarray}
R_{\mbox{\tiny NNLO}}^{v,\mbox{\tiny thr}}(q^2) & = &
\frac{4\,\pi}{q^2}\,C^v\,
\mbox{Im}\Big[\,
{\cal{A}}^v(q^2)
\,\Big] 
+ \frac{3\,N_c\,\Gamma_t}{2\,\pi\,M_t}\,\Big[\,
2\frac{M_t}{\Lambda}-(1+\sqrt{3})
\,\Big]
\,\bigg] 
\,,
\label{vectorcrosssectionexpandedgamma}
\\[3mm]
R_{\mbox{\tiny NNLO}}^{a,\mbox{\tiny thr}}(q^2) & = &
\frac{4\,\pi}{q^2}\,C^a\,
\mbox{Im}\Big[\,
{\cal{A}}^a(q^2)
\,\Big] + \frac{N_c\,\Gamma_t}{\pi\,M_t}\,\Big[\,
-2\frac{\Lambda}{M_t}+2\,(\sqrt{3}-1)
\,\Big]
\,\bigg]
\,.
\label{axialvectorcrosssectionexpandedgamma}
\end{eqnarray}
From Eqs.~(\ref{Sopticalunstableproper}) and
(\ref{Popticalunstableproper}) we can derive the
centre-of-mass three-momentum distributions of the top quarks
($k\equiv|\mbox{\boldmath $k$}|$), 
\begin{eqnarray}
\frac{d R_{\mbox{\tiny NNLO}}^{v,\mbox{\tiny thr}}(q^2)}
{d |\mbox{\boldmath$k$}|}
& = &
C^v\,
\frac{6\,N_c}{\pi\,q^2}\,
\Gamma_t^2\,\bigg(1 - \frac{\mbox{\boldmath $k$}^2}{M_t^2}
\,\bigg)\,
\mbox{\boldmath$k$}^2
\int\limits_{-(\frac{\sqrt{q^2}}{2}-\sqrt{k^2+M_W^2})}^{
  (\frac{\sqrt{q^2}}{2}-\sqrt{k^2+M_W^2})}\,\frac{d k^0}{2\pi}\,
|S(k^0,\mbox{\boldmath $k$})|^2
\,,
\label{Sopticalunstablepropergamma}
\\[5mm]
\frac{d R_{\mbox{\tiny NNLO}}^{a,\mbox{\tiny thr}}(q^2)}
{d |\mbox{\boldmath$k$}|} 
& = &
C^a\,
\frac{4\,N_c}{\pi\,q^2}\,
\Gamma_t^2\,
\frac{\mbox{\boldmath $k$}^4}{M_t^2}\,
\int\limits_{-(\frac{\sqrt{q^2}}{2}-\sqrt{k^2+M_W^2})}^{
  (\frac{\sqrt{q^2}}{2}-\sqrt{k^2+M_W^2})}\,\frac{d k^0}{2\pi}\,
|P(k^0,\mbox{\boldmath $k$})|^2
\,.
\label{Popticalunstablepropergamma}
\end{eqnarray}
Unless $|\mbox{\boldmath $k$}|$ is chosen close to the endpoint 
$(q^2/4-M_W^2)^{1/2}$ the numerical difference obtained by replacing the
physical limits of the $k^0$ integrations by $\pm\infty$ is
negligible.
We note that the three-momentum distributions shown in
Eqs.~(\ref{Sopticalunstablepropergamma}) and
(\ref{Popticalunstablepropergamma})
are not equal to the physical observable three-momentum distributions,
because the exchange of gluons between the top decay and production
processes leads to additional non-negligible corrections at
NLO~\cite{Harlander1,Sumino2} and NNLO. These ``interconnection'' 
effects belong
to the electroweak corrections, which are not treated in this work.  
The NNLO corrections to the three-momentum 
distribution calculated here are only a first step towards a
complete NNLO treatment of the three-momentum distribution. We also
want to mention that the three momentum distribution is strictly
speaking an ambiguous quantity since the three (as well as the four)
momentum of a coloured particle is an ambiguous concept. This is
in contrast to the total cross section, which describes the rate of
colour singlet top-antitop events.

We conclude this section with some remarks on the inconsistencies that
can arise if the Fadin-Khoze replacement rule ``$E\to E+i \Gamma_t$''
is employed at NNLO for the calculation of the top-antitop cross
section close to threshold. We emphasize that there is nothing wrong,
in principle, in calculating the current correlators for stable top
quarks via the Schr\"odinger equation~(\ref{NNLOSchroedinger}),
supplemented afterwards by the replacement $\sqrt{q^2}-2 M_t\to
\sqrt{q^2}-2 M_t + i \Gamma_t$. This corresponds essentially to the
modification
\begin{equation}
\sum_n\hspace{-5mm}\int\, 
\frac{|n\rangle\langle n|}{E_n-E-i \epsilon} \, \to \,
\sum_n\hspace{-5mm}\int\, 
\frac{|n\rangle\langle n|}{E_n-E-i \Gamma_t}
\end{equation}
in the spectral representation of the Green function of
Eq.~(\ref{NNLOSchroedinger}) and is equivalent to keeping only the terms
$\psi^\dagger i\Gamma_t \psi$ and $\chi^\dagger i\Gamma_t \chi$ in the
modified version of the NRQCD Lagrangian. In this approach, also the
optical theorem remains valid in the form
\begin{eqnarray}
\mbox{Im}\Big[\,{\cal{A}}^v(q^2)\,\Big]
& = &
6\,N_c\,
\int\frac{d^3 \mbox{\boldmath $k$}}{(2\pi)^3}\,
|S(\mbox{\boldmath $k$})|^2\,
\Gamma_t
\,,
\label{Sopticalunstableconsistent}
\end{eqnarray}
for the vector current correlator, as an example.
(For simplicity, we neglect the subtleties of the phase space effects,
because they are irrelevant to this discussion.) 
However, there is a caveat, since it is possible, for the case of zero
width, to simplify the form of Eq.~(\ref{NNLOSchroedinger}) in a way
that for the stable top quark case the results remain correct, whereas
inconsistencies arise if the results undergo the replacement rule
``$E\to E+i \Gamma_t$''. Such simplifications, based on the assumption
that certain singular terms, which arise during the simplification, can
be neglected, have in fact been carried out in
Refs.~\cite{Hoang2,Hoang3,Melnikov3,Yakovlev1}. In
Refs.~\cite{Hoang2,Melnikov3,Hoang5} it was shown that the NNLO kinetic
energy corrections
and the Breit-Fermi potential in Eq.~(\ref{NNLOSchroedinger}), if they
are treated as a perturbation to first order, can be rewritten in
terms of an energy dependent Coulomb potential $\propto
C_F\alpha_s/\mbox{\boldmath$Q$}^2 \times (p_0^2/M_t^2)$, a Darwin-like
constant potential, and a potential $\propto
\alpha_s^2/|\mbox{\boldmath$Q$}|$. 
Neglecting all NNLO corrections except the energy-dependent
corrections to the Coulomb potential, the simplified version of
the NNLO Schr\"odinger equation has
the form\footnote{
We emphasize that the neglect of the rest of the NNLO corrections does
not affect the validity of the following arguments, because they are
independent of the top quark width after the replacement rule ``$E\to
E+i \Gamma_t$'' has been applied.
} 
\begin{eqnarray}
\bigg[\,
 \frac{\mbox{\boldmath $k$}^2}{M_t} -
\frac{p_0^2}{M_t} 
\,\bigg]\,
\tilde G(\mbox{\boldmath $k$},\mbox{\boldmath $k$}^\prime) \,+\, 
\int\frac{d^3 \mbox{\boldmath $p$}^\prime}{(2\,\pi)^3}\,
\tilde V^{sim}(\mbox{\boldmath $k$}-\mbox{\boldmath $p$}^\prime)\,
\tilde G(\mbox{\boldmath $p$}^\prime,\mbox{\boldmath $k$}^\prime)
\, = \, 
(2\,\pi)^3\,\delta^{(3)}(\mbox{\boldmath $k$}-\mbox{\boldmath $k$}^\prime) 
\,,
\label{NNLOSchroedingersimplified}
\end{eqnarray}
where
\begin{equation}
\tilde V^{sim}(\mbox{\boldmath $Q$}) \, = \,
-\,\frac{C_F\,4\,\pi\,\alpha_s}
{\mbox{\boldmath $Q$}^2}\,
\bigg(\,
1+\frac{3\,p_0^2}{2\,M_t^2}
\,\bigg)
\,.
\end{equation}
Equation~(\ref{NNLOSchroedingersimplified})
is much easier to solve than the original Schr\"odinger
equation~(\ref{NNLOSchroedinger}). For real energies, and if the
corrections from the NNLO terms in Eqs.~(\ref{NNLOSchroedinger}) and
(\ref{NNLOSchroedingersimplified}) are treated as a perturbation to
first order only, the result obtained from 
Eqs.~(\ref{NNLOSchroedinger}) and (\ref{NNLOSchroedingersimplified})
are indeed equivalent, after a proper renormalization has been carried
out. This was in fact the case for which the form of
Eq.~(\ref{NNLOSchroedingersimplified}) has been derived in
Refs.~\cite{Hoang2,Melnikov3,Hoang5}.
However, Eq.~(\ref{NNLOSchroedingersimplified}) leads to
inconsistencies for
complex energies. This can be seen from the fact that for the total
cross section calculated from Eq.~(\ref{NNLOSchroedingersimplified}),
after applying the replacement rule $\frac{p_0^2}{M_t}\to
\frac{p_0^2}{M_t}+i \Gamma_t$, the actual
form of the optical theorem relation reads
\begin{eqnarray}
\mbox{Im} {\cal{A}}^v & = & 
6\,N_c\,\mbox{Im}\,\bigg[\,
\int\frac{d^3 \mbox{\boldmath $k$}}{(2\pi)^3}
\int\frac{d^3 \mbox{\boldmath $k$}^\prime}{(2\pi)^3}\,
\tilde G({\mbox{\boldmath $k$}},{\mbox{\boldmath $k$}^\prime})
\,\bigg]
\nonumber
\\[3mm] & = &
6\,N_c\,
\int\frac{d^3 \mbox{\boldmath $k$}}{(2\pi)^3}\,
|S(\mbox{\boldmath $k$})|^2\,
\Gamma_t
 \, - \,
6\,N_c\,
\int\frac{d^3 \mbox{\boldmath $k$}}{(2\pi)^3}\,
\int\frac{d^3 \mbox{\boldmath $k$}^\prime}{(2\pi)^3}\,
S^*(\mbox{\boldmath $k$})\,
\mbox{Im}\,\bigg[\,
\tilde V^{sim}(\mbox{\boldmath $k$}-\mbox{\boldmath $k$}^\prime)
\,\bigg]\,
S(\mbox{\boldmath $k$}^\prime)
\,,
\label{Sopticalunstableinconsistent}
\end{eqnarray}
rather than Eq.~(\ref{Sopticalunstableconsistent}).
The additional term on the RHS of
Eq.~(\ref{Sopticalunstableinconsistent}) originates from the energy
dependent Coulomb-type potential in
Eq.~(\ref{NNLOSchroedingersimplified}). However, the additional term
does not correspond
to any physical final state, because it corresponds to an absorption
process in the potential. In other words, it is impossible to
recover the total cross section from the momentum distribution, if one
defines it as the integral over the physical final states
represented by the first term on the RHS of
Eq.~(\ref{Sopticalunstableinconsistent}). 
Including also the rest of the
NNLO corrections not displayed in
Eq.~(\ref{NNLOSchroedingersimplified}), we have checked, with the
numerical methods described in the next section, that the size of the
second term on the RHS of
Eq.~(\ref{Sopticalunstableinconsistent}) is between about $5\%$ (for
$\sqrt{q^2}-2M_t\sim 5$~GeV) and $20\%$ (for $\sqrt{q^2}-2M_t\sim
-5$~GeV) for the choices of parameters employed in the analysis of
Sec.~\ref{sectionpolescheme}. (Similar results can already be obtained
by analysing the known analytic solutions of the non-relativistic
Coulomb problem~\cite{Wichmann1,Hostler1,Schwinger1} for a Coulomb
potential with a complex coupling.)
Thus for the determination of the total top-antitop cross section
close to threshold, there is an unacceptable discrepancy between the
integrated momentum distribution over physical final states and the
absorptive part of the non-relativistic current correlator. We believe
that the size of the second term on the RHS of
Eq.~(\ref{Sopticalunstableinconsistent})
should in principle be taken as an estimate for the inherent
uncertainties of using the simplified NNLO Schr\"odinger
equation~(\ref{NNLOSchroedingersimplified}) supplemented by the
replacement rule of Fadin and Khoze. We have checked, however,
that the LHS
of Eq.~(\ref{Sopticalunstableinconsistent}) is much closer to the
correct result, obtained from the original Schr\"odinger
equation~(\ref{NNLOSchroedingergamma}), than the integrated momentum
distribution. From this
point of view the use of the simplified Schr\"odinger
equation~(\ref{NNLOSchroedingersimplified}) might be justified for the
total cross section, but is questionable for the momentum
distribution. 

\vspace{1.5cm}
\section{Numerical Implementation}
\label{sectionnumerics}
In this work we use numerical methods described in
Refs.~\cite{Jezabek1,Teubnerdip,Harlanderdip} to determine the vertex
functions $S$ and $P$, which are the building blocks for the
calculation of the total top-antitop production cross section and the
three momentum distribution. Because the three- and four-dimensional
versions of the vertex functions are related through
Eq.~(\ref{amputatedrelation}), it is sufficient to determine the
amputated vertex functions $S/G^f$ and $P/G^f$ from the 
three-dimensional integral
equations~(\ref{Sintegral}) and (\ref{Pintegral}).
The amputated vertex function are spherically symmetric and depend
only on the modulus of
the three momentum $\mbox{\boldmath $k$}$. It is therefore possible to
reduce Eqs.~(\ref{Sintegral}) and (\ref{Pintegral}) to one-dimensional
integral equations. 

Obviously, when solving Eqs.~(\ref{Sintegral}) and (\ref{Pintegral}), the
singular behaviour of the potentials 
$\tilde V(\mbox{\boldmath $k$},\mbox{\boldmath $p$}^\prime)$
and $\tilde V_c^{\tiny\mbox{LO}}(\mbox{\boldmath $k$},
\mbox{\boldmath $p$}^\prime)$ for $\mbox{\boldmath $p$}^\prime \to 
\mbox{\boldmath $k$}$ requires special treatment. To avoid numerical
problems we rewrite the integral equations for the amputated
vertex functions as
\begin{eqnarray}
{\cal K}^v(\mbox{\boldmath $k$}) & \equiv &  
\frac{S(\mbox{\boldmath $k$})}{G^f(\mbox{\boldmath $k$})} 
\nonumber\\[2mm]
& = & 
1+\frac{{\mbox{\boldmath $k$}^2}}{6 M_t^2}
\,-\, 
\int\frac{d^3 \mbox{\boldmath $p$}^\prime}{(2\pi)^3}\,
\tilde V(\mbox{\boldmath $k$},\mbox{\boldmath $p$}^\prime)\,
G^f(\mbox{\boldmath $p$}^\prime)\,
{\cal K}^v(\mbox{\boldmath $p$}^\prime)
\nonumber 
\\[2mm]
 & = & 1+\frac{{\mbox{\boldmath $k$}^2}}{6 M_t^2}
\,-\, 
\int\frac{d^3 \mbox{\boldmath $p$}^\prime}{(2\pi)^3}\,
\tilde V(\mbox{\boldmath $k$},\mbox{\boldmath $p$}^\prime)\,
G^f(\mbox{\boldmath $p$}^\prime)\,
\Big({\cal K}^v(\mbox{\boldmath $p$}^\prime) - 
      {\cal K}^v(\mbox{\boldmath $k$}) \Big) - 
{\cal B}^v(\mbox{\boldmath $k$}) \, {\cal K}^v(\mbox{\boldmath $k$})
\,,
\label{Samputatednew}
\\[4mm]
{\cal K}^a(\mbox{\boldmath $k$}) & \equiv & 
\frac{P(\mbox{\boldmath $k$})}{G^f(\mbox{\boldmath $k$})} 
\nonumber\\[2mm] 
& = & 
1\,-\,
\int\frac{d^3 \mbox{\boldmath $p$}^\prime}{(2\pi)^3}\,
\frac{\mbox{\boldmath $k$} \mbox{\boldmath $p$}^\prime}{k^2}\,
\tilde V_c^{\tiny \mbox{LO}}(\mbox{\boldmath $k$},
      \mbox{\boldmath $p$}^\prime)\,
G^f(\mbox{\boldmath $p$}^\prime)\,
{\cal K}^a(\mbox{\boldmath $p$}^\prime)
\nonumber
\\[2mm]
 & = & 1\,-\,
\int\frac{d^3 \mbox{\boldmath $p$}^\prime}{(2\pi)^3}\,
\tilde V_c^{\tiny \mbox{LO}}(\mbox{\boldmath $k$},
      \mbox{\boldmath $p$}^\prime)\,
G^f(\mbox{\boldmath $p$}^\prime)\,
\left(\frac{\mbox{\boldmath $k$} \mbox{\boldmath $p$}^\prime}{k^2}\,
      {\cal K}^a(\mbox{\boldmath $p$}^\prime) - 
      {\cal K}^a(\mbox{\boldmath $k$}) \right) - 
{\cal B}^a(\mbox{\boldmath $k$}) \, {\cal K}^a(\mbox{\boldmath $k$})
\,,
\label{Pamputatednew}
\end{eqnarray}
where ${\cal B}^v$ and ${\cal B}^a$ are defined as
\begin{eqnarray}
{\cal B}^v(\mbox{\boldmath $k$}) & \equiv & 
\int\frac{d^3 \mbox{\boldmath $p$}^\prime}{(2\pi)^3}\,
\tilde V(\mbox{\boldmath $k$},\mbox{\boldmath $p$}^\prime)\,
G^f(\mbox{\boldmath $p$}^\prime)\,,
\label{Bvintegral}
\\[2mm]
{\cal B}^a(\mbox{\boldmath $k$}) & \equiv & 
\int\frac{d^3 \mbox{\boldmath $p$}^\prime}{(2\pi)^3}\,
\tilde V_c^{\tiny \mbox{LO}}(\mbox{\boldmath $k$},
      \mbox{\boldmath $p$}^\prime)\,
G^f(\mbox{\boldmath $p$}^\prime)\,.
\label{Baintegral}
\end{eqnarray}
As mentioned above, both ${\cal K}^v$ and ${\cal K}^a$ depend only on the
modulus of the three momentum. The angular dependence of the
integrand in
Eqs.~(\ref{Samputatednew}) and (\ref{Pamputatednew}) (including ${\cal
  B}^v$ and ${\cal B}^a$) is only coming from the potentials and the dot
product $\mbox{\boldmath $k$} \mbox{\boldmath $p$}^\prime$. 
The angular integration can be carried out analytically. The
remaining one-dimensional integral equations are then solved
numerically by discretization: the integrals $\int d p^\prime$ 
($p^\prime\equiv |\mbox{\boldmath $p$}^\prime|$) are
transformed into sums $\sum_i$ over a fixed set of momenta ${p^\prime}^i$, and
the integral equations for ${\cal K}^v(\mbox{\boldmath $k$})$ and
${\cal K}^a(\mbox{\boldmath $k$})$ are
each reduced to a system of linear equations, where the same
set of momenta has to be used for the $k^i$.\footnote{At
  this point 
  the subtraction
  carried out in Eqs.~(\ref{Samputatednew}) and (\ref{Pamputatednew})
  becomes crucial. Without it the singularities in the potentials for
  $p^\prime \to k$ would be manifest for ${p^i}^\prime = k^i$, even in
  the case of integrable singularities.} 
The resulting (complex) matrices are then inverted numerically to give the
amputated vertex functions ${\cal K}^v$ and ${\cal K}^a$ for the momenta
$k^i$.

In practice it turned out that the use of the Gaussian quadrature
formulae is very efficient for the discretization. A surprisingly
small number of points (of the order of 100) already leads to a high
numerical accuracy. In addition, integrals were split into two (or
more) parts and a suitable transformation of integration variables was
applied wherever needed. It should also be noted that the finite
width of the top quark is essential for the numerical stability of the
method. It makes potentially dangerous denominators in
the integrands of Eqs.~(\ref{Samputatednew}) and (\ref{Pamputatednew}),
which originate from the free Green function, well behaved (compare
Eqs.~(\ref{Gfree}) and (\ref{Gfreegamma})). Clearly, the UV regularization
by a momentum cutoff as discussed in Sec.~\ref{sectionregularization}
is most naturally implemented in our
numerical approach. It is sufficient to choose the momenta of the 
(Gauss-Legendre) grid to be limited by the value of the cutoff. Such a
cutoff is, in principle, not needed in the case of pure Coulomb
potentials. There the solution of the integral equations is possible
without any cutoff and would correspond to a different
regularization scheme with different short-distance
coefficients. However, the potentials $\tilde
V_{\mbox{\tiny BF}}$, $\tilde V_{\mbox{\tiny NA}}$ and the kinematic
corrections introduced at NNLO require a UV regularization already
for purely numerical reasons, which can be seen from naive power counting
in Eq.~(\ref{Samputatednew}).

\vspace{1.5cm}
\section{A First Analysis in the Pole Mass Scheme}
\label{sectionpolescheme}
In this section we carry out a first brief analysis of the total cross
sections $Q_t^2 R^v$ and $R^a$  and their three-momentum
distributions in the pole mass scheme. We
do this even though it is known that in the pole mass scheme
there are uncomfortably large NNLO corrections in the location of
the $1S$ peak position as well as in the normalization of the total
vector-current-induced cross
section $R^v$~\cite{Hoang3,Melnikov3,Yakovlev1}. We will show in
Sec.~\ref{sectionuncertainties}
that these large corrections are a consequence of the pole mass
scheme, and that the pole mass definition has to be
abandoned as far as a precise extraction of a top quark mass from
experimental data is concerned. Nevertheless, the pole mass is a well
defined quantity in the framework of perturbation
theory~\cite{Kronfeld1,Tarrach1},
and, despite all its problems at larger orders of perturbation theory, 
remains a very convenient mass parameter to use for the formulation of
(P)NRQCD and for calculations of the cross section. Thus a brief analysis
in the pole mass scheme serves as a reference point with which
results obtained with different approaches can be compared and from
which we can visualize in which way alternative top quark mass
definitions can improve the situation and in which way they cannot.

\begin{figure}[t!] 
\begin{center}
\leavevmode
\epsfxsize=3.8cm
\epsffile[200 400 400 530]{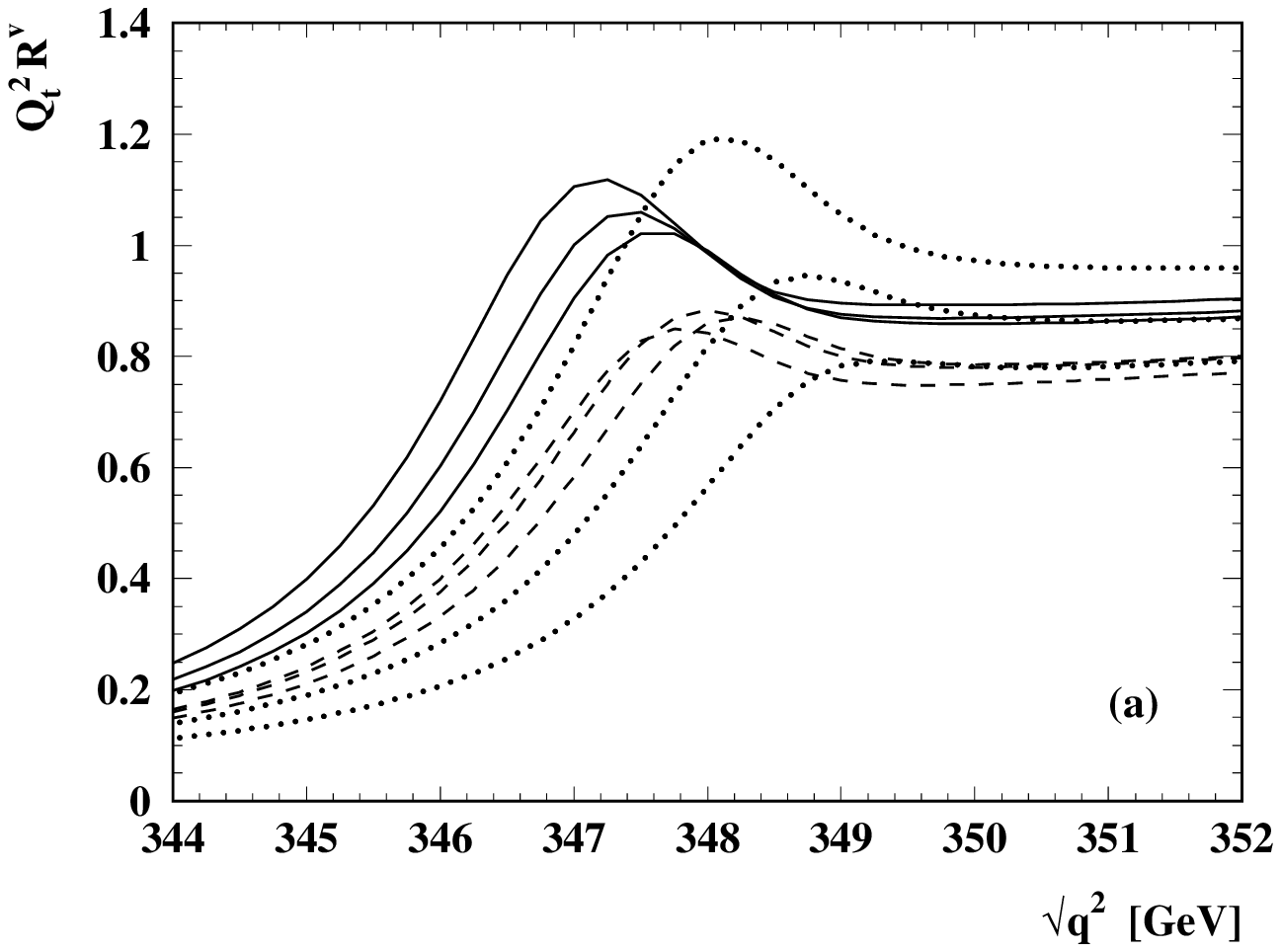}
\hspace{4.cm}
\epsfxsize=3.8cm
\leavevmode
\epsffile[200 400 400 530]{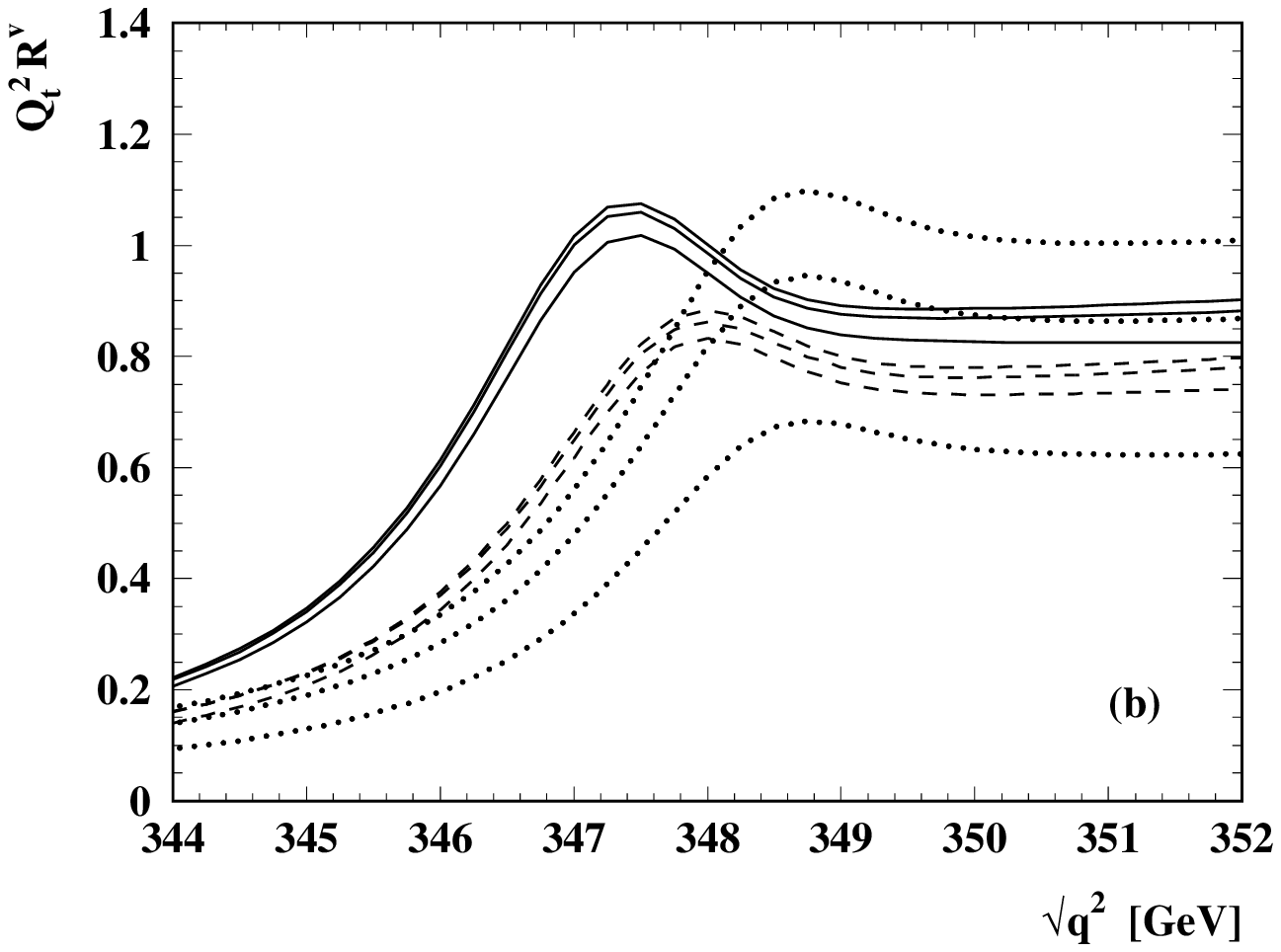}\\[3cm]
\leavevmode
\epsfxsize=3.8cm
\epsffile[200 400 400 530]{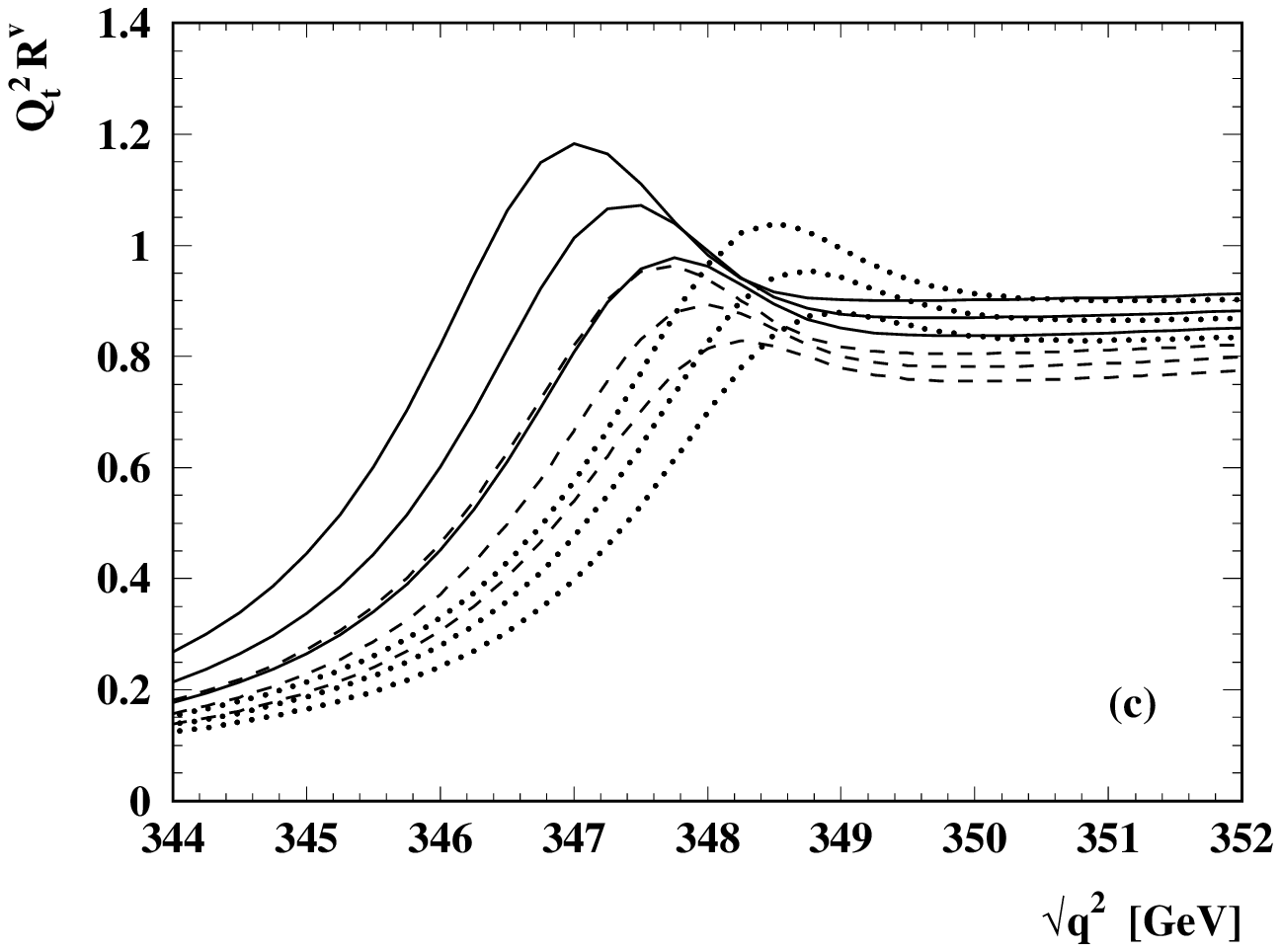}
%
%
\vskip  2.7cm
 \caption{\label{figtotpole} 
The total vector-current-induced cross section $Q_t^2 R^v$ for
centre-of-mass energies $344\,\mbox{GeV}< \sqrt{q^2}< 352$~GeV in the
pole mass scheme.
The dependence on the renormalization scale $\mu$ (a),
on the cutoff $\Lambda$ (b) and on $\alpha_s(M_Z)$ (c) is displayed.
More details and the choice of parameters are given in the text.
}
 \end{center}
\end{figure}
\begin{figure}[t!] 
\begin{center}
\leavevmode
\epsfxsize=3.8cm
\epsffile[200 400 400 530]{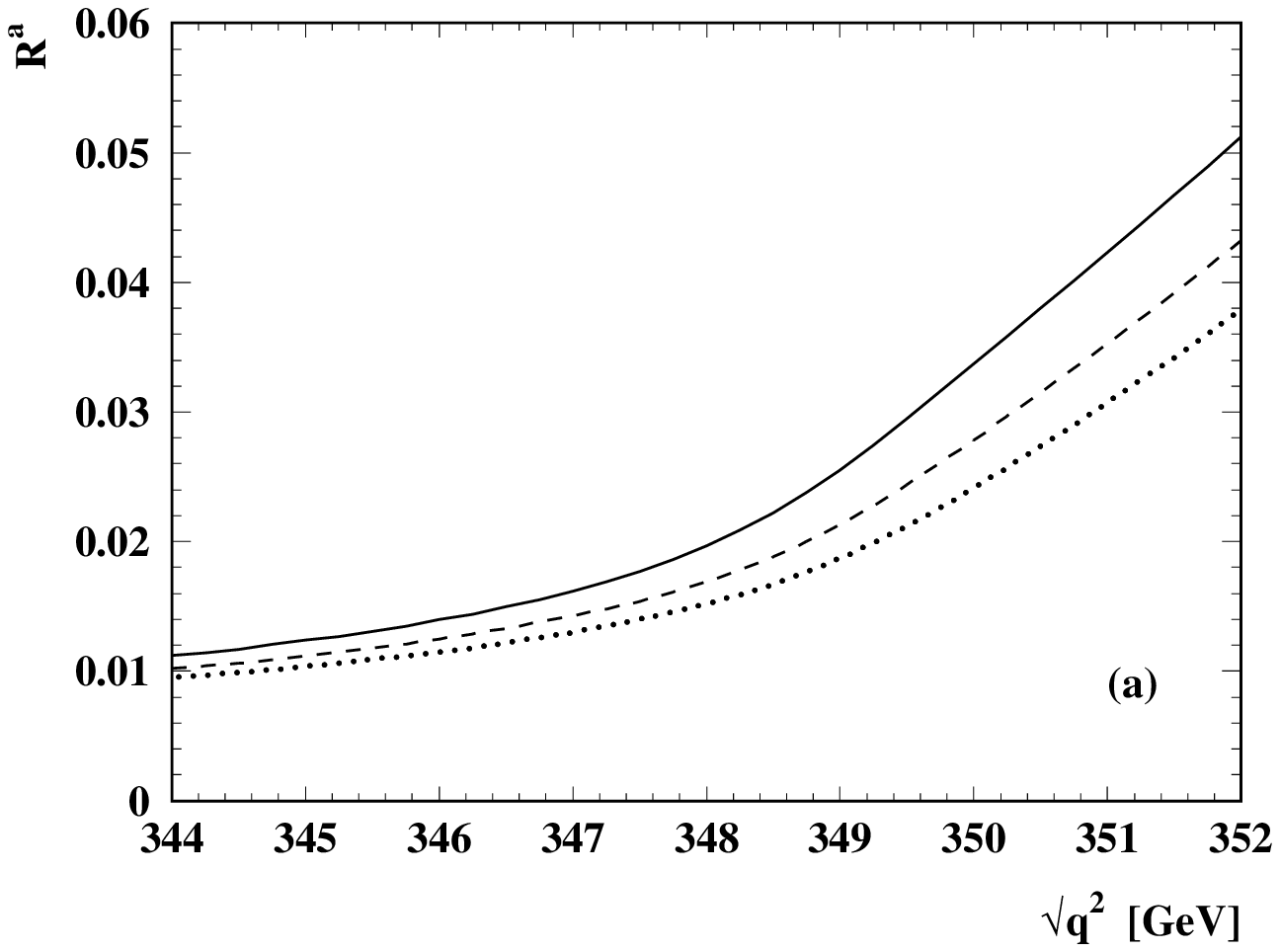}
\hspace{4.2cm}
\epsfxsize=3.8cm
\leavevmode
\epsffile[200 400 400 530]{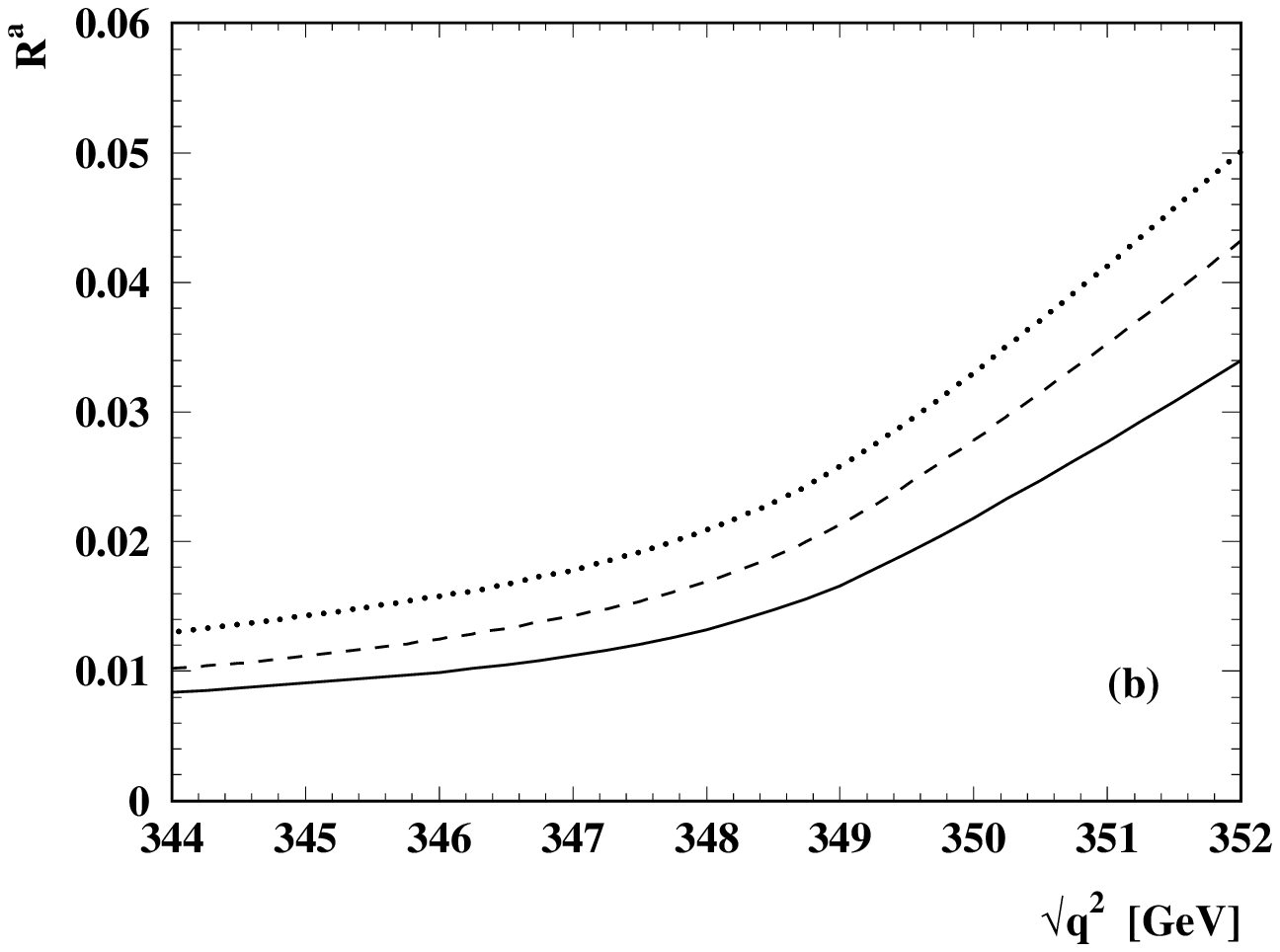}\\[3cm]
\leavevmode
\epsfxsize=3.8cm
\epsffile[200 400 400 530]{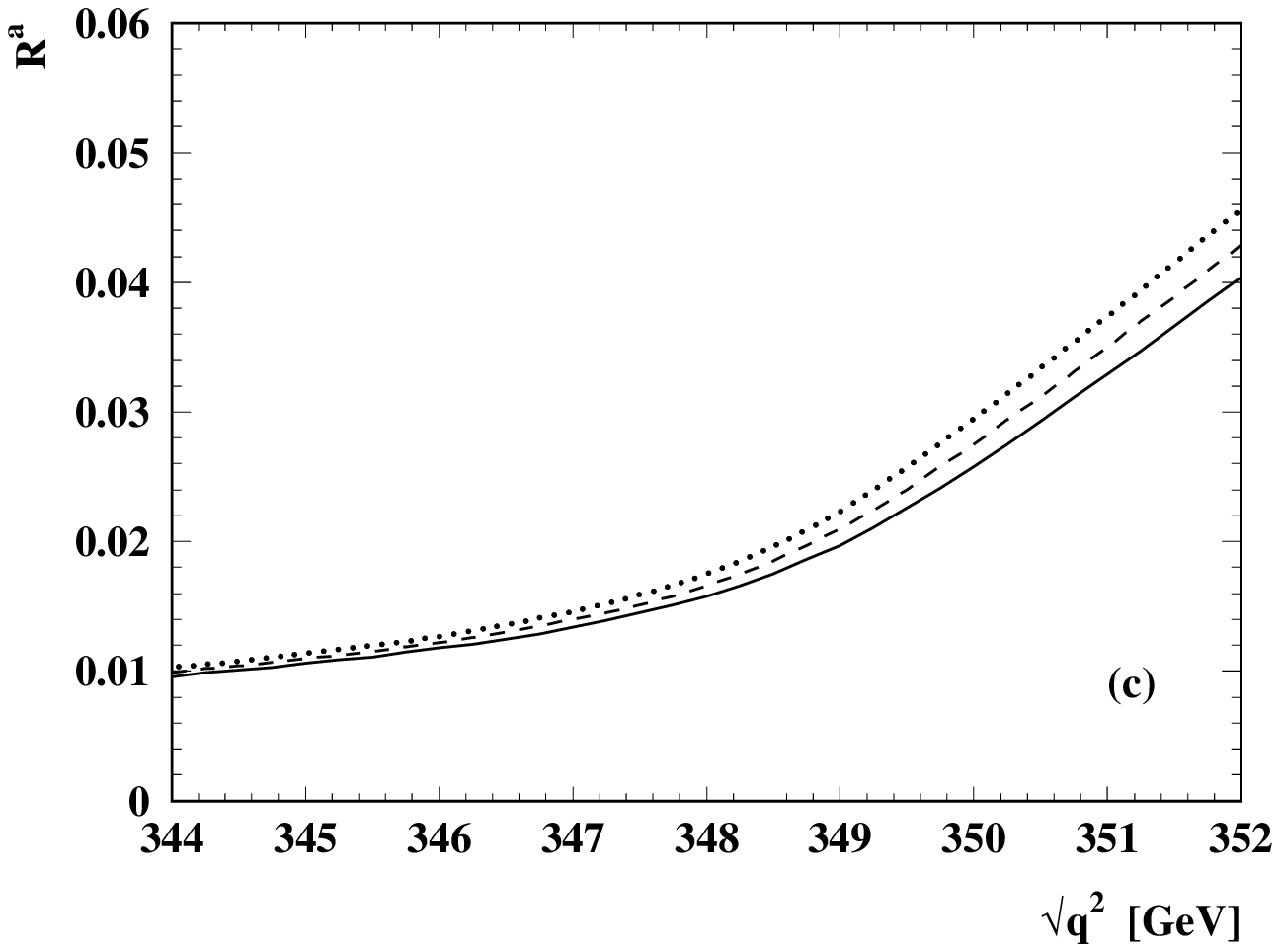}
%
%
\vskip  2.7cm
 \caption{\label{figtotpoleax} 
The total axial-vector-current induced cross section $R^a$ for
centre-of-mass energies  $344\,\mbox{GeV}< \sqrt{q^2}< 352$~GeV in the
pole mass scheme. 
The dependence on the renormalization scale $\mu$ (a),
on the cutoff $\Lambda$ (b) and on $\alpha_s(M_Z)$ (c) is
displayed. More details and the choice of parameters are given in
the text.
}
 \end{center}
\end{figure}
\begin{figure}[t!] 
\begin{center}
\leavevmode
\epsfxsize=3.8cm
\epsffile[200 400 400 530]{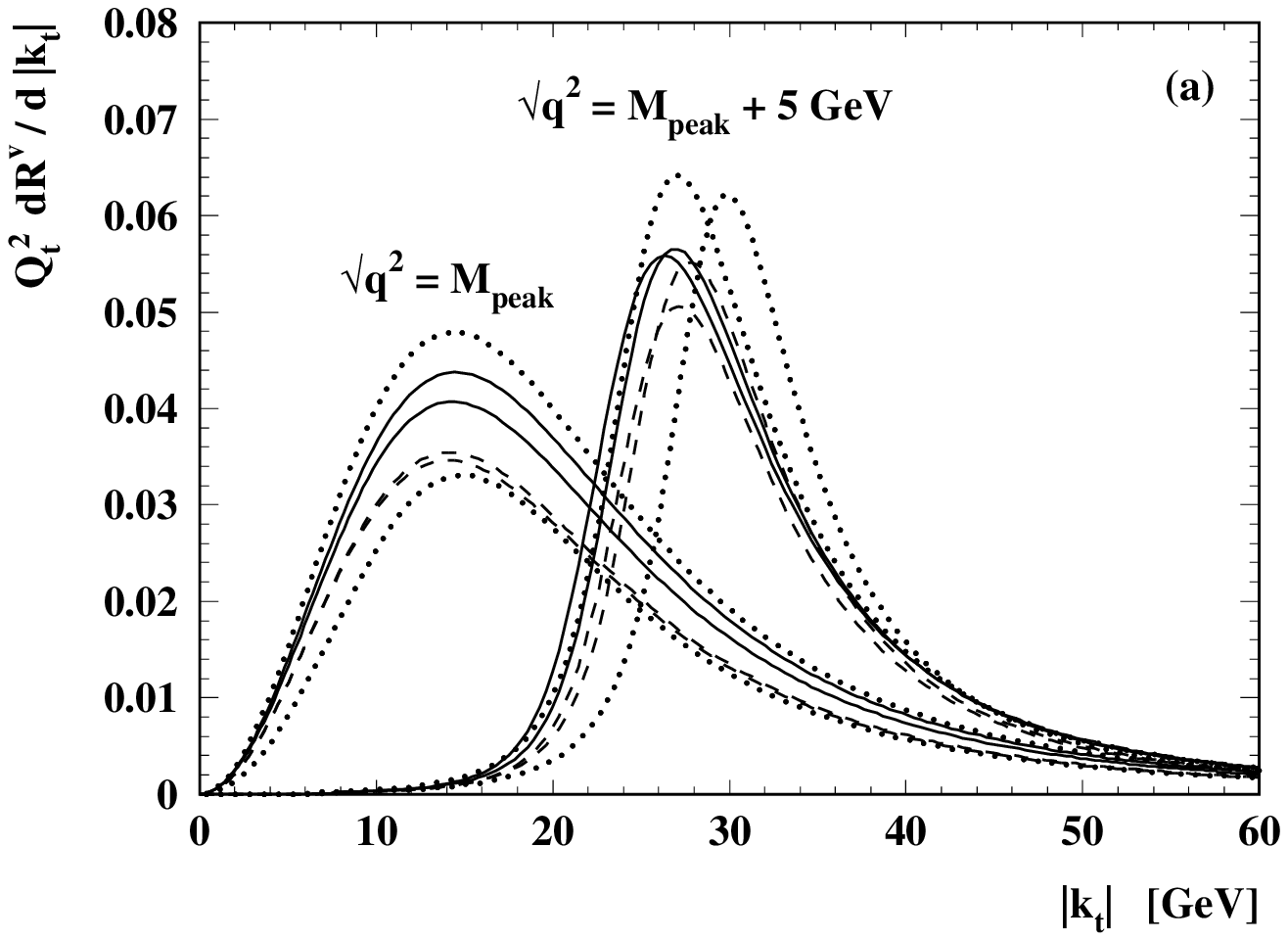}
\hspace{4.4cm}
\epsfxsize=3.8cm
\leavevmode
\epsffile[200 400 400 530]{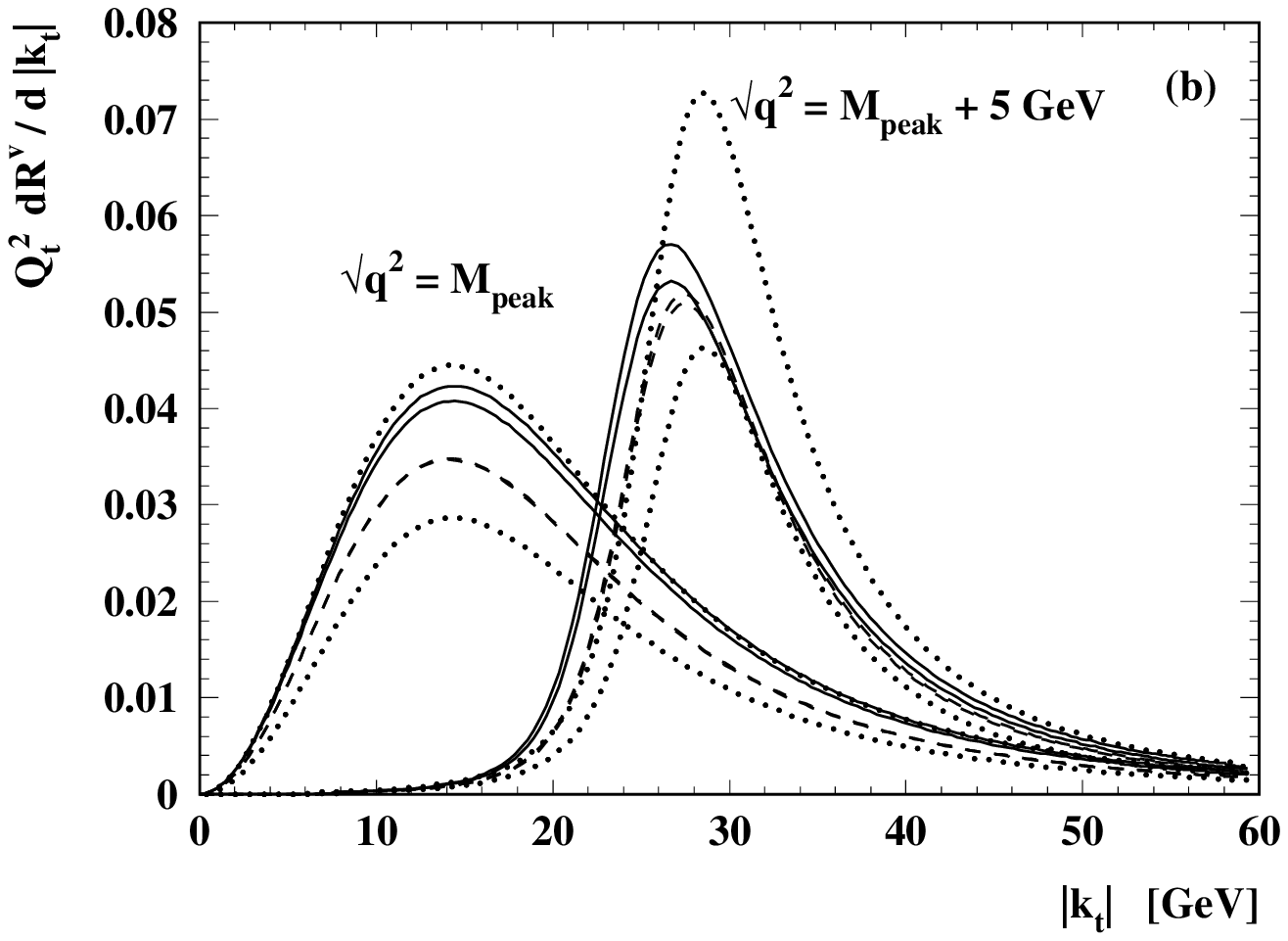}\\[3cm]
\leavevmode
\epsfxsize=3.8cm
\epsffile[200 400 400 530]{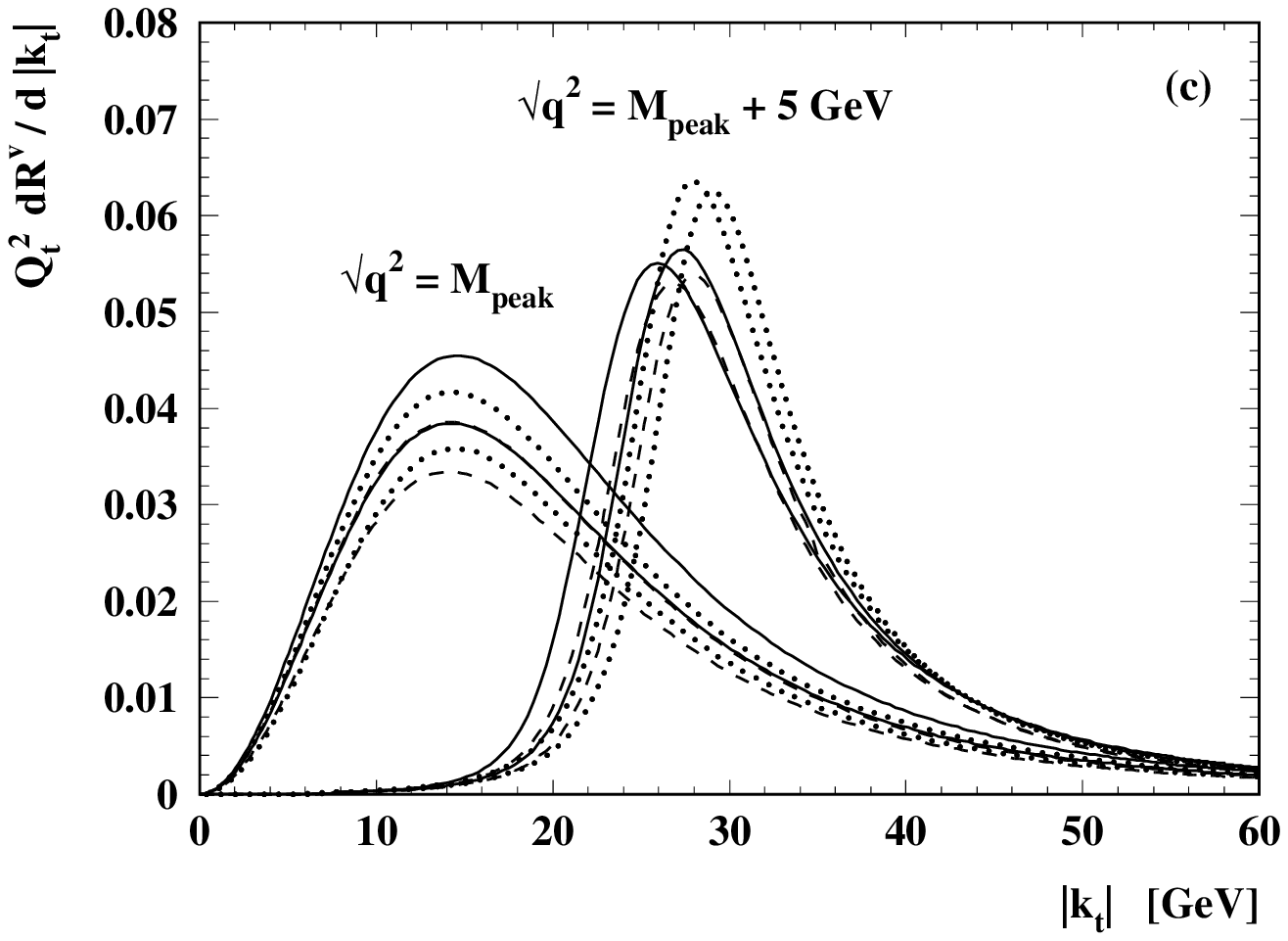}
%
%
\vskip  2.7cm
 \caption{\label{figdistpole} 
The three-momentum distribution of the vector-current-induced cross
section 
$Q_t^2 R^v$ for centre-of-mass energies $\sqrt{q^2}=M_{peak}$ and
$M_{peak}+5$~GeV in the pole mass scheme.
The dependence on the renormalization scale $\mu$ (a),
on the cutoff $\Lambda$ (b) and on $\alpha_s(M_Z)$ (c) is displayed.
More details and the choice of parameters are given in the text.
}
 \end{center}
\end{figure}
\begin{figure}[t!] 
\begin{center}
\leavevmode
\epsfxsize=3.8cm
\epsffile[200 400 400 530]{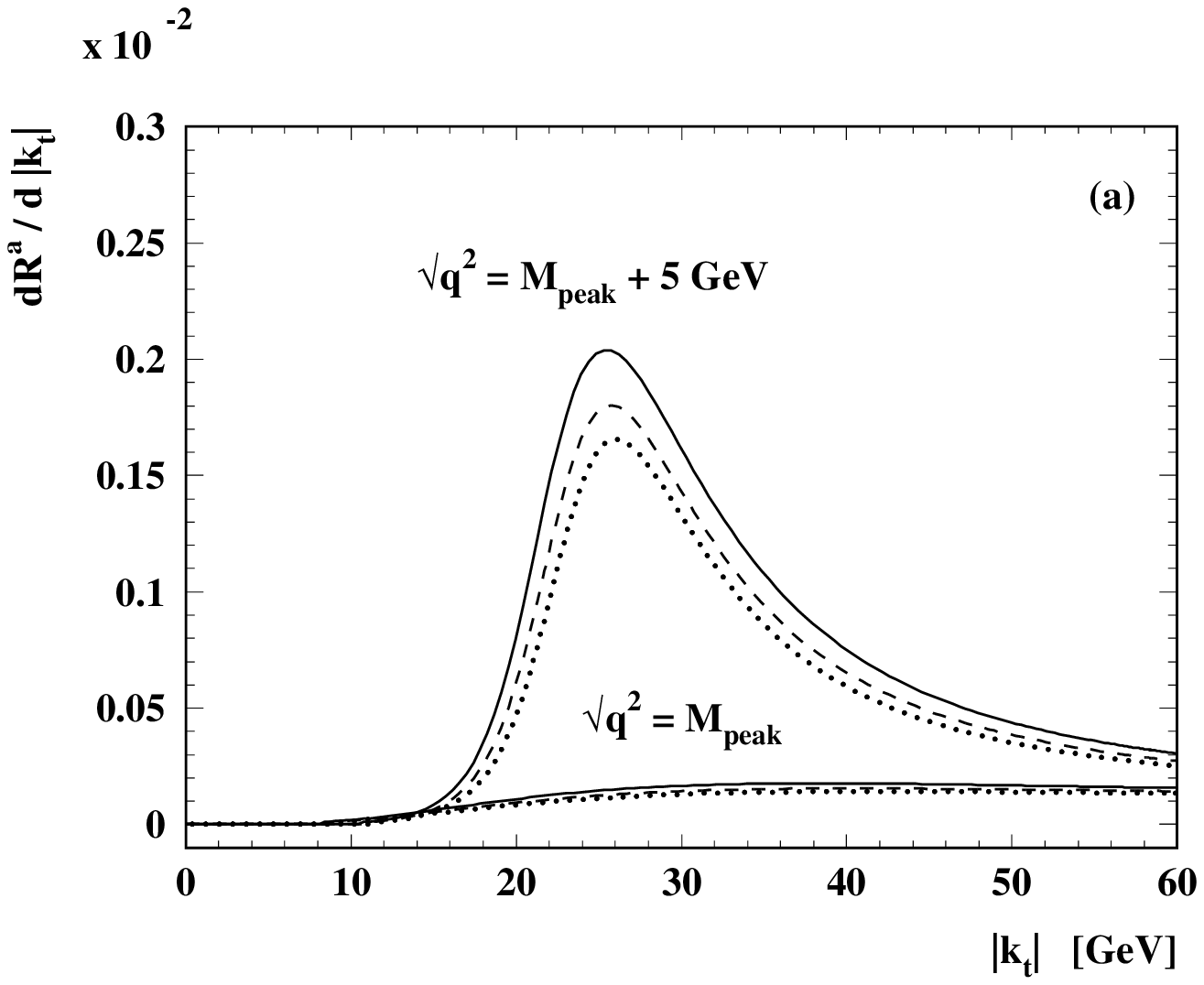}
\hspace{4.4cm}
\epsfxsize=3.8cm
\leavevmode
\epsffile[200 400 400 530]{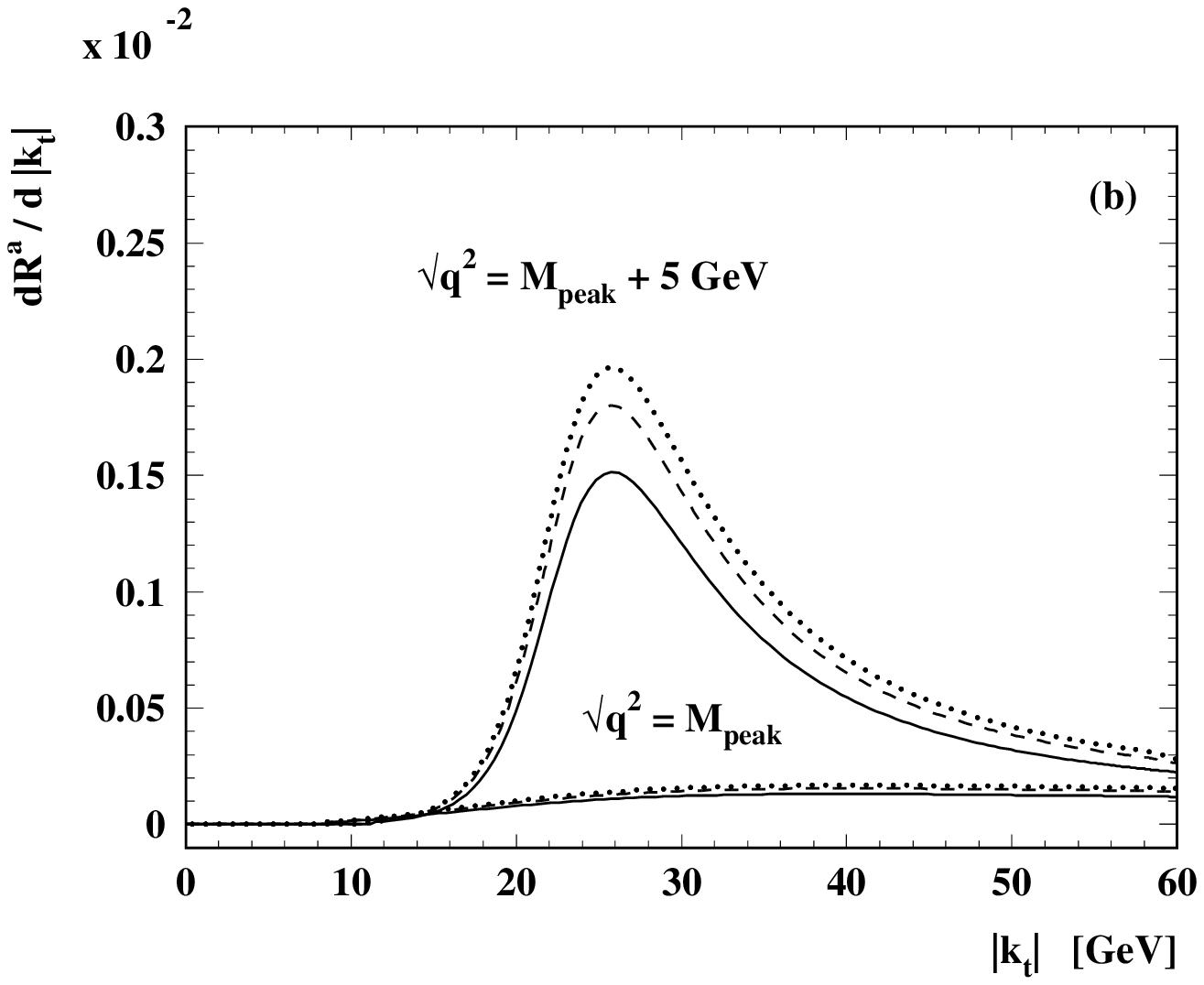}\\[3cm]
\leavevmode
\epsfxsize=3.8cm
\epsffile[200 400 400 530]{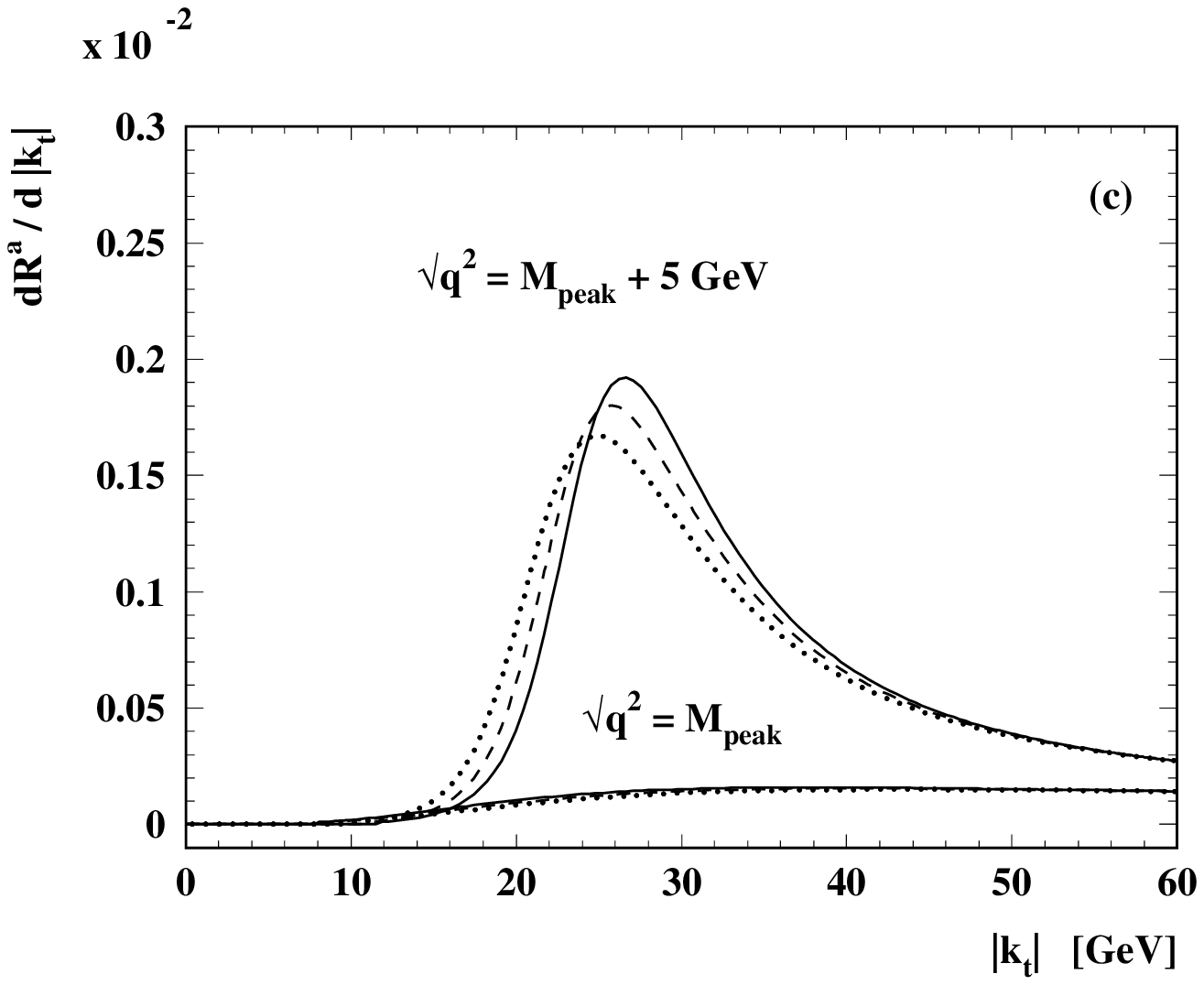}
%
%
\vskip  2.7cm
 \caption{\label{figdistpoleax} 
The three-momentum distribution of the axial-vector-current-induced
cross section $R^a$ for centre-of-mass energies
$\sqrt{q^2}=M_{peak}$ and $M_{peak}+5$~GeV
in the pole mass scheme.
The dependence on the renormalization scale $\mu$ (a),
on the cutoff $\Lambda$ (b) and on $\alpha_s(M_Z)$ (c) is displayed.
More details and the choice of parameters are given in the text.
}
 \end{center}
\end{figure}

In Figs.~\ref{figtotpole} the total vector-current-induced cross
section $Q_t^2 R^v$ is displayed for
$344\,\mbox{GeV}<\sqrt{q^2}<352\,\mbox{GeV}$
at LO (dotted lines), NLO (dashed lines) and NNLO (solid lines).
The LO curves are determined from 
Eq.~(\ref{NNLOSchroedingergamma}), excluding all relativistic
corrections and taking into account only the ${\cal{O}}(\alpha_s)$
contribution to the Coulomb potential and setting $C^v=1$. In addition,
the NLO curves also include the NLO corrections to $C^v$ and
the ${\cal{O}}(\alpha_s^2)$ contributions of the Coulomb potential.
At NNLO all contributions mentioned in this paper are included.
Only at NNLO have we taken into account the phase space counter-terms
displayed in Eqs.~(\ref{vectorcrosssectionexpandedgamma}) and
(\ref{axialvectorcrosssectionexpandedgamma}).
In all figures shown in this section the top quark width is chosen as
$\Gamma_t=1.43$~GeV and the top quark pole mass as $M_t=175$~GeV.
Figure~\ref{figtotpole}a displays the dependence on the renormalization
scale for $\mu=15$, $30$ and $60$~GeV for
$\alpha_s(M_Z)=0.118$, $\Lambda=175$~GeV. For $\sqrt{q^2}\lsim
347$~GeV, $\mu=15$, $30$ and $60$~GeV corresponds to the upper, middle
and lower curves, respectively. 
We note that in Refs.~\cite{Hoang3,Melnikov3,Yakovlev1} the
renormalization scale governing
the strong coupling of the potentials in the Schr\"odinger
equation~(\ref{NNLOSchroedinger}) has been chosen between $50$ and
$100$~GeV. Regarding the fact that the inverse Bohr radius $\sim
M_t\alpha_s$ is the scale that governs the cross section at NNLO,
the natural renormalization scale is of order $30$~GeV. This causes
logarithms of the ratio $\Lambda/\mu\sim 1/\alpha_s$ in the vector
current correlator (see App.~\ref{appendixshortdistance}), which,
however, are connected only to the running of the strong coupling.  
As has already been demonstrated in previous publications on the NNLO
corrections~\cite{Hoang3,Melnikov3,Yakovlev1},
the position of the $1S$ peak varies considerably at the different
orders of approximation and for the different choices of the
renormalization scale $\mu$. (For explicit numbers see
Table~\ref{tabpeakpositionpole}.)
It is also evident that the normalization
of the total cross section is subject to large corrections. We see
that, in general, the NNLO corrections to the normalization are of
order $20\%$ and as large as the NLO ones. Further, the dependence on
the renormalization 
scale is even larger at NNLO than at NLO. In Fig.~\ref{figtotpole}b
the dependence of $Q_t^2 R^v$ on the choice of the cutoff $\Lambda$ is
shown for $\alpha_s(M_Z)=0.118$, $\mu=30$~GeV and $\Lambda=90$, $175$
and $350$~GeV. For LO and NNLO $\Lambda=90$, $175$ and $350$~GeV
correspond to the lower, middle and upper curves, respectively. For
NLO $\Lambda=90$, $175$ and $350$~GeV
correspond to the lower, upper and middle curve, respectively. 
Whereas the dependence of the LO cross section on the choice of
$\Lambda$ is quite dramatic, because at LO there is no short-distance
correction
that could compensate for a variation in $\Lambda$, the variations at
NLO and NNLO are significantly smaller (or order 5--10\%). However,
there is again no reduction of the variation from NLO to NNLO.
The variation of the cross section with respect to $\Lambda$ is
small compared with the variation with respect to the renormalization
scale $\mu$ for centre-of-mass energies closer and below the
peak. Figure~\ref{figtotpole}c displays the dependence of $Q_t^2
R^v$ on the choice of $\alpha_s(M_Z)$ for $\alpha_s(M_z)=0.113, 0.118$
and $0.123$ and $\Lambda=175$~GeV, $\mu=30$~GeV. As for the variations
in the renormalization scale we see that the position of the peak depends
considerably on the choice of $\alpha_s$. We observe a strong positive
correlation between the choice of $M_t$ and $\alpha_s(M_Z)$. 
Because the peak in the
total cross section is the most pronounced feature of the total cross
section, its behaviour directly reflects the quality of a top mass
extraction from experimental data. Thus, if one would like to fit the
pole mass to data for the cross section line-shape from a threshold
scan, one finds a strong positive correlation between the pole
mass and the strong coupling~\cite{Orange1} and a strong dependence of
$M_t$ on the choice of the renormalization scale leading to quite
large theoretical uncertainties in the pole mass measurements.

In Figs.~\ref{figtotpoleax}a,b,c the total
axial-vector-current-induced cross
section $R^a$ is displayed for the same input parameters as in
Figs.~\ref{figtotpole}. Figure~\ref{figtotpoleax}a shows the dependence
on the renormalization scale for $\mu=15$ (solid line), $30$ (dashed
line) and $60$~GeV (dotted line).
Figure~\ref{figtotpoleax}b exhibits the dependence of the
cutoff for  $\Lambda=90$ (solid line), $175$ (dashed line) and
$350$~GeV (dotted line). Figure~\ref{figtotpoleax}c shows $R^a$ for 
$\alpha_s(M_z)=0.113$ (solid line), $0.118$ (dashed line) and $0.123$
(dotted line). We observe that due to the $v$ suppression of the
axial-vector currents no peak-line enhancement as in the vector
current case is visible. 
The variations of $R^a$ with respect to the
renormalization scale and the cutoff are quite large, because $R^a$
has only been determined at leading order, and the
short-distance coefficient $C^a$ does not contain any corrections
that could compensate for the variations. We note that the NLO
corrections to $R^a$ can be implemented in the same way as for
$R^v$. They are, however, beyond NNLO and have therefore not been
included into our analysis. From the phenomenological point of view
the next-to-leading order corrections are irrelevant if one takes into
account the small normalization of $R^a$ compared to $R^v$.

In Figs.~\ref{figdistpole}a,b and c the LO (dotted lines), NLO (dashed
line) and NNLO (solid lines) top-antitop
vector-current-induced three momentum distribution
$Q_t^2 d R^v/d |\mbox{\boldmath$k$}_t|$ is shown for
$0 < |\mbox{\boldmath$k$}_t| < 60$~GeV and both for centre-of-mass
energies exactly on top of the visible peak, $\sqrt{q^2}=M_{peak}$, and
for $\sqrt{q^2}=M_{peak}+5$~GeV. The input parameters have been chosen
as in Figs.~\ref{figtotpole}. Figure~\ref{figdistpole}a shows
the distributions for $\mu=15$ and $60$~GeV. At LO and NNLO
$\mu=15$~GeV corresponds to the upper curves below the peak and
$\mu=60$~GeV to the lower curves. At NLO $\mu=60$~GeV corresponds to
the higher peak and $\mu=15$~GeV to the lower. 
Figure~\ref{figdistpole}b displays the dependence of the distributions
on the cutoff for $\Lambda=90$ (lower curves) and $350$~GeV (upper
curves), and Fig.~\ref{figdistpole}c exhibits the dependence of the
distributions on the strong coupling for $\alpha_s(M_Z)=0.113$ and
$0.123$. Below the peak, the larger value of $\alpha_s(M_Z)$
always corresponds to the upper curve. 
As for the total cross
section, we observe a strong dependence of the normalization of the
distributions on the renormalization scale and the cutoff. For
$\sqrt{q^2}=M_{peak}+5$~GeV, the dependence of the peak 
position on the renormalization scale is particularly
strong.\footnote{
The peak position is always located approximately at
$|\mbox{\boldmath$k$}_t|\approx (p_0^4+M_t^2\Gamma_t^2)^{1/4}$, which
can be regarded as the effective three momentum of the top quarks.
}

Finally, in Figs.~\ref{figdistpoleax}a,b and c the
top-antitop axial-vector-current-induced three momentum distribution
$d R^a/d |\mbox{\boldmath$k$}_t|$ is shown for
$0 < |\mbox{\boldmath$k$}_t| < 60$~GeV and for centre-of-mass
energies exactly on top of the visible peak, $\sqrt{q^2}=M_{peak}$, and
for $\sqrt{q^2}=M_{peak}+5$~GeV. The input parameters have been chosen
as before. Figure~\ref{figdistpoleax}a shows
the distributions for $\mu=15$ (solid curves), $30$ (dashed curves)
and $60$~GeV (dotted curves).
Figure~\ref{figdistpoleax}b displays the dependence of the distributions
on the cutoff for $\Lambda=90$ (solid curves) $175$ (dashed curves)
and $350$~GeV (dotted curves), and Fig.~\ref{figdistpoleax}c exhibits
the dependence of the distributions on the strong coupling for
$\alpha_s(M_Z)=0.113$ (solid curves), $0.118$ (dashed curves) and
$0.123$ (dotted curves).
As expected, the momentum distribution is strongly suppressed for
smaller centre-of-mass energies. The variations of the normalization
of the distributions are comparable to the variations of the
normalization of the total cross sections.

\vspace{1.5cm}
\section{Theoretical Uncertainties}
\label{sectionuncertainties}
In the previous section we have seen that the location of the
peak position and the normalization of the vector-current-induced 
total cross section, as well as of its three-momentum distribution,
receive large NNLO corrections in the 
pole mass scheme. In the following two subsections we discuss the
origin of the large corrections in the vector-current-induced total
cross section. For the peak position of the total cross section we 
propose a solution, which allows for a considerable stabilization.

\vspace{1cm}
\subsection{The $1S$ Top Quark Mass and the Peak Position}
\label{subsectionmass}
From past experience in the theoretical description of $B$-meson
decays~\cite{Beneke3,Bigi1}, it is well known that the pole mass,
defined
as the pole of the perturbative quark propagator, although 
infrared-finite to all orders in perturbation 
theory~\cite{Kronfeld1,Tarrach1}, is a concept that is
ambiguous to an amount of order $\Lambda_{QCD}$. This might be a
reflection of the fact that the perturbative quark pole does not exist
in reality because of confinement. Within the framework of perturbation
theory this ambiguity is caused by an $n$-factorial increase of the
coefficients in the perturbative relation between the pole mass and a
short-distance mass such as $\overline{\mbox{MS}}$. The large corrections
are caused by an increasing infrared sensitivity of the perturbative
coefficients for large orders. It has also been shown that the large
top width does not lead to a suppression of these large
corrections~\cite{Willenbrock1}. From this point of
view the unstable behaviour of the peak position in the total cross
section is not unexpected and it would be quite appealing
conceptually to conclude that the use of a short-distance mass instead
of the pole definition would cure the problem. 

In fact, it has been demonstrated in Refs.~\cite{Hoang6,Beneke4} that
the dominant
source of infrared sensitivity in the Green function of the
Schr\"odinger equation~(\ref{NNLOSchroedinger}) comes from the terms
in the static
energy $2 M_t + V(\mbox{\boldmath $r$})$. Whereas the rest (pole) mass
energy $2 M_t$ and
the potential energy $V$ (which has traditionally been calculated in
the pole mass scheme~\cite{Fischler1,Billoire1,Schroeder1,Peter1}) are
individually ambiguous to an amount
of order $\Lambda_{QCD}$~\cite{Beneke3,Bigi1,Aglietti1}, the sum of
both is not~\cite{Hoang6,Beneke4}. This shows that quantities such as
spectra or the total
cross section calculated from Eq.~(\ref{NNLOSchroedinger}) are much
less sensitive to infrared momenta than the pole mass itself,
rendering it an irrelevant mass parameter. Thus, any
sensible mass definition used to parametrize the top-antitop cross
section close to threshold should have no ambiguity of order
$\Lambda_{QCD}$, i.e. it should be a short-distance mass.

However, we emphasize that, in practice, large corrections at lower
orders in perturbation theory in the pole scheme do not necessarily
come from the ambiguity in the pole mass. This is because the
cancellations of infrared sensitive contributions in static and rest
mass energy is a phenomenon that is relevant in large orders where
the corresponding series are dominated by the most infrared sensitive
contributions in the loop integrals. Thus, large corrections in 
perturbation theory at low orders could very well come from scales
which are
not infrared. To get a clearer picture for the case of the peak
position in the total cross section, let us have a look at the size of
the individual corrections to the peak position. In
Table~\ref{tabpeakpositionpole} we have displayed the LO, NLO and NNLO
contributions to the peak position with respect to the pole rest mass
energy
\begin{eqnarray}
M_{peak} & = &
2M_t \, - \, 
\delta M_{peak}^{LO} \, - \, 
\delta M_{peak}^{NLO} \, - \, 
\delta M_{peak}^{NNLO,\beta_0} \, - \, 
\delta M_{peak}^{NNLO,rest} 
\nonumber
\\[3mm] & = &
2M_t \, - \, 
\delta M_{peak}
\,,
\end{eqnarray}
for  $M_t=175$~GeV, $\Gamma_t=1.43$~GeV, $\alpha_s(M_Z)=0.113, 0.118$
and $0.123$, and $\mu=15, 30$ and $60$~GeV.
\begin{table}[t]  
\vskip 7mm
\begin{center}
\begin{tabular}{|c||c||c|c|c|c|c|} \hline
$\mu [\mbox{GeV}]$ & $\alpha_s(M_Z)$ & $\delta M_{peak}^{LO} $ 
                        & $\delta M_{peak}^{NLO} $ 
                        & $\delta M_{peak}^{NNLO,\beta_0} $ 
                        & $\delta M_{peak}^{NNLO,rest} $ 
                        & $\delta M_{peak} $ \\ 
\hline\hline
$15$ & $0.113$ & $1.60$ & $0.28$ & $0.20$ & $0.32$ & $2.40$
\\\hline
$30$ &  & $1.02$ & $0.70$ & $0.27$ & $0.24$ & $2.23$
\\\hline
$60$ &  & $0.35$ & $1.18$ & $0.36$ & $0.18$ & $2.07$
\\ \hline\hline
$15$ & $0.118$ & $1.89$ & $0.31$ & $0.24$ & $0.37$ & $2.80$ 
\\\hline
$30$ &  & $1.26$ & $0.75$ & $0.31$ & $0.27$ & $2.58$
\\\hline
$60$ &  & $0.69$ & $1.09$ & $0.40$ & $0.20$ & $2.39$
\\\hline\hline
$15$ & $0.123$ & $2.17$ & $0.33$ & $0.29$ & $0.43$ & $3.23$
\\\hline
$30$ &  & $1.49$ & $0.81$ & $0.36$ & $0.31$ & $2.96$
\\\hline
$60$ &  & $0.92$ & $1.12$ & $0.45$ & $0.23$ & $2.72$
\\ \hline
\end{tabular}
\caption{\label{tabpeakpositionpole} 
LO, NLO and NNLO contributions to the peak position of the total
vector current induced cross section $R^v$ in GeV in the pole mass scheme
for $M_t=175$~GeV, $\Gamma_t=1.43$~GeV, $\alpha_s(M_Z)=0.113,
0.118$ and $0.123$, and $\mu=15, 30$ and $60$~GeV,
respectively. For the strong coupling two-loop running has been
employed. The results are insensitive to the choice of the cutoff
scale $\Lambda\sim M_t$.
}
\end{center}
\vskip 3mm
\end{table}
At NNLO we have separated from the rest the contributions with the
highest power of
$\beta_0$, which represent the contributions most sensitive to
infrared momenta\footnote{
These contributions are determined by using the replacement rule
$n_f\to -\frac{3}{2}\beta_0$ for the terms with the highest power of
$n_f$,
where $n_f$ is the number of light quark species. This method is
called ``naive non-Abelianization'' and accounts for the most infrared
sensitive contributions in perturbative series relating the pole mass
to a short-distance mass~\cite{Beneke5}. The result of this replacement is
referred to as the ``large-$\beta_0$'' limit.
}. 
All the large-$\beta_0$ terms originate from the Coulomb
potential $V_c$, Eq.~(\ref{NNLOCoulomb}).
From the numbers presented in Table~\ref{tabpeakpositionpole} we see
that depending on the choice of the renormalization scale the
large-$\beta_0$ contributions at NNLO contribute between about $30$ and
$60\%$ to the total NNLO corrections to the peak position. Thus at NNLO
the large shift in the peak position consists to approximately equal
parts of corrections very sensitive to infrared momenta and
corrections coming from subleading infrared terms and relativistic
corrections. From this we can conclude that using some unspecified
short-distance
mass definition instead of the pole mass does not necessarily
lead to smaller NNLO corrections to the peak position because the
most infrared sensitive terms are not yet dominating at NNLO.
For the same reason, we cannot conclude that some unspecified
short-distance mass
definition necessarily leads to a significantly reduced
renormalization scale dependence of the NNLO peak position or to a
smaller correlation between the peak position and the strong coupling.
Thus, the question of which mass definition one should use is not only a
conceptual issue, but also a practical one. 

We formulate two requirements for a proper top mass definition for the
total cross section close to threshold:
\begin{itemize}
\item[A)] it must be a short-distance mass, and 
\item[B)] it must lead to a considerable stabilization
  of the peak location with respect to the order of approximation used
  and also to variations of parameters such as the strong coupling or the
  renormalization scale. 
\end{itemize}
Requirement A reflects the necessity that, if a top mass determination
at the linear collider with uncertainties of $200$~MeV or even
better is intended, the corresponding mass parameter must be free of
intrinsic ambiguities of order $\Lambda_{QCD}$. In addition, only a
short-distance mass can be reliably related to the
$\overline{\mbox{MS}}$ top quark mass, which is the preferred mass
parameter used in calculations at high energies and for top quark
corrections to electroweak precision observables. 
Requirement B ensures that the mass parameter can be extracted from
experimental data with small systematic uncertainties. 

\begin{figure}[t!] 
\begin{center}
\leavevmode
\epsfxsize=3.8cm
\epsffile[200 400 400 530]{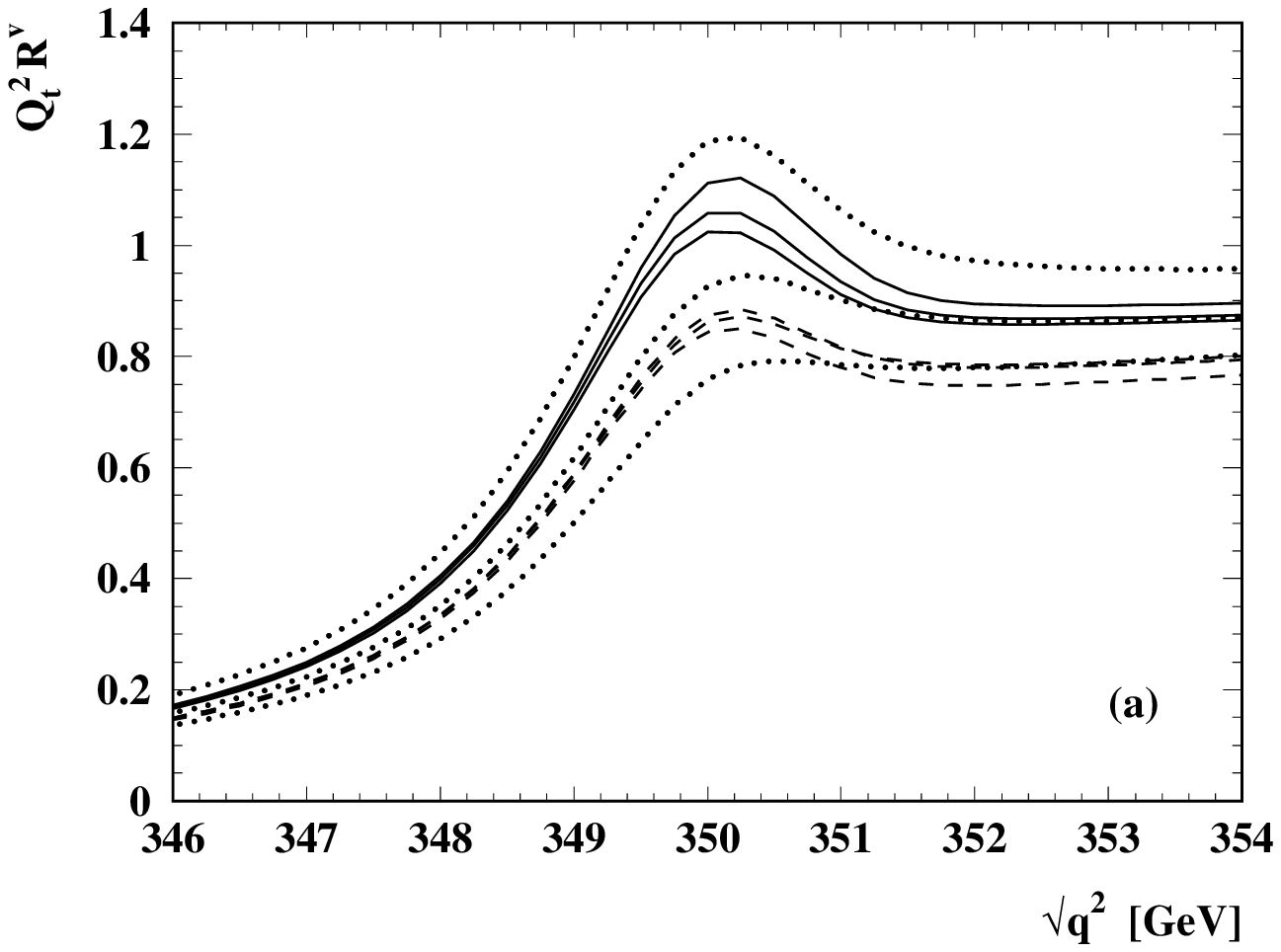}
\hspace{4.2cm}
\epsfxsize=3.8cm
\leavevmode
\epsffile[200 400 400 530]{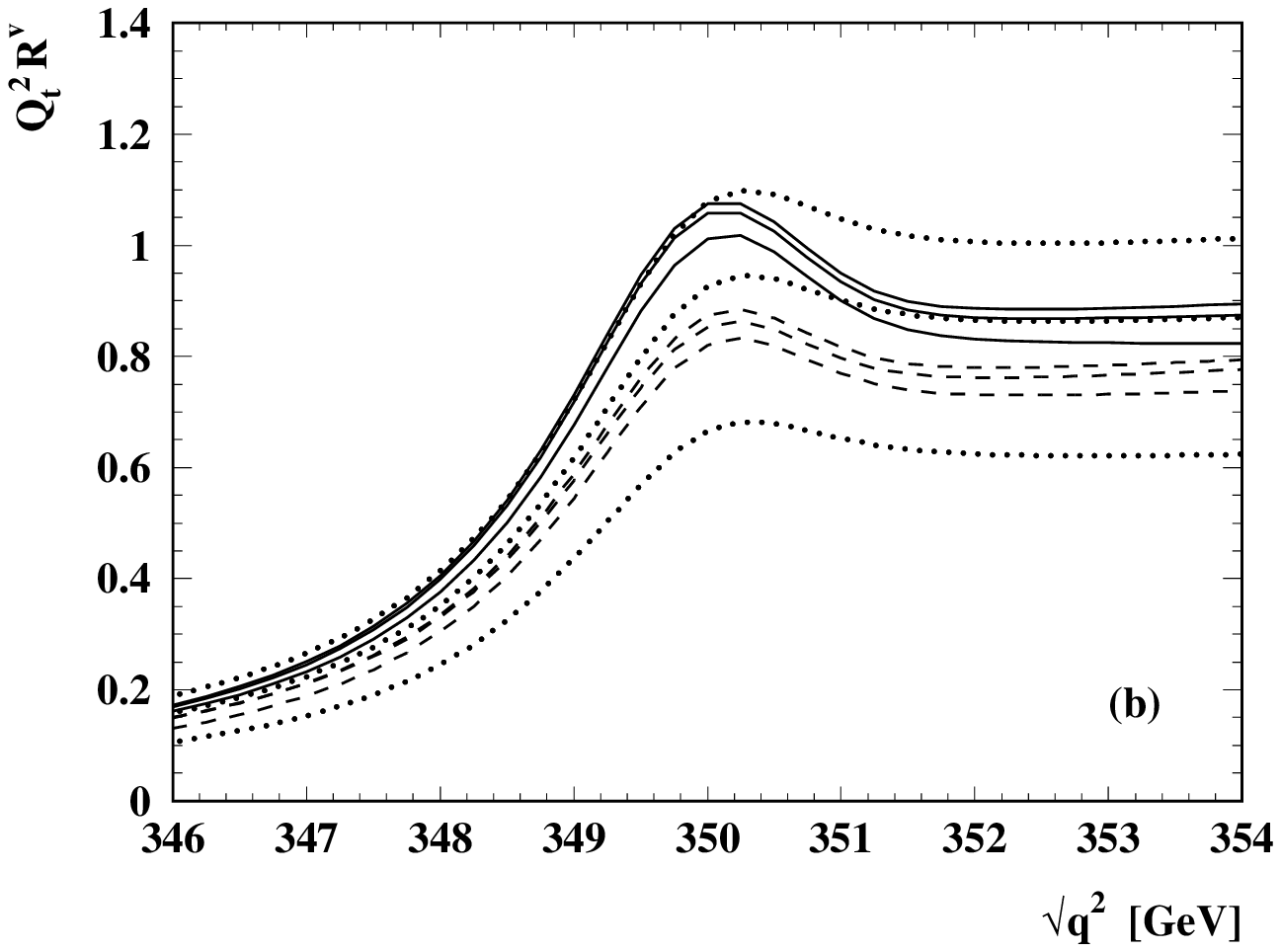}\\[3cm]
\leavevmode
\epsfxsize=3.8cm
\epsffile[200 400 400 530]{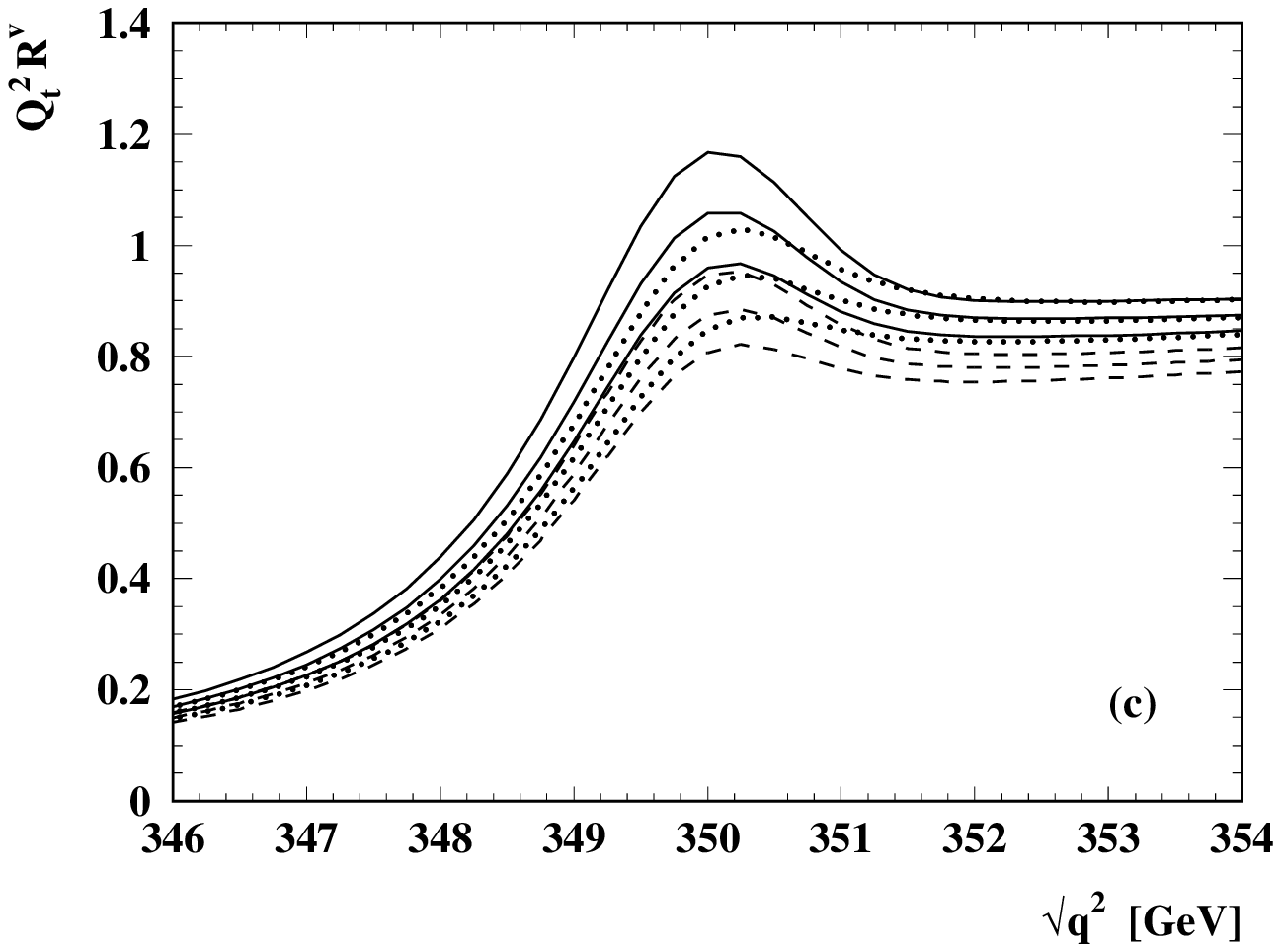}
%
%
\vskip  2.7cm
 \caption{\label{figtotm1S}
The total vector-current-induced cross section $Q_t^2 R^v$ for
centre-of-mass energies $346\,\mbox{GeV}< \sqrt{q^2}< 354$~GeV in the
$1S$ mass scheme.
The dependence on the renormalization scale $\mu$ (a),
on the cutoff $\Lambda$ (b) and on $\alpha_s(M_Z)$ (c) is displayed.
More details and the choice of parameters are given in the text.
}
 \end{center}
\end{figure}

The mass that seems to be most appropriate to us to fulfil this
task, because it is closely related to the peak position in the 
vector-current-induced cross section, is what we call the $1S$ mass,
$M_{1S}$. The $1S$ mass is defined as 
half the perturbative mass of the fictitious toponium $1\,{}^3\!S_1$
ground state, where the top quark is assumed to be stable. 
Expressed in terms of the pole mass, the $1S$ mass reads
($a_s=\alpha_s(\mu)$)~\cite{Melnikov5,Pineda2}: 
\begin{eqnarray}
M_{1S} & = &
M_t \, - \, \epsilon
\frac{M_t\,C_F^2\,a_s^2}{8}
\nonumber
\\ & & 
-  \,\epsilon^2 \,
\frac{M_t\,C_F^2\,a_s^2}{8}\, 
\Big(\frac{a_s}{\pi}\Big)\,\bigg[\,
\beta_0\,\bigg( L + 1 \,\bigg) + \frac{a_1}{2} 
\,\bigg]
\nonumber
\\ & & 
-  \,\epsilon^3
\frac{M_t\,C_F^2\,a_s^2}{8}\, \Big(\frac{a_s}{\pi}\Big)^2\,
\bigg[\,
\beta_0^2\,\bigg(\, \frac{3}{4} L^2 +  L + 
                             \frac{\zeta_3}{2} + \frac{\pi^2}{24} +
                             \frac{1}{4} 
\,\bigg) + 
\beta_0\,\frac{a_1}{2}\,\bigg(\, \frac{3}{2}\,L + 1
\,\bigg)
\nonumber\\[3mm]
& & \hspace{1.5cm} +
\frac{\beta_1}{4}\,\bigg(\, L + 1
\,\bigg) +
\frac{a_1^2}{16} + \frac{a_2}{8} + 
\bigg(\, C_A - \frac{C_F}{48} \,\bigg)\, C_F \pi^2 
\,\bigg]
\,,
\label{1Sdef}
\end{eqnarray}
where 
\begin{eqnarray}
L & \equiv & \ln\Big(\frac{\mu}{C_F\,a_s\,M_t}\Big)
\,,
\end{eqnarray}
and  the contributions at LO, NLO and NNLO are labelled by the powers
$\epsilon$, $\epsilon^2$ and $\epsilon^3$, respectively, of
the auxiliary parameter $\epsilon=1$. 
In general, the electroweak corrections not calculated in this work
can lead to further corrections in Eq.~(\ref{1Sdef}).
We note that $2 M_{1S}$ is not equal
to the actual peak position visible in the vector-current-induced
total cross section because the top quark width leads to an additional
shift of the peak by about $+200$~MeV.\footnote{
The difference between $2 M_{1S}$ and the peak location of the
vector-current-induced total cross section is proportional to
$\Gamma_t$ times a function of $\Gamma_t/(M_t \alpha_s^2)$. For
$\Gamma_t\ll M_t \alpha_s^2$ the difference 
$ M_{peak}-2 M_{1S}$ is proportional to $\Gamma_t^4/(M_t C_F^2
\alpha_s^2)^3$. The size of the difference between $2 M_{1S}$ and the
peak location of about $+200$~MeV can be seen in Figs.~\ref{figtotm1S}.
}
In principle it would also be possible to define a mass that would be
equal to half the actual visible peak position. Except for additional
corrections coming from the top quark width, this would also require
the inclusion of an additional shift coming from the
axial-vector-induced cross section. Such a
definition would, however, not necessarily be more useful,
since the experimentally measurable line-shape of the total cross
section at a future $e^+e^-$ or $\mu^+\mu^-$ collider will 
be distorted by initial state radiation and beamstrahlung. In the
case of a muon collider these effects lead to an additional
shift and in the 
case of the $e^+e^-$ linear collider even to the disappearance of the
peak~\cite{Orange1}. Thus one has to consider the $1S$ mass, like the
$\overline{\mbox{MS}}$ mass, as a fictitious mass
parameter, which to NNLO is defined
through the perturbative series given in
Eq.~(\ref{1Sdef}). Nevertheless, twice the $1S$ mass is
quite close to the peak location and, by construction, leads to a
considerable reduction of the variation of the peak position with
respect to the order of approximation, the strong coupling and the
renormalization scale. 

To show that $M_{1S}$ is indeed a short-distance mass we recall that
the static energy $2M_t + V_c(\mbox{\boldmath r})$ 
represents the dominant infrared sensitive contribution in the
Schr\"odinger equation~(\ref{NNLOSchroedinger}). The difference
between $M_{1S}$ and the pole mass is\footnote{
Strictly speaking, the simple form of Eq.~(\ref{1Spolefull}) is 
true only up to NNLO because 
of retardation effects, which set in at N$^3$LO. Thus, in general,
there would also be a non-trivial integration over time components. 
The form of our proof also depends on the assumption that the static
potential is an infrared finite quantity. That this is most probably
not the case was already pointed out some time ago in
Ref.~\cite{Appelquist1}, because the perturbative
static potential might become sensitive to scales below the inverse
Bohr radius at ${\cal{O}}(\alpha_s^4)$. Some contributions at
${\cal{O}}(\alpha_s^4)$ have recently been calculated in
Ref.~\cite{Pineda3}. Up to ${\cal{O}}(\alpha_s^3)$,
i.e. NNLO in the non-relativistic expansion, the perturbative potential
has been proven to be finite by complete
calculations~\cite{Schroeder1,Peter1}.
Because $M_{1S}$ is defined as a physical quantity this would
not affect the final conclusion that it is a short-distance mass,
but it would change the form of the proof considerably.
}    
\begin{eqnarray}
M_{1S}-M_t & = & \frac{1}{2}\,
\int\frac{d^3\mbox{\boldmath $p$}}{(2\pi)^3}
\frac{d^3\mbox{\boldmath $q$}}{(2\pi)^3}\,
\tilde\Phi_{1S}^*(\mbox{\boldmath $p$})\,
{\cal{H}}(\mbox{\boldmath $p$},\mbox{\boldmath $q$})\,
\tilde\Phi_{1S}(\mbox{\boldmath $q$})
\,,
\label{1Spolefull}
\end{eqnarray}
where ${\cal{H}}$ is the Hamiltonian of Eq.~(\ref{NNLOSchroedinger})
and $\tilde\Phi_{1S}$ the normalized wave function of the $1S$ state
in momentum
space representation. The dominant infrared sensitive contribution in
relation~(\ref{1Spolefull}) reads
\begin{eqnarray}
\Big(\,M_{1S}-M_t\,\Big)^{IR} & \sim & \frac{1}{2}\,
\int\frac{d^3\mbox{\boldmath $p$}}{(2\pi)^3}
\frac{d^3\mbox{\boldmath $q$}}{(2\pi)^3}\,
\tilde\Phi_{1S}^*(\mbox{\boldmath $p$})\,
\tilde V_c(\mbox{\boldmath $p$}-\mbox{\boldmath $q$})\,
\tilde\Phi_{1S}(\mbox{\boldmath $q$})
\,.
\label{1SpoleIR1}
\end{eqnarray}
Because the infrared region in Eq.~(\ref{1SpoleIR1}) is given by
$|\mbox{\boldmath $p$}-\mbox{\boldmath $q$}|<\mu_f$, where $\mu_f$ is
much smaller than the inverse Bohr radius $\sim M_t\alpha_s$, the
characteristic scale governing the dynamics described by the wave
function, we can simplify relation~(\ref{1SpoleIR1}),
\begin{eqnarray}
\Big(\,M_{1S}-M_t\,\Big)^{IR} & \sim & \frac{1}{2}\,
\int\limits^{|\mbox{\boldmath $p$}-\mbox{\boldmath $q$}|<\mu_f}
\frac{d^3\mbox{\boldmath $p$}}{(2\pi)^3}
\frac{d^3\mbox{\boldmath $q$}}{(2\pi)^3}\,
|\tilde\Phi_{1S}(\mbox{\boldmath $p$})|^2\,
\tilde V_c(\mbox{\boldmath $p$}-\mbox{\boldmath $q$})
\nonumber \\ & \sim &
\frac{1}{2}\,\int\limits^{|\mbox{\boldmath $q$}|<\mu_f}\,
\frac{d^3\mbox{\boldmath $q$}}{(2\pi)^3}\,
\tilde V_c(\mbox{\boldmath $q$})
\,.
\label{1SpoleIR2}
\end{eqnarray}
It has been shown in Refs.~\cite{Beneke4,Bigi1} that the RHS of
Eq.~(\ref{1SpoleIR2}) is equivalent to the dominant infrared contributions
of the difference between the $\overline{\mbox{MS}}$ and pole
mass. Therefore the relation between $M_{1S}$ and the
$\overline{\mbox{MS}}$ mass $\bar m_t$ only contains subleading
infrared
contributions, which are suppressed by at least one power of
$1/M_t$. In other words the ambiguity in the relation between $M_{1S}$
and $\bar m_t$ is parametrically of order $\Lambda_{QCD}^2/M_t$. This
proves that $M_{1S}$ is a short-distance mass.
We also see from Eq.~(\ref{1SpoleIR2}) that, if the pole mass is
expressed in terms of the $1S$ mass and if the resulting mass
difference $2(M_t-M_{1S})$ is absorbed into 
the potential, the rest mass and the potential
energy term contained in the total static energy are individually free
of ambiguities of order $\Lambda_{QCD}$. The RHS of Eq.~(\ref{1SpoleIR2})
just subtracts the low momentum (i.e. dominant infrared sensitive)
contribution from the Coulomb potential $V_c(\mbox{\boldmath $x$})$. 

We note that in order to implement the $1S$ mass definition into our
numerical codes, which solve the integral equations~(\ref{Sintegral}) and
(\ref{Pintegral}), we have to invert relation~(\ref{1Sdef}). It has
been shown in Refs.~\cite{Hoang7,Hoang8} that a consistent way to 
achieve this task is to carry out the inversion with respect to the
auxiliary parameter $\epsilon$. For the reason that this modified
perturbative expansion has been applied for the first time to express
inclusive $B$ decays in terms of the $\Upsilon(1S)$ mass, it has
been called the ``Upsilon expansion''.~\cite{Hoang7,Hoang8} If the
$1S$ mass is expressed in terms of the $\overline{\mbox{MS}}$ mass,
which is related to the pole 
mass by a series of the form $\bar m_t-M_t = M_t \sum_{n=1}^\infty
a_n\alpha_s^n$,
one has to consider a term $\propto \alpha_s^n$ in this relation of
order $\epsilon^n$ in the Upsilon expansion. In other words, if one
relates the $1S$ mass to  a mass which is different from the pole
mass, one must combine terms of different order in
$\alpha_s$. As an example, this means that in order to relate the
N$^k$LO $1S$ mass to the $\overline{\mbox{MS}}$ mass one needs to know
its relation to the pole mass to ${\cal{O}}(\alpha_s^{k+1})$.
This is necessary because this is the only way in which the high order
large perturbative corrections coming from infrared-sensitive terms
are cancelled.

\begin{figure}[t!] 
\begin{center}
\leavevmode
\epsfxsize=3.8cm
\epsffile[200 400 400 530]{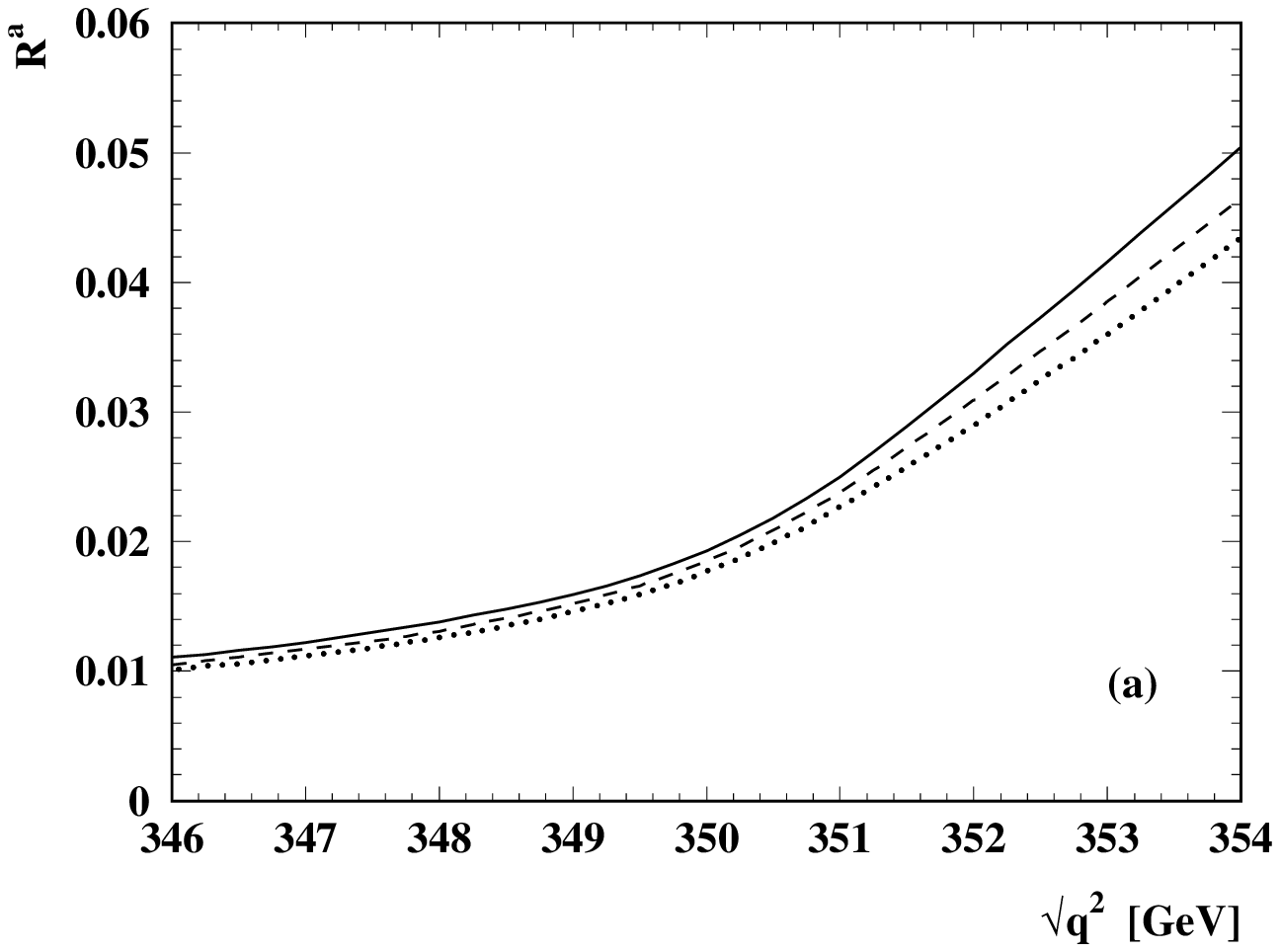}
\hspace{4.2cm}
\epsfxsize=3.8cm
\leavevmode
\epsffile[200 400 400 530]{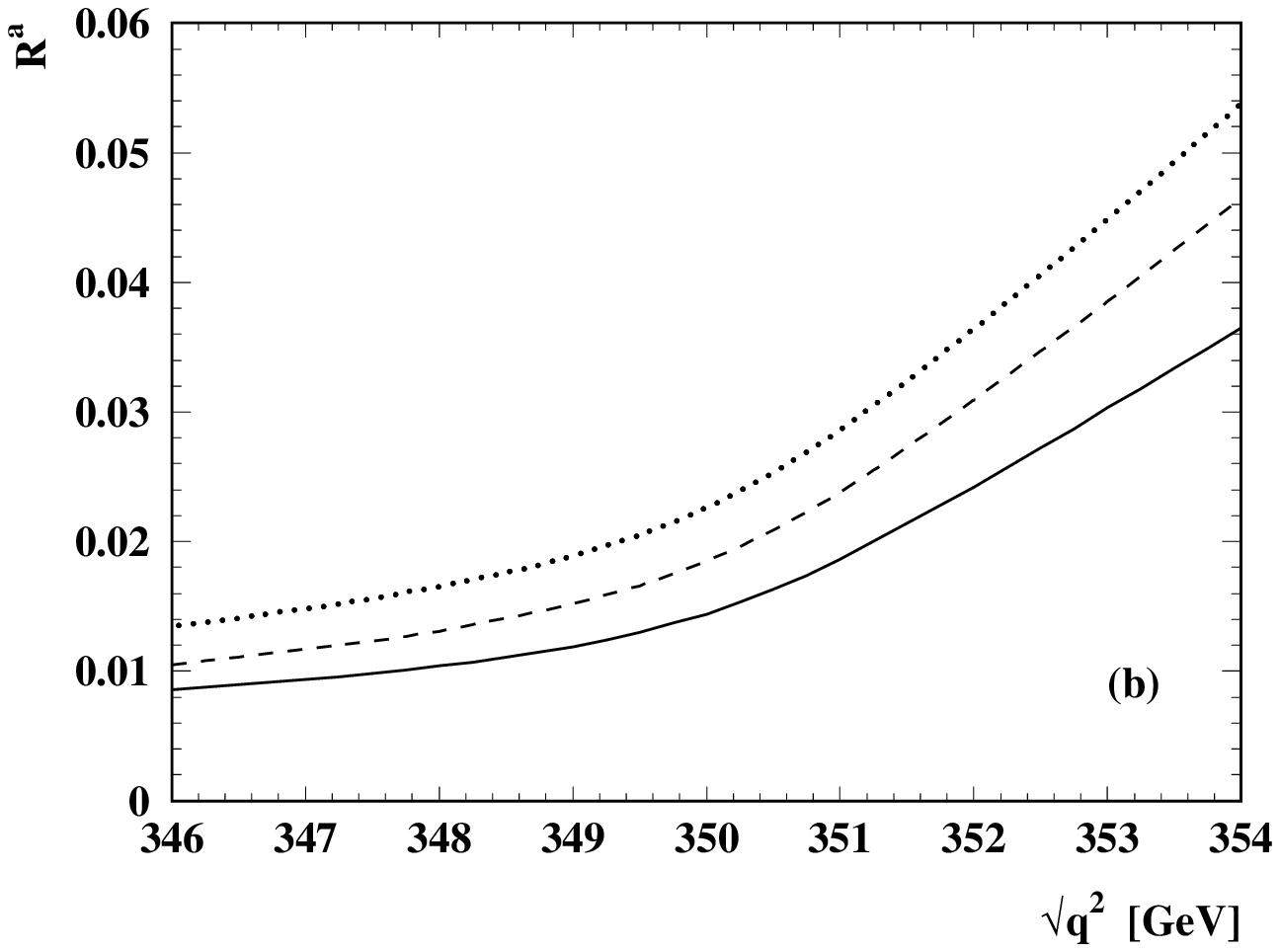}\\[3cm]
\leavevmode
\epsfxsize=3.8cm
\epsffile[200 400 400 530]{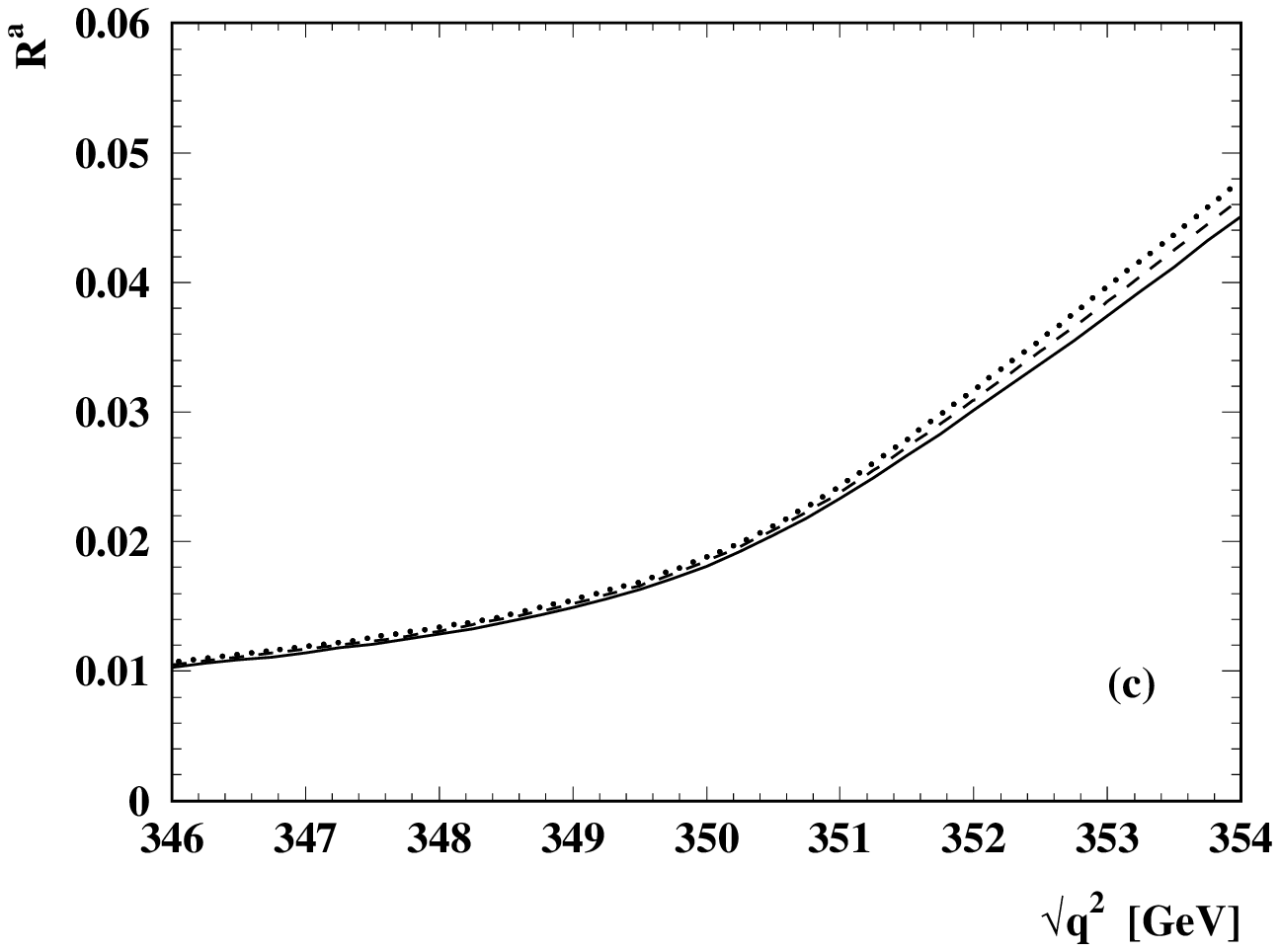}
%
%
\vskip  2.7cm
 \caption{\label{figtotm1Sax} 
The total axial-vector-current-induced cross section $R^a$ for
centre-of-mass energies $346\,\mbox{GeV}< \sqrt{q^2}< 354$~GeV in the
$1S$ mass scheme.
The dependence on the renormalization scale $\mu$ (a),
on the cutoff $\Lambda$ (b) and on $\alpha_s(M_Z)$ (c) is
displayed. More details and the choice of parameters are given in
the text.
}
 \end{center}
\end{figure}
\begin{figure}[t!] 
\begin{center}
\leavevmode
\epsfxsize=3.8cm
\epsffile[200 400 400 530]{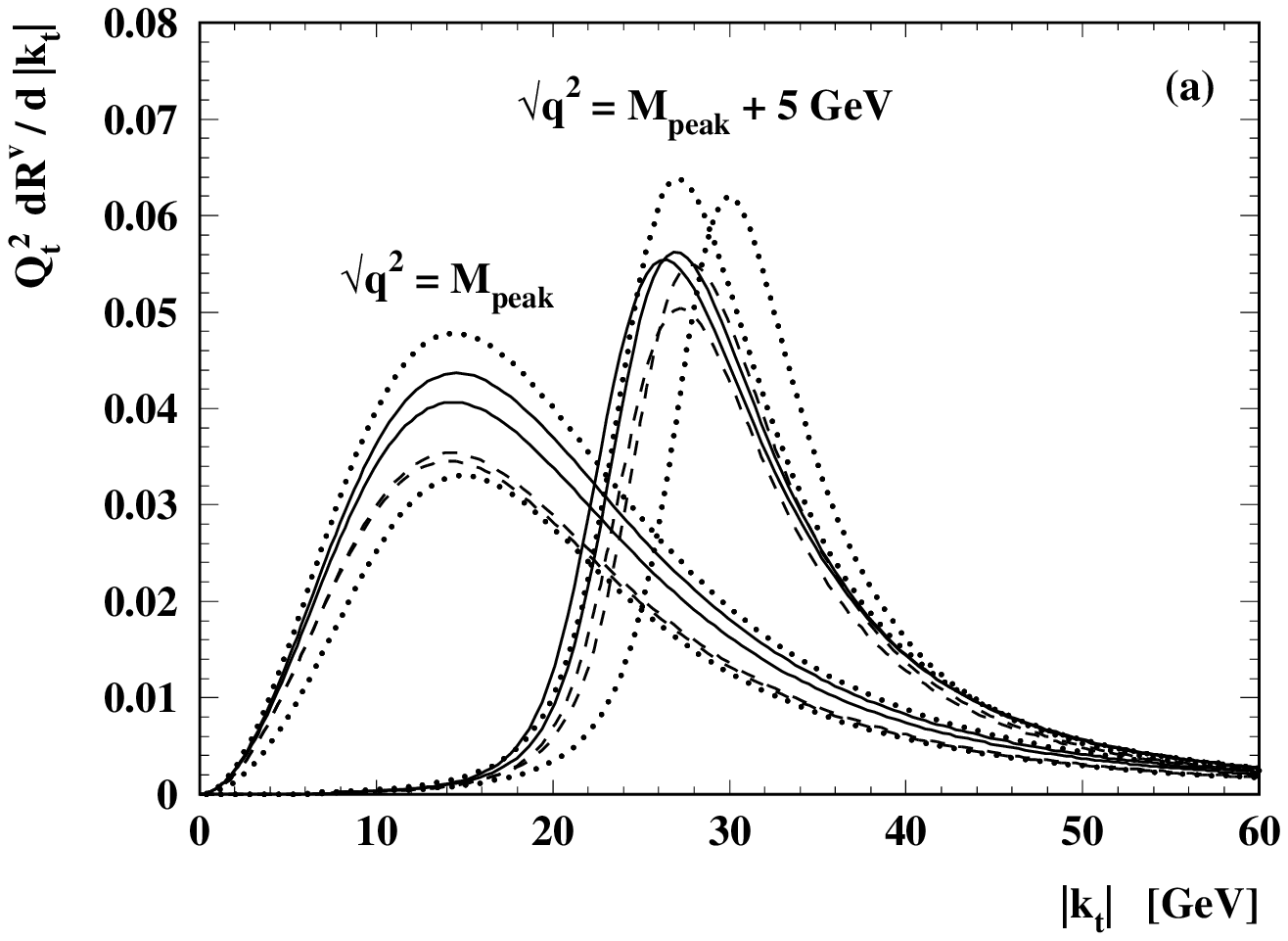}
\hspace{4.4cm}
\epsfxsize=3.8cm
\leavevmode
\epsffile[200 400 400 530]{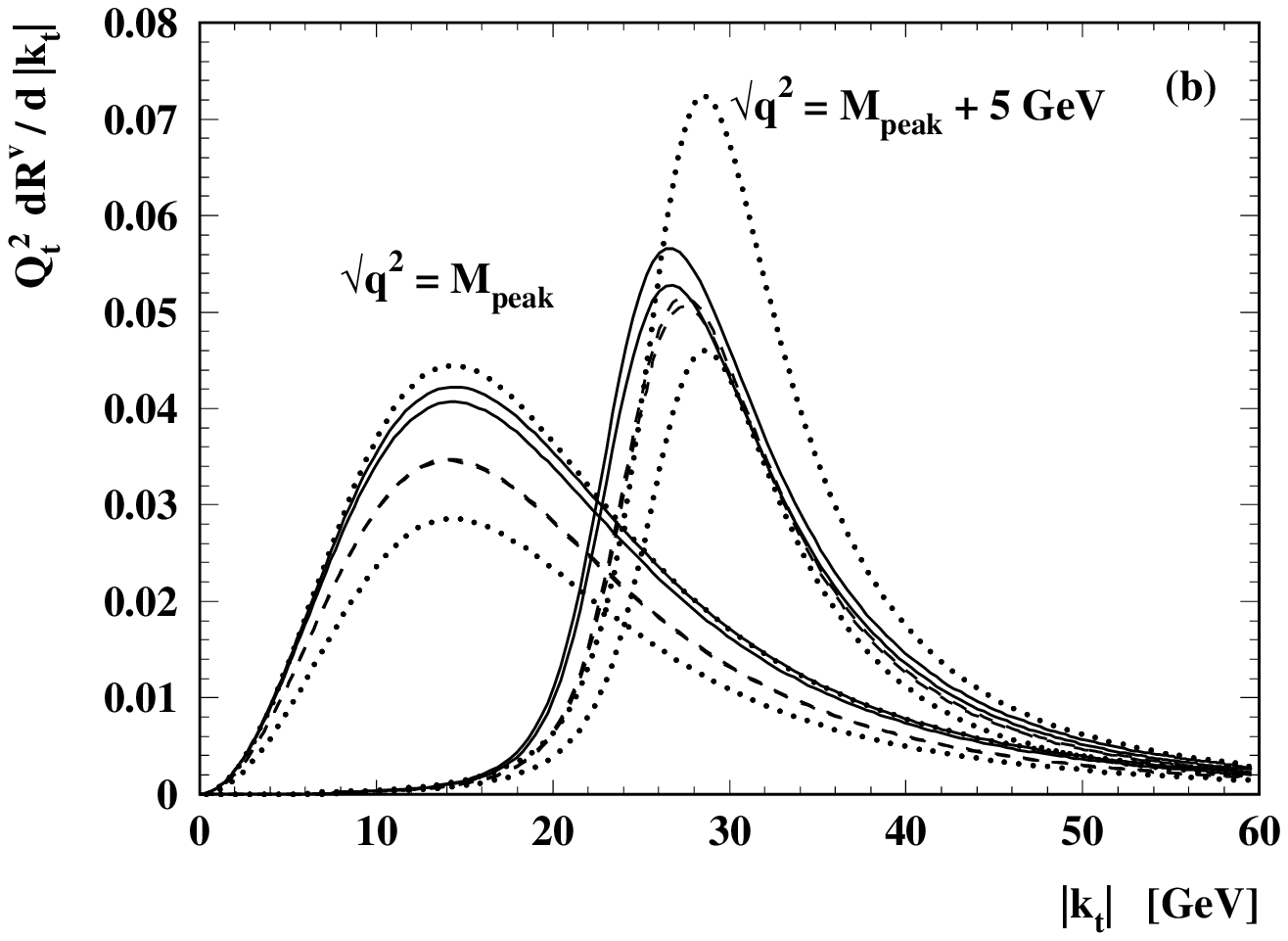}\\[3cm]
\leavevmode
\epsfxsize=3.8cm
\epsffile[200 400 400 530]{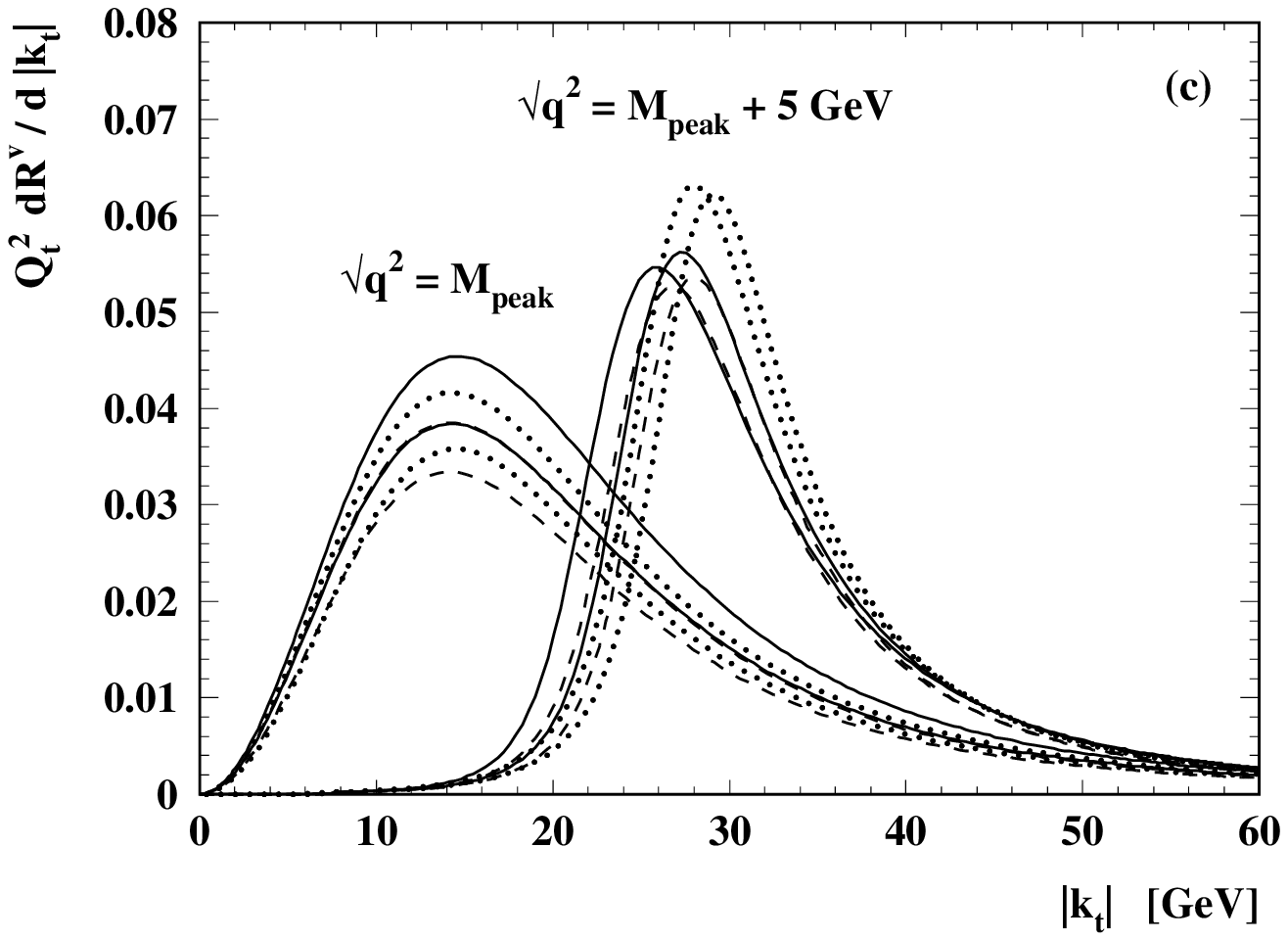}
%
%
\vskip  2.7cm
 \caption{\label{figdistm1S} 
The three-momentum distribution of the vector-current-induced cross
section $Q_t^2 R^v$ for centre-of-mass energies $\sqrt{q^2}=M_{peak}$
and $M_{peak}+5$~GeV in the $1S$ mass scheme.
The dependence on the renormalization scale $\mu$ (a),
on the cutoff $\Lambda$ (b) and on $\alpha_s(M_Z)$ (c) is displayed.
More details and the choice of parameters are given in the text.
}
 \end{center}
\end{figure}
\begin{figure}[t!] 
\begin{center}
\leavevmode
\epsfxsize=3.8cm
\epsffile[200 400 400 530]{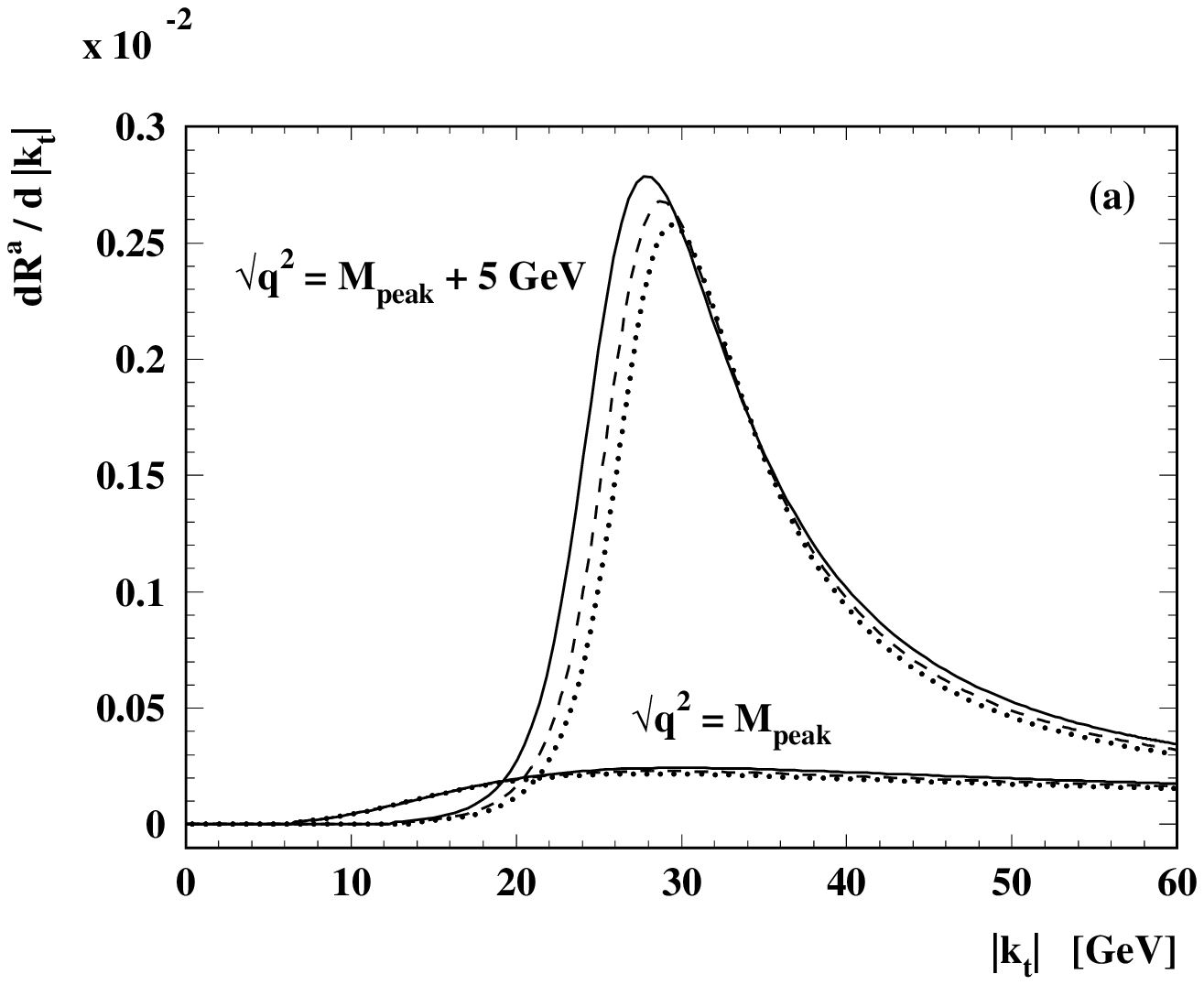}
\hspace{4.4cm}
\epsfxsize=3.8cm
\leavevmode
\epsffile[200 400 400 530]{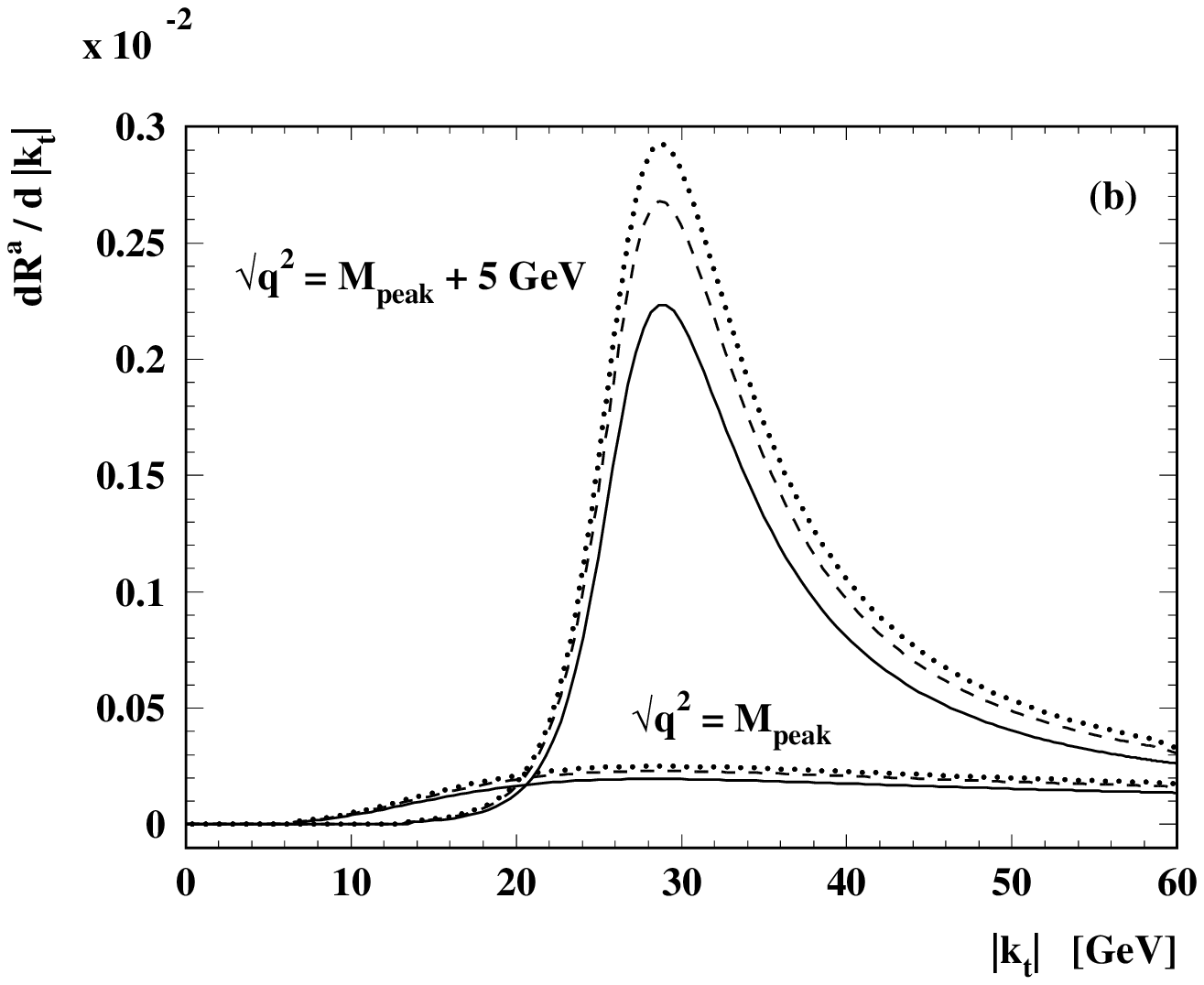}\\[3cm]
\leavevmode
\epsfxsize=3.8cm
\epsffile[200 400 400 530]{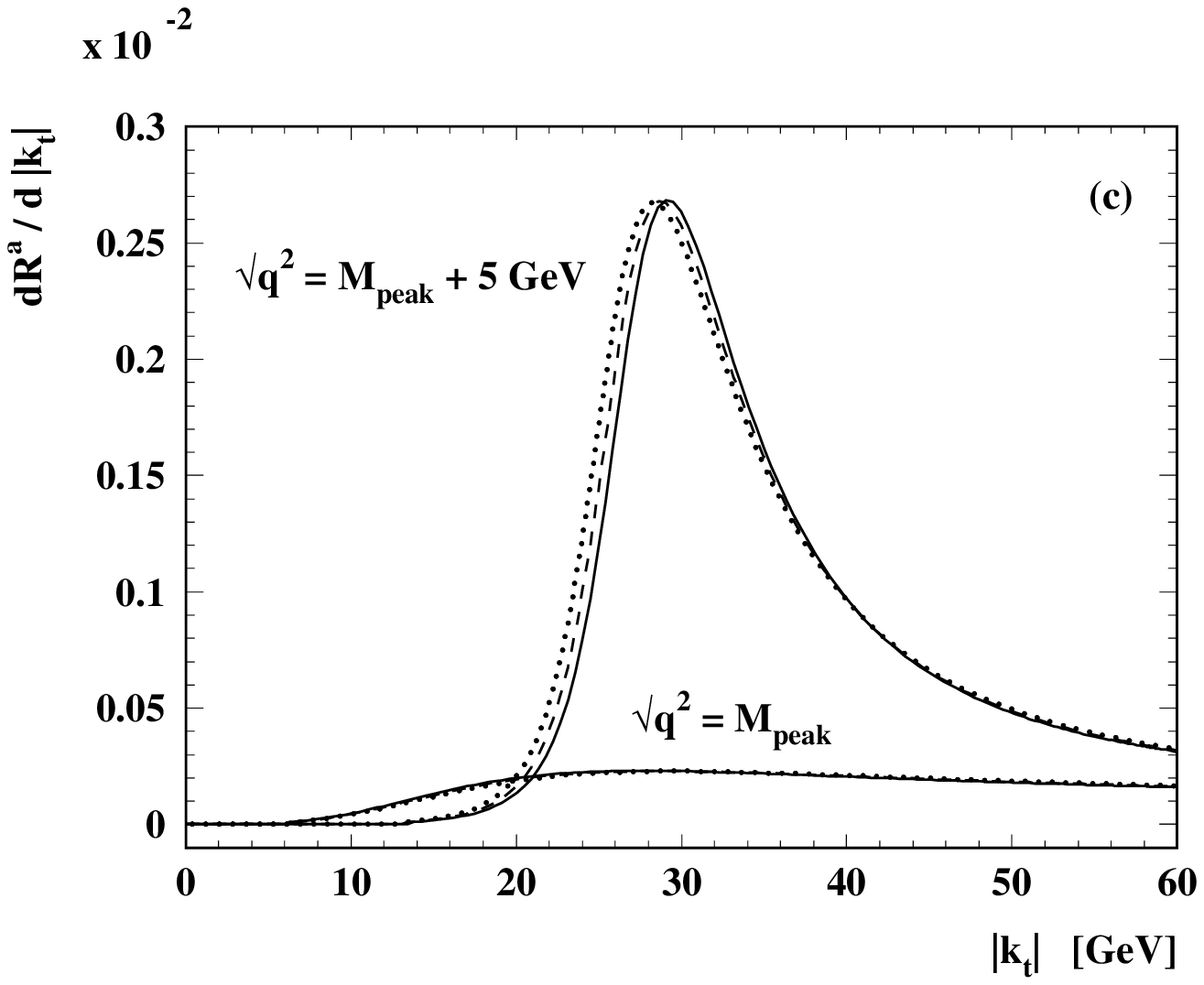}
%
%
\vskip  2.7cm
 \caption{\label{figaxdistm1S} 
The three-momentum distribution of the axial-vector-current-induced
cross section $R^a$ for centre-of-mass energies
$\sqrt{q^2}=M_{peak}$ and $M_{peak}+5$~GeV in the $1S$ mass scheme.
The dependence on the renormalization scale $\mu$ (a),
on the cutoff $\Lambda$ (b) and on $\alpha_s(M_Z)$ (c) is displayed.
More details and the choice of parameters are given in the text.
}
 \end{center}
\end{figure}

In Figs.~\ref{figtotm1S} the total vector-current-induced cross
section $Q_t^2 R^v$ is displayed in the $1S$ scheme for
$346\,\mbox{GeV}<\sqrt{q^2}<354\,\mbox{GeV}$
at LO (dotted lines), NLO (dashed lines) and NNLO (solid lines).
In all figures shown in this section the top quark width is chosen as
$\Gamma_t=1.43$~GeV and the top quark $1S$ mass as $M_{1S}=175$~GeV.
Figure~\ref{figtotm1S}a displays the dependence on the renormalization
scale for $\mu=15$, $30$ and $60$~GeV for
$\alpha_s(M_Z)=0.118$ and $\Lambda=175$~GeV. 
At LO and NNLO the choices $\mu=15$, $30$ and $60$~GeV correspond to
the upper, middle and lower curves. At NLO the choices $\mu=15$, $30$
and $60$~GeV correspond to the lower, middle and upper curves for
centre-of-mass energies below the peak position.
In Fig.~\ref{figtotm1S}b the dependence of $Q_t^2 R^v$ on the choice of
the cutoff $\Lambda$ is shown for $\alpha_s(M_Z)=0.118$, $\mu=30$~GeV
and $\Lambda=90$ (lower curves), $175$ (middle curves) and $350$~GeV
(upper curves). Figure~\ref{figtotm1S}c displays the dependence of $Q_t^2
R^v$ on the choice of $\alpha_s(M_Z)$ for $\alpha_s(M_z)=0.113$ (lower
curves), $0.118$ (middle curves) and $0.123$ (upper curves) and
$\Lambda=175$~GeV, $\mu=30$~GeV.  
Comparing the result with the curves displayed in
Figs.~\ref{figtotpole}, the improvement of the stability of the peak
position is evident. The strong dependence on the renormalization
scale and the strong correlation with $\alpha_s(M_Z)$ have vanished.
However, we also observe that the large
corrections in the normalization of the curves are essentially not
affected at all.

In Figs.~\ref{figtotm1Sax}a,b,c the total axial-vector-current-induced cross
section $R^a$ is displayed in the $1S$ mass scheme for the same input
parameters as in Figs.~\ref{figtotm1S}. Figure~\ref{figtotm1Sax}a shows
the dependence on the renormalization scale for $\mu=15$ (solid line),
$30$ (dashed line) and $60$~GeV (dotted
line), respectively. Figure~\ref{figtotm1Sax}b exhibits the dependence
of the cutoff for  $\Lambda=90$ (solid line), $175$ (dashed line) and
$350$~GeV (dotted line). Figure~\ref{figtotpoleax}c shows $R^a$ for 
$\alpha_s(M_z)=0.113$ (solid line), $0.118$ (dashed line) and $0.123$
(dotted line). Compared to the plots in the pole mass scheme, we find a
slightly smaller variation in the normalization with respect to the
renormalization scale and the choice of $\alpha_s(M_Z)$.  
Clearly the effects of using the $1S$ scheme instead of the pole one
are much smaller in the axial-vector case because no peak is
visible there.

In Figs.~\ref{figdistm1S}a,b and c the LO (dotted lines), NLO (dashed
lines)  and NNLO (solid lines) top-antitop
vector-current-induced three-momentum distribution
$Q_t^2 d R^v/d |\mbox{\boldmath$k$}_t|$ is shown for
$0 < |\mbox{\boldmath$k$}_t| < 60$~GeV in the $1S$ mass scheme for
centre-of-mass energies exactly on top of the visible peak,
$\sqrt{q^2}=M_{peak}$ and 
for $\sqrt{q^2}=M_{peak}+5$~GeV. The input parameters have been chosen
as in Figs.~\ref{figtotm1S}. Figure~\ref{figdistm1S}a shows
the distributions for $\mu=15$ and $60$~GeV. At LO and NNLO
$\mu=15$~GeV corresponds to the upper curves and
$\mu=60$~GeV to the lower curves for centre-of-mass energies below the
peak. At NLO $\mu=60$~GeV corresponds to
the higher peak and $\mu=15$~GeV to the lower. 
Figure~\ref{figdistm1S}b displays the dependence of the distributions
on the cutoff for $\Lambda=90$ (lower curves) and $350$~GeV (upper
curves), and Fig.~\ref{figdistm1S}c exhibits the dependence of the
distributions on the strong coupling for $\alpha_s(M_Z)=0.113$ and
$0.123$. Below the peak the larger value of $\alpha_s(M_Z)$
always corresponds to the upper curve. 

In Figs.~\ref{figaxdistm1S}a,b and c the
top-antitop axial-vector-current-induced three-momentum distribution
$d R^a/d |\mbox{\boldmath$k$}_t|$ is shown in the $1S$ scheme for
$0 < |\mbox{\boldmath$k$}_t| < 60$~GeV and both for centre-of-mass
energies exactly on top of the visible peak, $\sqrt{q^2}=M_{peak}$, and
for $\sqrt{q^2}=M_{peak}+5$~GeV. The input parameters have been chosen
as before. Figure~\ref{figaxdistm1S}a shows
the distribution for $\mu=15$ (solid curves), $30$ (dashed curves)
and $60$~GeV (dotted curves). Figure~\ref{figaxdistm1S}b displays the
dependence of the distribution on the cutoff for $\Lambda=90$ (solid
curves), $175$ (dashed curves) and $350$~GeV (dotted curves), and
Fig.~\ref{figaxdistm1S}c exhibits the dependence of the distribution
on the strong coupling for $\alpha_s(M_Z)=0.113$ (solid curves)
$0.118$ (dashed curves) and $0.123$ (dotted curves). 
The curves shown in Figs.~\ref{figaxdistm1S} are somewhat higher than
in Figs.~\ref{figdistpoleax}, because the choice of $175$~GeV for the
top quark mass corresponds to a higher value for $M_{peak}$ in the
$1S$ scheme. From  
Figs.~\ref{figdistm1S} and \ref{figaxdistm1S} it is evident that the
$1S$ scheme does not essentially affect at all the three-momentum
distributions. Compared to the results in the pole mass scheme the
variations of the peak position remain unchanged. 
This can be understood from the fact that a mass redefinition
corresponds to a shift in the centre-of-mass energy, but leaves the
definition of the off-shell top quark three-momentum unchanged. 

In Table~\ref{tabpeakposition1S} we have displayed the LO, NLO and NNLO
corrections to the peak position with respect to $2 M_{1S}$: 
\begin{eqnarray}
M_{peak} & = &
2M_{1S} \, + \, 
\delta M_{peak,1S}^{LO} \, + \, 
\delta M_{peak,1S}^{NLO} \, + \, 
\delta M_{peak,1S}^{NNLO} 
\nonumber
\\[3mm] & = &
2M_{1S} \, + \, 
\delta M_{peak,1S}
\,,
\end{eqnarray}
in the
$1S$ mass scheme for $M_{1S}=175$~GeV, $\Gamma_t=1.43$~GeV,
$\alpha_s(M_Z)=0.113, 0.118$ and $0.123$, and $\mu=15, 30$ and
$60$~GeV for
various choices of the renormalization scale $\mu$ and the strong
coupling $\alpha_s(M_Z)$.
\begin{table}[t]  
\vskip 7mm
\begin{center}
\begin{tabular}{|c||c||c|c|c|c|} \hline
$\mu [\mbox{GeV}]$ & $\alpha_s(M_Z)$ & $\delta M_{peak,1S}^{LO} $ 
                        & $\delta M_{peak,1S}^{NLO} $ 
                        & $\delta M_{peak,1S}^{NNLO} $  
                        & $\delta M_{peak,1S} $ \\ 
\hline\hline
$15$ & $0.113$ & $0.21$ & $0.03$ & $-0.03$ & $0.20$
\\\hline
$30$ &  & $0.38$ & $-0.11$ & $-0.09$ & $0.17$
\\\hline
$60$ &  & $0.78$ & $-0.50$ & $-0.11$ & $0.17$
\\ \hline\hline
$15$ & $0.118$ & $0.16$ & $0.02$ & $-0.00$ & $0.17$
\\\hline
$30$ &  & $0.30$ & $-0.09$ & $-0.08$ & $0.12$
\\\hline
$60$ &  & $0.54$ & $-0.32$ & $-0.10$ & $0.11$
\\\hline\hline
$15$ & $0.123$ & $0.12$ & $-0.00$ & $0.04$ & $0.16$
\\\hline
$30$ &  & $0.23$ & $-0.07$ & $-0.08$ & $0.08$
\\\hline
$60$ &  & $0.42$ & $-0.26$ & $-0.10$ & $0.07$ 
\\ \hline
\end{tabular}
\caption{\label{tabpeakposition1S} 
LO, NLO and NNLO contributions to the peak position of the total
vector-current-induced cross section $R^v$ in GeV in the $1S$ mass
scheme for $M_{1S}=175$~GeV, $\Gamma_t=1.43$~GeV,
$\alpha_s(M_Z)=0.113, 0.118$ and $0.123$, and
$\mu=15, 30$ and $60$~GeV, respectively. For the strong coupling
two-loop running has
been employed. The results are insensitive to the choice of the cutoff
scale $\Lambda\sim 175$~GeV. 
}
\end{center}
\vskip 3mm
\end{table}
Taking the size of the NLO and NNLO corrections as a measure for the
present theoretical uncertainty in the peak position, and assuming that
the latter can be used to estimate the theoretical uncertainty in
the determination of $M_{1S}$, we find that this uncertainty is
approximately $200$~MeV.
If the effects of beamstrahlung and initial state radiation
at a future $e^+e^-$ or muon pair collider do not spoil a precise
determination of the $1S$ mass with an uncertainty of $200$~MeV
one has to ask the question how $M_{1S}$ is related to the top mass
parameters usually used for calculations of physical observables
that are not related to the threshold regime. In principle, one could
use $M_{1S}$ as a new top mass parameter in its own right. This would,
of course, require that all formulae be expressed in
terms of $M_{1S}$, using the Upsilon expansion discussed after
Eq.~(\ref{1SpoleIR2}). A more economical way is to relate
the $1S$ mass to the $\overline{\mbox{MS}}$ top quark mass, which is 
a common mass parameter for perturbative calculations involving heavy
quarks and which, in a number of cases, even leads to improved
convergence properties of the perturbative series.\footnote{
Prominent cases are the top quark QCD corrections to the
$\rho$-parameter~\cite{Chetyrkin1} and the massive quark pair production cross
section at large energies~\cite{HoangKT1}. 
}
Because the $1S$ mass is a short-distance mass, its perturbative
relation to the $\overline{\mbox{MS}}$ mass is much better behaved at
large orders than the corresponding relation of the pole mass. The
relation between $M_{1S}$ and
$\bar m_t(\bar m_t)$ can be derived from Eq.~(\ref{1Sdef}) and the
relation between pole mass and $\bar m_t(\bar m_t)$ using the Upsilon
expansion discussed above. We emphasize that the
three-loop relation between the pole and the  $\overline{\mbox{MS}}$
mass is needed
to relate $M_{1S}$ and $\bar m_t(\bar m_t)$ at NNLO accuracy. Assuming
that those 3-loop corrections can be approximated by the known
corrections in the large-$\beta_0$ limit~\cite{Beneke5}, we find the
following numerical value for  $\bar m_t(\bar m_t)$ for
$M_{1S}=175\pm 0.2$~GeV and $\alpha_s(M_Z)=0.118\pm x\,0.001$,
\begin{eqnarray}
\bar m_t(\bar m_t)  & = & 
\bigg[\,
175 - 7.58\, \epsilon (\mbox{LO}) - 
0.96\, \epsilon^2 (\mbox{NLO}) - 
0.23\, \epsilon^3 (\mbox{NNLO})
\nonumber
\\[3mm] & & \hspace{3cm}
\pm 0.2 (\delta M_{1S})
\pm x\,0.07 (\delta \alpha_s)
\,\bigg]~\mbox{GeV}
\,.
\label{MSmassestimate}
\end{eqnarray}
For the numbers given in Eq.~(\ref{MSmassestimate}) we have assumed an
uncertainty in the value of the strong coupling at $M_Z$ of $x \,0.001$
in order to demonstrate the importance of $\alpha_s$ for the
determination of $\bar m_t(\bar m_t)$. This uncertainty is independent
of the order to which the relation between the pole and the
$\overline{\mbox{MS}}$ mass is known because it comes from the LO
term. We note that this fact shows that the strong correlation of the
peak position 
to the strong coupling, which was visible in the pole mass scheme,
is not necessarily eliminated by adopting the $1S$ scheme. This
correlation might come back whenever the $1S$ mass is related to
another short-distance mass or is used as a parameter in other
quantities. However, the use of the $1S$ mass has the advantage to
free the process of the mass extraction from the total cross section
close to threshold also from strong dependences on other parameters
such as the renormalization scale or the order of approximation
used. Therefore systematic uncertainties are expected to be
smaller if the $1S$ scheme is used for the threshold calculations.
Equation~(\ref{MSmassestimate}) shows that the knowledge of the
3-loop corrections in the relation of pole and $\overline{\mbox{MS}}$
mass and a small uncertainty in $\alpha_s(M_Z)$ are crucial for a
determination of  $\bar m_t(\bar m_t)$ with
uncertainties comparable to $\delta M_{1S}$.  

In recent literature there have been two other proposals for
alternative short-distance mass definitions, which can also be used for
a measurement of the top quark mass from the total cross section. In
Refs.~\cite{Voloshin1,Bigi2} the ``low scale running mass'' was
proposed to subtract 
the infrared behaviour from the heavy quark self energy. The ``low
scale running mass'' was devised in order to improve the convergence
of the perturbative series describing the contributions leading in
$1/M_b$ in inclusive $B$-meson decays. Due to the
universality of the dominant infrared sensitive contribution, the low
scale running mass can also serve as a top mass definition, which leads
to an improved stability of the peak position in the total cross
section. The low scale running mass depends on the cutoff $\mu_{LS}$,
which limits the momenta that are subtracted from the self energy. At
order $\alpha_s$ (i.e. at LO) its relation to the pole mass
reads~\cite{Voloshin1,Bigi2}
\begin{eqnarray}
m_t^{LS}(\mu_{LS}) - M_t & = & 
-\,\frac{16}{9}\,\frac{\alpha_s}{\pi}\,\mu_{LS}\,
\bigg[\,
1 + {\cal{O}}(\alpha_s) + {\cal{O}}\Big(\frac{\mu_{LS}}{M_t}\Big)
\,\bigg]
\,.
\label{lowscalepole}
\end{eqnarray}
By adjusting the scale $\mu_{LS}$ in such a way that the RHS of
Eq.~(\ref{lowscalepole}) is comparable in size to the RHS
of Eq.~(\ref{1Sdef}) the position of the peak in the total cross
section can be stabilized. In Ref.~\cite{Beneke4} the 
``potential-subtracted'' mass was proposed. It subtracts the dominant
infrared-sensitive contribution
in the Schr\"odinger equation~(\ref{NNLOSchroedinger}), which is
contained in the static potential $V_c$. The subtraction is in fact
equal to the RHS of Eq.~(\ref{1SpoleIR2}). Like the low scale running
mass, the potential-subtracted mass depends on a cutoff, $\mu_{PS}$. At
order $\alpha_s$ (LO) the relation to the pole mass reads~\cite{Beneke4}
\begin{eqnarray}
m_t^{PS}(\mu_{PS}) - M_t & = & 
-\,\frac{4}{3}\,\frac{\alpha_s}{\pi}\,\mu_{PS}\,
\bigg[\,
1 + {\cal{O}}(\alpha_s)
\,\bigg]
\,.
\label{polesubtractedpole}
\end{eqnarray}
As for the low scale running mass, the scale $\mu_{PS}$ can be
adjusted in such a way that the RHS of Eq.~(\ref{polesubtractedpole})
is comparable in size to the RHS of Eq.~(\ref{1Sdef}). To achieve this,
$\mu_{PS}$ has to be chosen of the order of the inverse Bohr radius 
$\sim M_t\alpha_s$, which is much larger than the scale $\mu_f\ll
M_t\alpha_s$ introduced in
Eq.~(\ref{1SpoleIR2}). For $\mu_{PS}=\frac{4}{3}\mu_{LS}$ the low
scale running and the potential-subtracted mass lead to approximately
equivalent results. However, the stabilization of the peak position can
be expected to be slightly worse than for the $1S$ mass if $\mu_{PS}$
or $\frac{4}{3}\mu_{LS}$ are not fine-tuned. In addition, the results
that could finally be obtained for the $\overline{\mbox{MS}}$ top
mass can depend on the value that is chosen for the cutoff scale
$\mu_{LS}$ and $\mu_{PS}$.

\vspace{1cm}
\subsection{Normalization of the Total Cross Section}
\label{subsectionnormalization}
In the previous subsection we have demonstrated that a proper
redefinition of the top quark mass leads to a considerable improvement
in the stability of the peak position in the vector-current-induced
total cross section $R^v$. However, there have been only marginal
changes in the size of
the NNLO corrections to the overall normalization of the
line-shape. Compared to the NLO normalization of the total 
vector-current-induced cross section, 
the NNLO corrections are between $15$ and $25\%$, which is rather
large if one recalls that the NNLO corrections 
are parametrically of order $v^2\sim\alpha_s^2$. In this
section we try to find some answers to the question, whether the large
NNLO corrections to the normalization of the total 
vector-current-induced cross section have
to be interpreted as a sign that the non-relativistic expansion for the
top-antitop cross section close to threshold breaks down. Clearly,
this question can only be answered reliably after the complete
N${^3}$LO corrections have been determined, which are, unfortunately,
beyond the capabilities of present technology. We therefore analyse the
NNLO corrections to the normalization of the total cross section with
respect to their sensitivity to infrared momenta and carry out a
comparison to the one- and two-loop cross section for energies far
above the 
top-antitop threshold, where conventional perturbation theory in
$\alpha_s$ is believed to be reliable. We provide arguments that the
large NNLO corrections to the normalization are genuine
${\cal{O}}(v^2,\alpha_s^2)$ relativistic corrections, which cannot be
removed by changing the definition of $\alpha_s$ or the top quark
mass, and that their size does not necessarily indicate a breakdown of
the non-relativistic expansion used in this work.

\begin{figure}[t!] 
\begin{center}
\leavevmode
\epsfxsize=4.5cm
\epsffile[200 400 400 530]{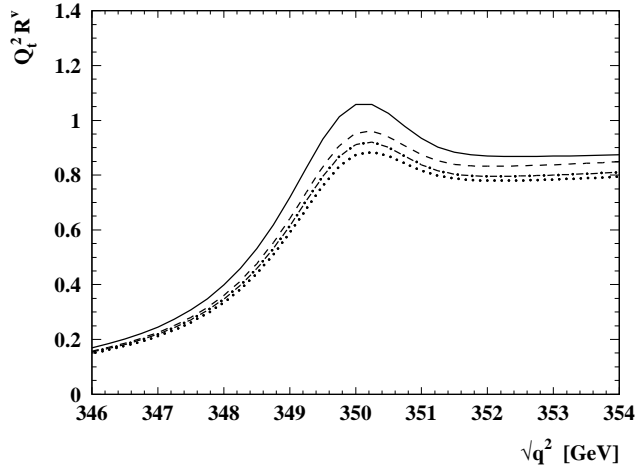}
%
%
\vskip  2.7cm
 \caption{\label{figcompareinfrared}
The total vector-current-induced
cross section $Q_t^2 R^v$ close to threshold for $M_{1S}=\Lambda=175$~GeV,
$\mu=30$~GeV, $\alpha_s(M_Z)=0.118$ and $\Gamma_t=1.43$~GeV at NLO
(dotted curve) and NNLO (solid curve). The dash-dotted curve is NLO
including also the NNLO corrections to the Coulomb potential $V_c$,
and the dashed line contains, in addition, all Abelian NNLO
corrections. The differences between the curves indicates the size of
individual NNLO relativistic corrections.
}
 \end{center}
\end{figure}

As far as a redefinition of the top quark mass is concerned, it is
quite obvious that it cannot significantly affect the normalization
of the total cross section because the dominant effect in a mass shift
is an energy shift of the entire line-shape. Nevertheless, it is quite
interesting that the normalization is at all insensitive to the
dominant infrared-sensitive terms in the Schr\"odinger
equation~(\ref{NNLOSchroedinger})\footnote{
In our case the two issues are in fact connected to each
other. But it is important to conceptually separate the issue of a
simple energy shift from the more fundamental question of infrared
sensitivity.
}, 
which, in the pole mass scheme,  would cause the corrections to the
peak position to grow factorially at large orders of perturbation
theory. To
show this let us recall that the total vector-current-induced cross
section is proportional
to the absorptive part of the Green function, with both arguments
evaluated at the origin in configuration space representation:
\begin{eqnarray}
R^v & \sim & \mbox{Im}\,\sum\hspace{-5mm}\int\limits_n\hspace{2mm}\,
\frac{|\Phi_n(0)|^2}{E_n-E-i\,\Gamma_t}
\,,
\end{eqnarray}
where the sum extends over discrete and continuum states with 
$S$ wave quantum numbers. Thus, for fixed energy the normalization
only depends on the wave function. Repeating the steps following
Eq.~(\ref{1Spolefull}) we find that the correction to the wave
function coming from the dominant infrared-sensitive
terms in the Schr\"odinger equation reads
\begin{eqnarray}
\bigg[\,\delta \Phi_n(0)\,\bigg]^{IR} & = &
\bigg[\, 
\int\frac{d^3\mbox{\boldmath $p$}}{(2\pi)^3}
\int\frac{d^3\mbox{\boldmath $q$}}{(2\pi)^3}\,
\sum\hspace{-6.5mm}\int\limits_{m\neq n}\,
\frac{\Phi_m(0)\,\tilde\Phi_m^*(\mbox{\boldmath $p$})}{E_m-E-i\Gamma_t}
\,\delta {\cal{H}}(\mbox{\boldmath $p$},\mbox{\boldmath $q$})\,
\tilde\Phi_{n}(\mbox{\boldmath $q$})
\,\bigg]^{IR}
\nonumber\\ & \sim &
\int\frac{d^3\mbox{\boldmath $p$}}{(2\pi)^3}
\int\limits^{|\mbox{\boldmath $q$}|<\mu_f}
  \frac{d^3\mbox{\boldmath $q$}}{(2\pi)^3}\,
\sum\hspace{-6.5mm}\int\limits_{m\neq n}\,
\frac{\Phi_m(0)\,\tilde\Phi_m^*(\mbox{\boldmath $p$})}{E_m-E-i\Gamma_t}\,
\tilde\Phi_{n}(\mbox{\boldmath $p$})
\,\delta\tilde V_c(\mbox{\boldmath $q$})
\nonumber \\[3mm] & = &
0
\,,
\end{eqnarray}
i.e. it vanishes because of the orthogonality of the wave
functions. Therefore the large corrections in the normalization of the
total cross section are not related to an infrared sensitivity of the
corrections, in particular at large orders. To demonstrate that this
is also the case for the NNLO corrections
calculated in this work, we have displayed in
Fig.~\ref{figcompareinfrared} the total vector-current-induced
cross section $Q_t^2 R^v$ close to threshold for $M_{1S}=\Lambda=175$~GeV,
$\mu=30$~GeV, $\alpha_s(M_Z)=0.118$ and $\Gamma_t=1.43$~GeV 
successively including various NNLO corrections. 
The dotted line represents the NLO cross section and the solid line
the NNLO one. The dash-dotted line is the NLO cross section
including also the NNLO corrections to the Coulomb potential $V_c$;
the dashed line contains, in addition, all Abelian NNLO
corrections, i.e. those that do not involve the SU(3) group
theoretical factor $C_A$. The separation of the NNLO corrections into
those coming from the Coulomb potential and from Abelian and non-Abelian
relativistic corrections is gauge-invariant. The difference between
the dashed and the solid curve represents the corrections of the
non-Abelian NNLO effects originating from the potential $V_{\mbox{\tiny
NA}}$ and those ${\cal{O}}(\alpha_s^2)$ contributions to the
short-distance coefficient $C^v$ that are proportional to $C_A$. From
the rather small difference between the dotted and the dash-dotted
curves (2--4\%) we see that the large NNLO corrections to the
normalization are not related to the corrections in the Coulomb
potential. Because a redefinition of the strong coupling would mainly
affect the size of the higher-order corrections in the Coulomb
potential, we can conclude that using a different 
scheme for the strong coupling (such as the
$V$-scheme~\cite{Brodsky1,Brodsky2}) will not
significantly affect the size of the NNLO corrections. The curves
plotted in Fig.~\ref{figcompareinfrared} demonstrate that the
${\cal{O}}(20\%)$ NNLO correction to the normalization is a sum of
corrections, each of which positive and individually either smaller
than or approximately equal to ${\cal{O}}(10\%)$. Although this
observation, of
course, cannot be taken as a proof that the still unknown N$^3$LO
corrections are smaller than the NNLO ones, it indicates that the size
of the latter does not necessarily have to be taken as an argument
for the non-relativistic expansion to break down for the normalization
of the total cross section. 

\begin{figure}[t!] 
\begin{center}
\leavevmode
\epsfxsize=3.8cm
\epsffile[200 400 400 530]{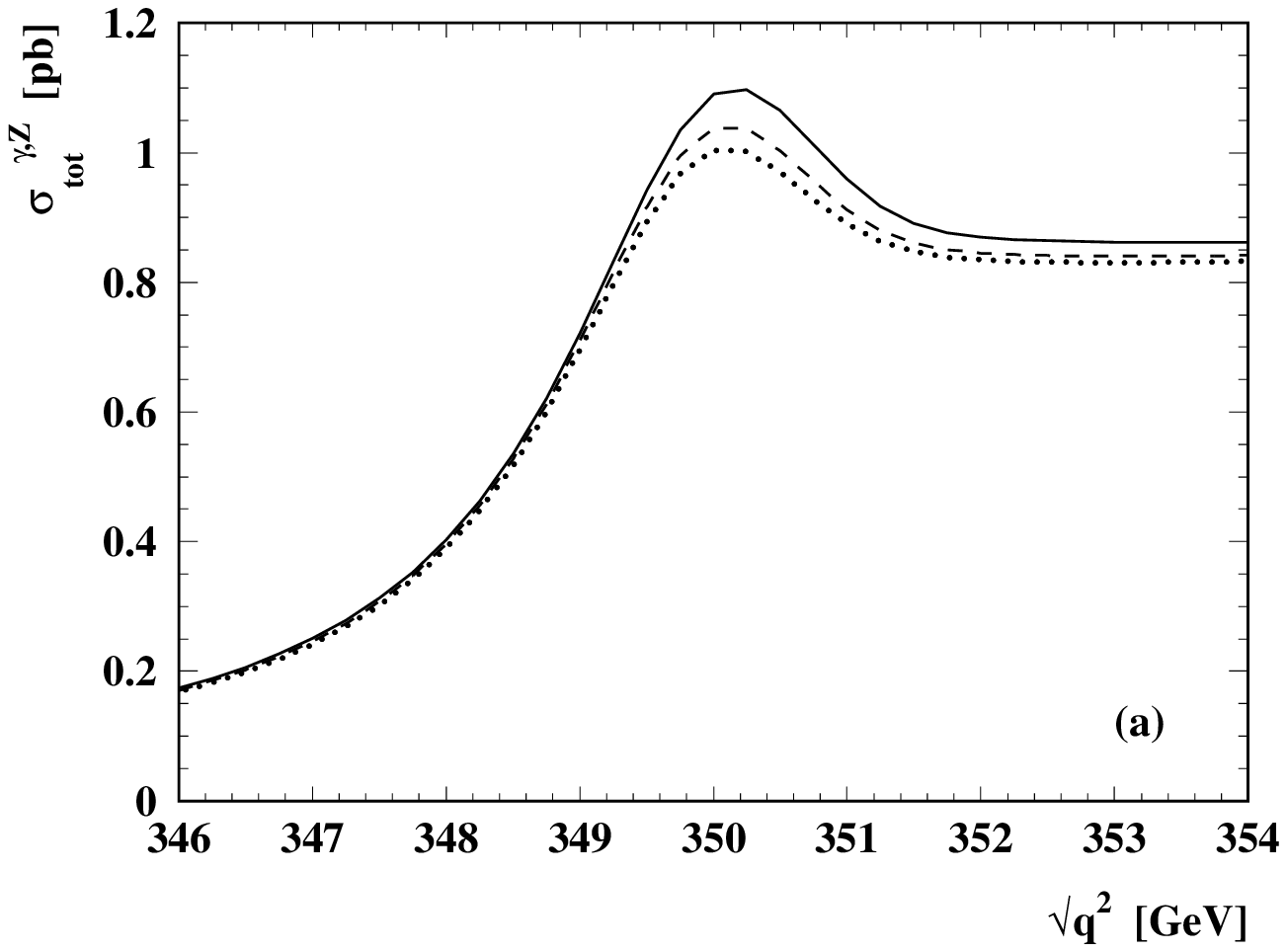}
\hspace{4.4cm}
\epsfxsize=3.8cm
\leavevmode
\epsffile[200 400 400 530]{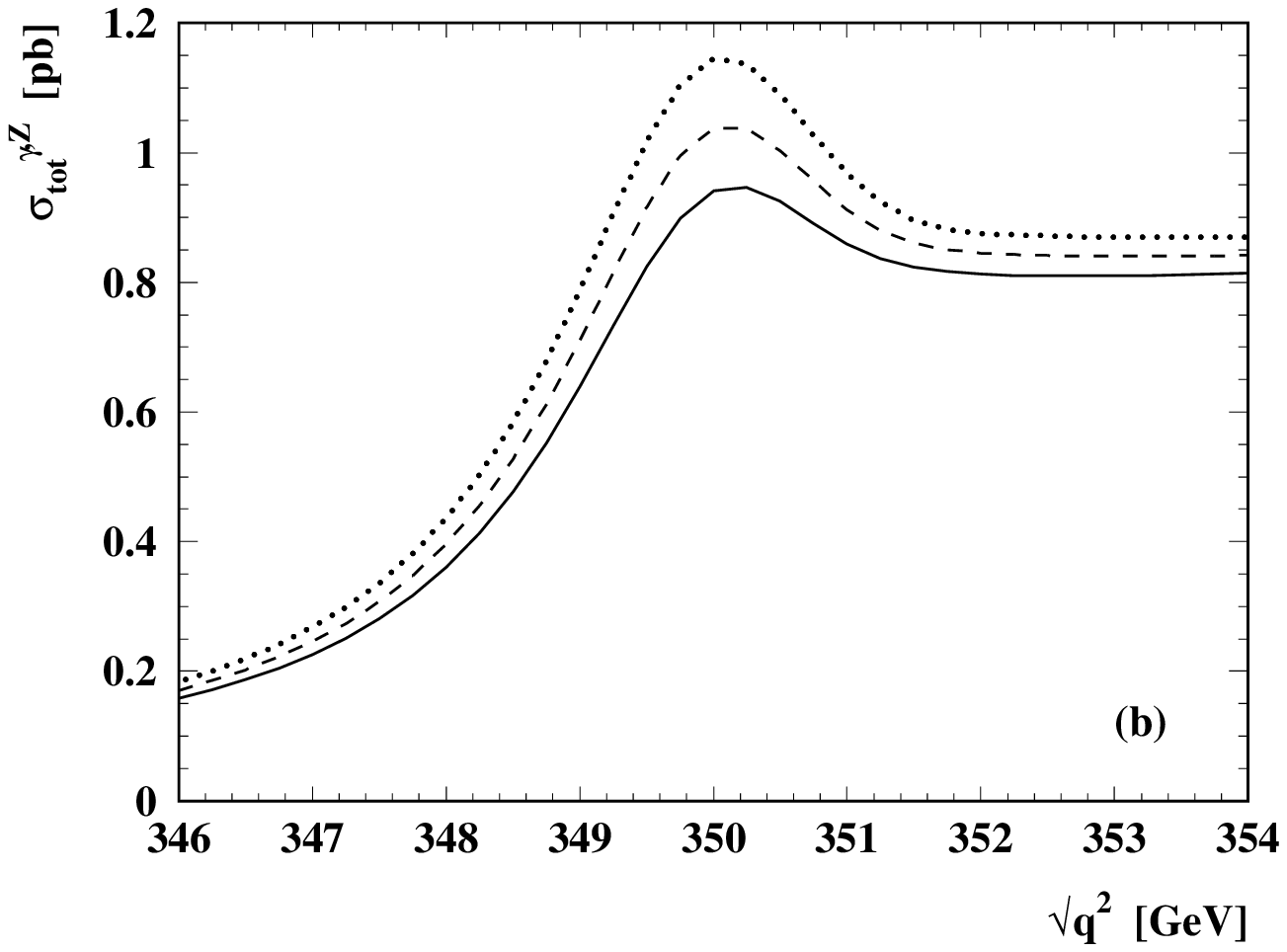}
%
%
\vskip  2.7cm
 \caption{\label{figalltotalm1S} 
The total cross section 
$\sigma_{tot}^{\gamma,Z}$, Eq.~(\ref{totalcrossfullQCD}), is plotted in
the $1S$ scheme at NNLO for $M_{1S}=\Lambda=175$~GeV,
$\Gamma_t=1.43$~GeV and $\alpha=1/125.7$. (a) shows the
renormalization scale dependence for $\mu=15$ (solid line), $30$
(dashed line) and $60$~GeV (dotted line), and $\alpha_s(M_Z)=0.118$;
(b) shows
the dependence on the strong coupling for
$\alpha_s(M_Z)=0.113$ (solid line), $0.118$ (dashed line) and $0.123$
(dotted line), and for $\mu=30$~GeV.
}
 \end{center}
\end{figure}

An interesting insight into the question of how to interpret the large
normalization corrections can also be obtained by comparing the total
cross section line-shape, which we have calculated in the threshold
regime, with earlier calculations of the total cross section for higher
energies, where a resummation of Coulomb singular terms is not yet
necessary and perturbation theory in $\alpha_s$ is believed to be
reliable.~\cite{Chetyrkin2} We would like to note that it is the large
mass of
the top quark that allows us to draw conclusions from a
comparison of the threshold cross section with the one calculated for
higher energies. To illustrate this we recall that our calculation of
the threshold cross section is valid if the hierarchy $\alpha_s, v\ll
1$ is satisfied, where the scale of the strong coupling is of the
order of 
the inverse Bohr radius, the kinetic energy, or the top width. This
means that the threshold cross section represents a simultaneous
expansion in $\alpha_s$ and $v$, where powers of $(\alpha_s/v)$ are
resummed to all orders in $\alpha_s$. The high energy cross section, on
the other hand, is valid if $\alpha_s\ll v, 1$, where the scale in the
strong coupling if of order the top-antitop relative momentum or the
centre-of-mass energy. Thus a comparison of the threshold results with
the high energy perturbative ones
is only sensible if there exists a kinematic regime where both
hierarchies are satisfied at the same time, i.e. if $\alpha_s\ll v\ll
1$. In this regime the effects of the resummation of powers of
$(\alpha_s/v)$ not contained in the high energy cross section should
be small as well as the effects of velocity corrections beyond NNLO,
which are not contained in the threshold cross section. Obviously this
relation is difficult or impossible to satisfy for bottom or charm
quarks, but it is possible for the top quark case. For
$\alpha_s(M_t\alpha_s)\sim 0.13$ we can argue that a meaningful
comparison between threshold and high energy cross section should be
possible for $v\approx$~0.3--0.4, which corresponds to $\sqrt{s}\approx
365$~GeV.
\begin{figure}[t!] 
\begin{center}
\leavevmode
\epsfxsize=5cm
\epsffile[220 360 400 550]{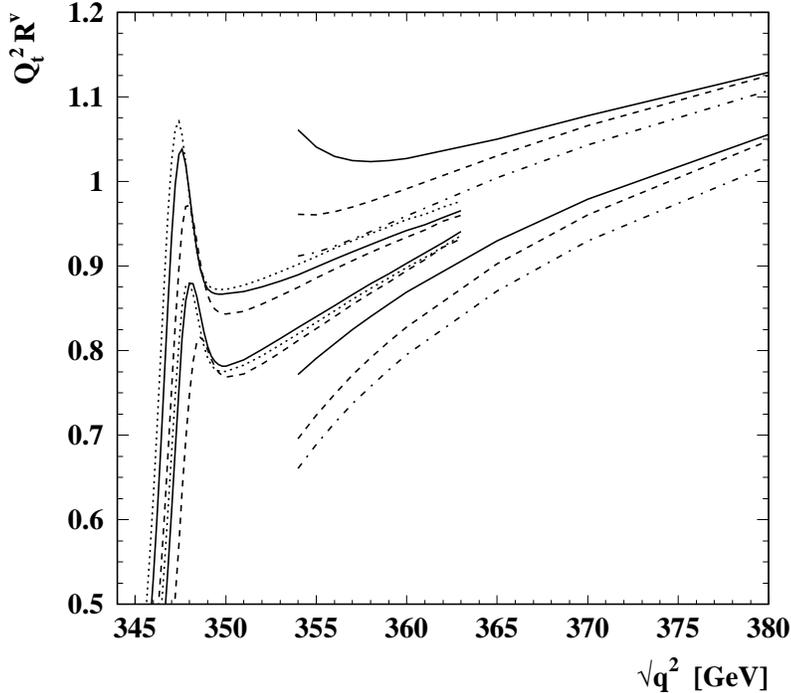}
%
%
\vskip  3.4cm
 \caption{\label{figthreshhigh} 
The vector-current-induced total cross section in the non-relativistic
expansion at NLO (lower bunch of threshold curves), NNLO (upper bunch)
and in conventional perturbation theory at ${\cal{O}}(\alpha_s)$
(lower bunch of high energy curves) and  ${\cal{O}}(\alpha_s^2)$
(upper bunch). The pole mass scheme has been used.
The curves have been plotted for $\alpha_s(M_z)=0.118$,
$\Gamma_t=1.43$~GeV, $M_t=\Lambda=175$~GeV and $\mu=25$~GeV (dotted
lines), $2 (p_0^4+M_t^2\,\Gamma_t^2)^{1/4}$ (solid lines), $175$~GeV
(dashed lines) and $\sqrt{q^2}$ (dash-dotted lines).
The formulae for the ${\cal{O}}(\alpha_s^2)$ high energy cross section
have been taken from Ref.~\cite{Chetyrkin2}.
}
 \end{center}
\end{figure}
In Fig.~\ref{figthreshhigh} we have plotted the threshold and the high
energy cross sections at NLO/NNLO and
${\cal{O}}(\alpha_s)$/${\cal{O}}(\alpha_s^2)$, respectively, for the
renormalization scales $\mu=25$~GeV (dotted lines), 
$2 (p_0^4+M_t^2\,\Gamma_t^2)^{1/4}$ (solid lines),
$175$~GeV (dashed lines) and
$\sqrt{q^2}$ (dash-dotted lines) for
$M_t=\Lambda=175$~GeV, $\alpha_s(M_Z)=0.118$ and $\Gamma_t=1.43$. 
The lower bunch of threshold curves (characterized by the peak at
around $\sqrt{q^2}=348$~GeV) is NLO and the upper bunch NNLO.
Likewise, the lower bunch of high energy curves is
${\cal{O}}(\alpha_s)$ and the upper bunch ${\cal{O}}(\alpha_s^2)$.
We note that we have not plotted the threshold curves for
$\mu=\sqrt{q^2}$ and the high energy curves not for $\mu=25$~GeV,
which seems to be a rather unnatural choice for each.
The formulae for the ${\cal{O}}(\alpha_s^2)$ high energy cross section
have been taken from Ref.~\cite{Chetyrkin2}.
For convenience we have plotted the curves in Fig.~\ref{figthreshhigh}
in the pole mass scheme. Because the choice of the mass definition
does not alter the behaviour of the cross section normalization for
energies above the peak position, this choice does not affect the
conclusions drawn below. 
For the threshold (high energy) cross section, we observe that the NNLO
(${\cal{O}}(\alpha_s^2)$) corrections decrease for energies further
away from the threshold region. However, the ${\cal{O}}(\alpha_s^2)$
corrections to the high energy cross sections are much larger than the
NNLO corrections to the threshold cross section at the same
centre-of-mass energy. For $\sqrt{s}=\,$360--370~GeV the
${\cal{O}}(\alpha_s^2)$ corrections to the high energy cross section
are, for equal choices of renormalization scales, between 10 and 20\%
compared to only around $5\%$ for the NNLO corrections
to the threshold cross section. We also see a much weaker
renormalization scale dependence of the threshold cross sections. The
curves show that the resummation of Coulomb singular terms contained
in the threshold calculation leads to a considerable stabilization of
the cross section determined in conventional perturbation theory in
$\alpha_s$ for energies below $\sqrt{s}=365$~GeV. If we 
believe that conventional perturbation theory is reliable down to
energies around $\sqrt{s}=360$~GeV, the results displayed in
Fig.~\ref{figthreshhigh} indicate that the non-relativistic
expansion does certainly not break down.
However, the curves of Fig.~\ref{figthreshhigh} also make it evident that
the small renormalization scale dependence of the NLO threshold cross
section does certainly not reflect the true size of the remaining
theoretical uncertainties at NLO. 
We believe that $10\%$ should be a fair estimate of the remaining
theoretical uncertainties contained in the normalization of the NNLO
total cross section close to threshold. As far as the top mass
determination at a future electron-positron linear or muon pair
collider is concerned, this rather large normalization uncertainty
might in fact lead to uncertainties in the determination of the
$1S$ mass that are larger than indicated in the previous subsection.
This is due to the effects
of beamstrahlung and initial state radiation that lead to a smearing
of the effective centre-of-mass energy of about 
1--2~GeV~\cite{Orange1}. Beamstrahlung and initial state radiation
render the visible peak in the total cross section either smaller
(at the muon pair collider) or completely invisible (at the linear
collider), which makes it possible that the uncertainty in the
normalization feeds into larger uncertainties in the determination of
$M_{1S}$. It is the task of realistic simulation studies to
determine how large this effect is for the various collider and
detector designs and to devise optimized strategies to minimize it.
If the effects of beamstrahlung and initial state radiation on the top
quark mass determination are small, the uncertainty in the
normalization will mainly affect the measurement of the strong
coupling (see Figs.~\ref{figtotm1S}c and \ref{figalltotalm1S}b).

In Figs.~\ref{figalltotalm1S} the total cross section 
$\sigma_{tot}^{\gamma,Z}(e^+e^-\to \gamma^*, Z^*\to t\bar
t)$, Eq.~(\ref{totalcrossfullQCD}), is plotted at NNLO in the $1S$
scheme for $M_{1S}=\Lambda=175$~GeV, $\Gamma_t=1.43$~GeV and
$\alpha=1/125.7$.
Figure~\ref{figalltotalm1S}a shows the renormalization scale dependence
for $\mu=15$ (solid line), $30$ (dashed line) and $60$~GeV (dotted
line), and $\alpha_s(M_Z)=0.118$. Figure~\ref{figalltotalm1S}b
displays the dependence on the strong coupling for
$\alpha_s(M_Z)=0.113$ (solid line), $0.118$ (dashed line) and $0.123$
(dotted line), and for $\mu=30$~GeV. 

\vspace{1.5cm}
\section{Summary and Conclusions}
\label{sectionconclusion}
Within the framework of the non-relativistic effective field theories
NRQCD and PNRQCD, we have calculated the vector-current-induced total
cross section of top-antitop pair production in electron-positron
annihilation close to threshold at NNLO in the non-relativistic
expansion. The corresponding  
NNLO QCD relativistic corrections have also been determined for the 
vector-current-induced top three-momentum distribution. 
In addition, the axial-vector-current-induced total cross section and
the three-momentum distribution have been calculated to fully account
for the $Z$-boson contributions in electron-positron
annihilation. For the
total cross section and the three-momentum distribution, the 
axial-vector-current-induced contributions are suppressed by $v^2$
with respect to the vector current contributions; they have therefore
been determined in leading
order in the non-relativistic expansion. The size of the
axial-vector-current-induced contributions is smaller than the
remaining theoretical uncertainties in the vector-current-induced
cross section (for unpolarized electrons and positrons).
In contrast with previous literature on the same subject, we have
implemented the top quark width   
by including electroweak corrections into the (P)NRQCD matching
conditions of the Lagrangian and the currents. This allows for a
straightforward generalization of the Fadin-Khoze prescription ``$E\to
E+i \Gamma_t$'' to implement the top quark width at NNLO in the
non-relativistic expansion, where $\Gamma_t/M_t$ is counted as order
$v^2$.
We have shown that at NNLO this cannot be achieved by a
simple shift of the centre-of-mass energy to complex values. Our
calculations have been carried out using numerical techniques to
solve the corresponding integral equations within a cutoff
regularization scheme and using analytic methods for the matching
procedure. We have addressed the question of large NNLO
corrections to the peak position and the normalization of the total
vector-current-induced cross section.
The position of the peak, which is observable in the
total vector-current-induced cross section, can be stabilized if the
cross section is
expressed in terms of the $1S$ mass, instead of the pole mass. 
The $1S$ mass, $M_{1S}$, is defined as half the mass of a fictitious
${}^3\!S_1$ toponium ground state for a stable top quark. The $1S$
mass is a short-distance mass and, by construction, reduces to a large
extent the dependence of the peak position on theoretical parameters
such as the
renormalization scale of the strong coupling. We have also shown that
the large NNLO corrections to the normalization of the total cross
section, of order $20\%$, are genuine NNLO corrections, which cannot be
removed by a redefinition of the top quark mass or the strong
coupling. The large size of the corrections to the normalization
originates from the fact that NNLO relativistic corrections from
several sources have the same sign. We believe that the remaining
theoretical uncertainties in the normalization are of order $10\%$. If
the effects of beamstrahlung and initial state radiation at the
$e^+e^-$ linear collider do not lead
to a significant cross feed of the uncertainties in the normalization
into $M_{1S}$, we expect that an uncertainty in the determination of
$M_{1S}$ of less than $200$~MeV will
be possible at the linear collider with an integrated luminosity of 
50--100~$fb^{-1}$. In order to determine the $\overline{\mbox{MS}}$
top quark
mass from the $1S$ mass with the same precision, the knowledge 
of the full three-loop relation between the pole and the
$\overline{\mbox{MS}}$ mass, and a small uncertainty in
$\alpha_s(M_Z)$ are crucial. 

After completion of this work, we received
Refs.~\cite{Beneke6,Sumino3,Penin1}. In
Ref.~\cite{Beneke6} the total vector-current-induced cross section has
been calculated analytically, using the $\overline{\mbox{MS}}$
regularization scheme 
based on the Schr\"odinger equation~(\ref{NNLOSchroedinger}).
The NLO and NNLO corrections have been treated perturbatively,
supplemented by a resummation of the energy denominators for the $n=1$
and $n=2$ states in the spectral representation of the
Green function. The renormalization scale dependence
of the cross section line-shape is considerably larger in
Ref.~\cite{Beneke6} than in our work.
This might be a consequence of the perturbative
treatment of the NLO and NNLO corrections. In addition, a
next-to-leading logarithmic resummation
of logarithms of the ratio $M_t/\mu$ in the short-distance coefficient
$C^v_{\tiny \overline{\mbox{MS}}}$ has been carried out, taking the
$\overline{\mbox{MS}}$ cutoff scale $\mu$ of order $M_t v$. The
effect of this resummation is around $5\%$ for the normalization of
the total cross section. 
In our cutoff scheme, where
the regularization scale is of order $M_t$, the corresponding
logarithm is contained in the non-relativistic current correlators.
In Ref.~\cite{Sumino3} the  
vector-current-induced cross total section and the three-momentum
distribution have been calculated at NNLO, based on the simplified
Schr\"odinger equation~(\ref{NNLOSchroedingersimplified}), which we
have discussed critically at the end of Sec.~\ref{sectionwidth}. For
the three-momentum distribution the authors of Ref.~\cite{Sumino3}
have included further corrections to account for the difference with
the results of the correct Schr\"odinger
equation~(\ref{NNLOSchroedingergamma}).   
In Refs.~\cite{Beneke6,Sumino3} the ``potential-subtracted'' mass has
been tested in different ways as an alternative mass parameter for
the total cross section. As far as the uncertainties in the top mass
determination at a future linear collider are concerned,
Ref.~\cite{Beneke6} arrives at conclusions similar to ours.
In Ref.~\cite{Penin1} the techniques used in
Refs.~\cite{Hoang2,Hoang3} have been 
employed to calculate the total cross section, the angular
distribution and the top quark polarization for top quark pair
production close to threshold in $e^+e^-$ and $\gamma\gamma$
collisions. The corrections originating from the higher-order
contributions in the Coulomb potential have been calculated analytically.
In Refs.~\cite{Beneke6,Sumino3,Penin1}
the top quark width has been implemented
by the replacement rule ``$E\to E+i\Gamma_t$'', where $E$ is the
centre-of-mass energy with respect to two times the top mass.

\vspace{1.5cm}
\section*{Acknowledgement}
We thank M.~Beneke for discussions and 
M.~Beneke, Z.~Ligeti and A.~V.~Manohar for reading the manuscript. The
work of A.~H.~H. is in part supported by the EU Fourth Framework Program
``Training and Mobility of Researchers'', Network ``Quantum
Chromodynamics and Deep Structure of Elementary Particles'', contract
FMRX-CT98-0194 (DG12-MIHT). 

\vspace{2cm}
\begin{appendix}
\section{Calculation of the Short-Distance Coefficient $C^v$}
\label{appendixshortdistance}
In this appendix we present details of the calculation of the
short-distance coefficient $C^v$ to order $\alpha_s^2$,
assuming that the top quarks are stable ($\Gamma_t=0$). We recall that
$C^v$ is the square of the short-distance coefficient $c_1^v$ of the
${}^3\!S_1$ NRQCD current ${\tilde \psi}^\dagger \mbox{\boldmath
  $\sigma$}  \tilde \chi$
(see Eq.~(\ref{vectorcurrentexpansion1})); $c_1^v$ contains those
contributions in the vector-current-induced top-antitop production
diagrams, which come from loop momenta $p=(p^0,\vec{p})$ with
$|\vec{p}|>\Lambda$ for $\sqrt{q^2}=2 M_t$. As explained in
Sec.~\ref{sectionregularization}, we have to determine $C^v$ by
employing the specific routing convention shown in
Fig.~\ref{figroutingladder}. In principle, it would be possible to
determine $C^v$ by calculating the diagrams for the 
vector-current-induced cross section in full QCD restricting the loop
momenta such
that the spatial components would be larger than $\Lambda$. However,
in a cutoff scheme it is more economical to first calculate the 
vector-current-induced cross section in NRQCD up to order $\alpha_s^2$
and
NNLO in the velocity expansion and then to adjust the coefficients of
$C^v$ such that the cross section in NRQCD is equal to the cross
section in full QCD, likewise calculated to order $\alpha_s^2$ and
NNLO in the velocity expansion. 

The expression of the total vector-current-induced cross section in
full QCD at order $\alpha_s^2$ and NNLO in the velocity expansion
reads ($a\equiv C_F\alpha_s(\mu)$):
\begin{eqnarray}
\lefteqn{
R_{\mbox{\tiny 2loop QCD}}^{v,\mbox{\tiny NNLO}} \, = \,
N_c\,\bigg\{\,\bigg[\,
\frac{3}{2}\,\frac{p_0}{M_t}-\frac{5}{4}\,\frac{p_0^3}{M_t^3}
\,\bigg] +
\frac{a}{\pi}\,\bigg[\,
  \frac{3\,\pi^2}{4}-6\,\frac{p_0}{M_t}+
  \frac{\pi^2}{2}\,\frac{p_0^2}{M_t^2}
\,\bigg]
}
\nonumber\\[2mm] & & 
+\, a^2\,\bigg[\,
\frac{\pi^2\,M_t}{8\,p_0} 
- \frac{3}{2}\,\bigg(\,
2 
+ \frac{1}{8\,C_F} \,\Big(\,
\beta_0\,\ln\frac{4\,p_0^2}{\mu_{\rm hard}^2} - a_1
\,\Big)
\,\bigg) 
\nonumber\\[2mm] & & \qquad
+ \bigg(\, 
\frac{13\,\pi^2}{48}  
  + \frac{3}{2\,C_F^2}\,\kappa 
  +  \frac{3\,\beta_0}{2\,C_F\,\pi^2}
        \, \ln\frac{M_t^2}{\mu_{\rm hard}^2} 
  - \Big(1 +\frac{3}{2}\, \frac{C_A}{C_F} \Big)\,\ln\Big(\frac{p_0}{M_t}\Big)
\,\bigg)\,\frac{p_0}{M_t}
\,\bigg]
\,\bigg\}
\,,
\label{RphotonfullQCD}
\end{eqnarray}
where 
\begin{eqnarray}
\kappa & = &
C_F^2\,\bigg[\, \frac{1}{\pi^2}\,\bigg(\,
\frac{39}{4}-\zeta_3 \,\bigg) +
\frac{4}{3} \ln 2 - \frac{35}{18}
\,\bigg] 
- C_A\,C_F\,\bigg[\,  \frac{1}{\pi^2} \,\bigg(
\frac{151}{36} + \frac{13}{2} \zeta_3 \,\bigg) +
\frac{8}{3} \ln 2 - \frac{179}{72} \,\bigg] 
\nonumber\\[2mm] & & +\,
C_F\,T\,\bigg[\,
\frac{4}{9}\,\bigg(\, \frac{11}{\pi^2} - 1\,\bigg)
\,\bigg] +
C_F\,T\,n_l\,\bigg[\, \frac{11}{9\,\pi^2} \,\bigg] 
\,.
\label{kappadef}
\end{eqnarray}
The Born and ${\cal{O}}(\alpha_s)$~\cite{Kallensabry1} contributions
are standard.
At order $\alpha_s^2$ the contributions in Eq.~(\ref{RphotonfullQCD})
that are 
proportional to $C_F^2$, $C_A C_F$, $C_F T n_l$ and $C_F T$ have been
calculated in~\cite{Hoang4}, \cite{Melnikov4}, \cite{HoangKT1,Voloshin2}
and \cite{HoangKT1,Karshenboim1}, respectively.
(See also Refs.~\cite{Chetyrkin2,Chetyrkin3}.)

\begin{figure}[tb] 
\begin{center}
\leavevmode
\epsfxsize=1.7cm
\epsffile[220 410 420 540]{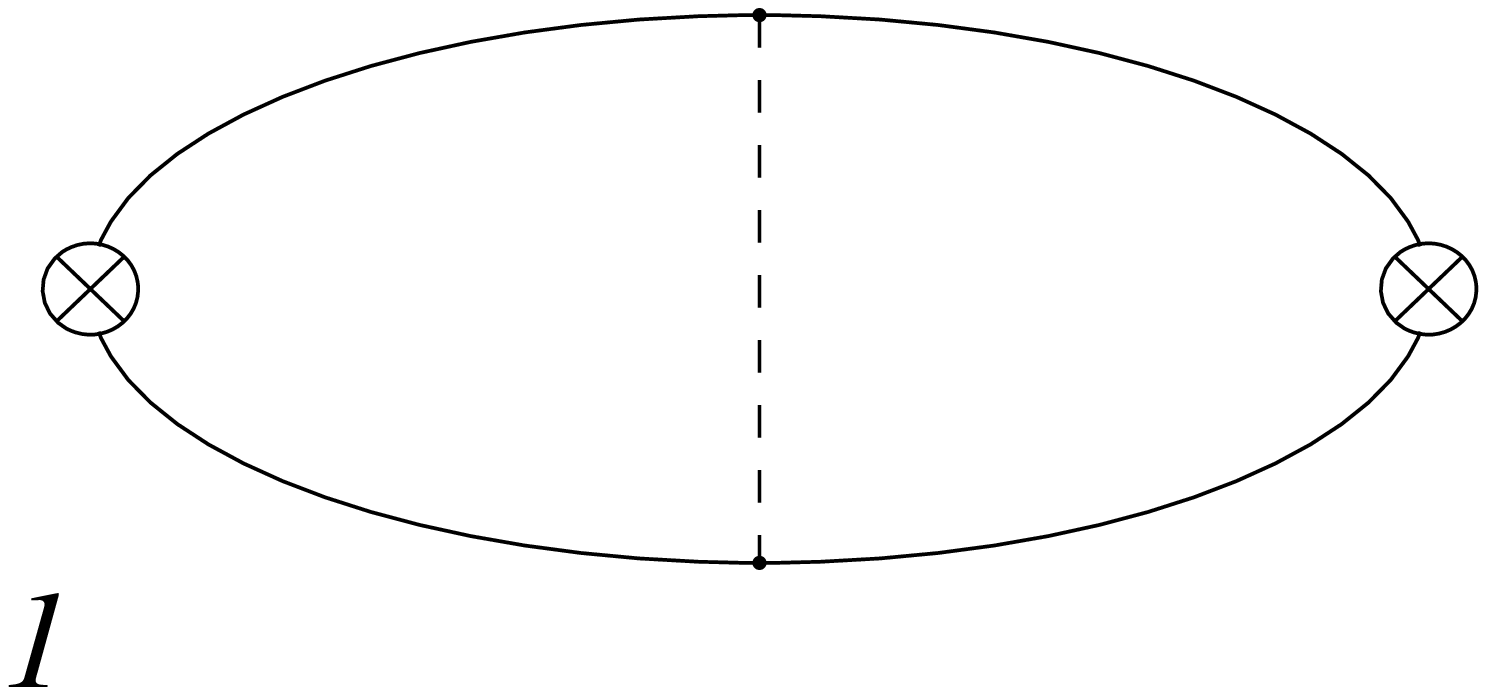}
\hspace{2.2cm}
\leavevmode
\epsfxsize=1.7cm
\epsffile[220 410 420 540]{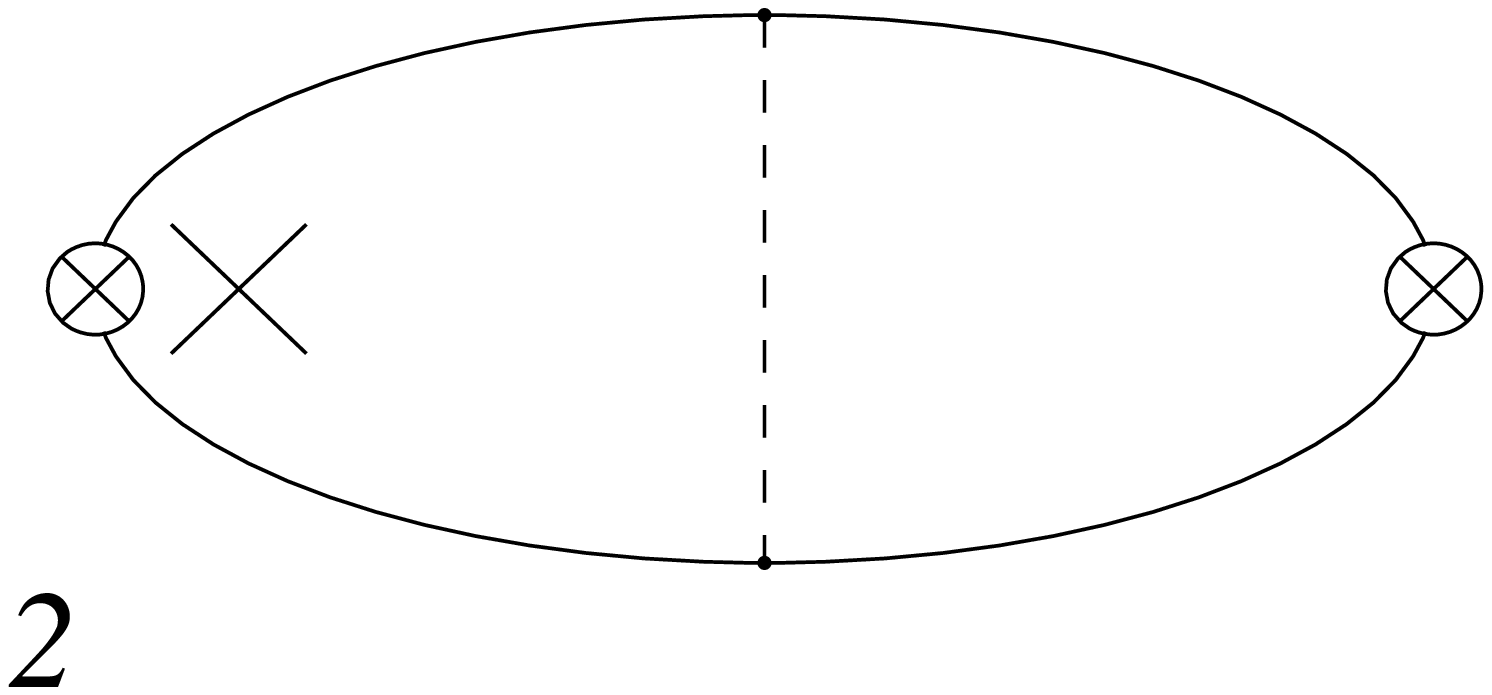}
\hspace{2.2cm}
\leavevmode
\epsfxsize=1.7cm
\epsffile[220 410 420 540]{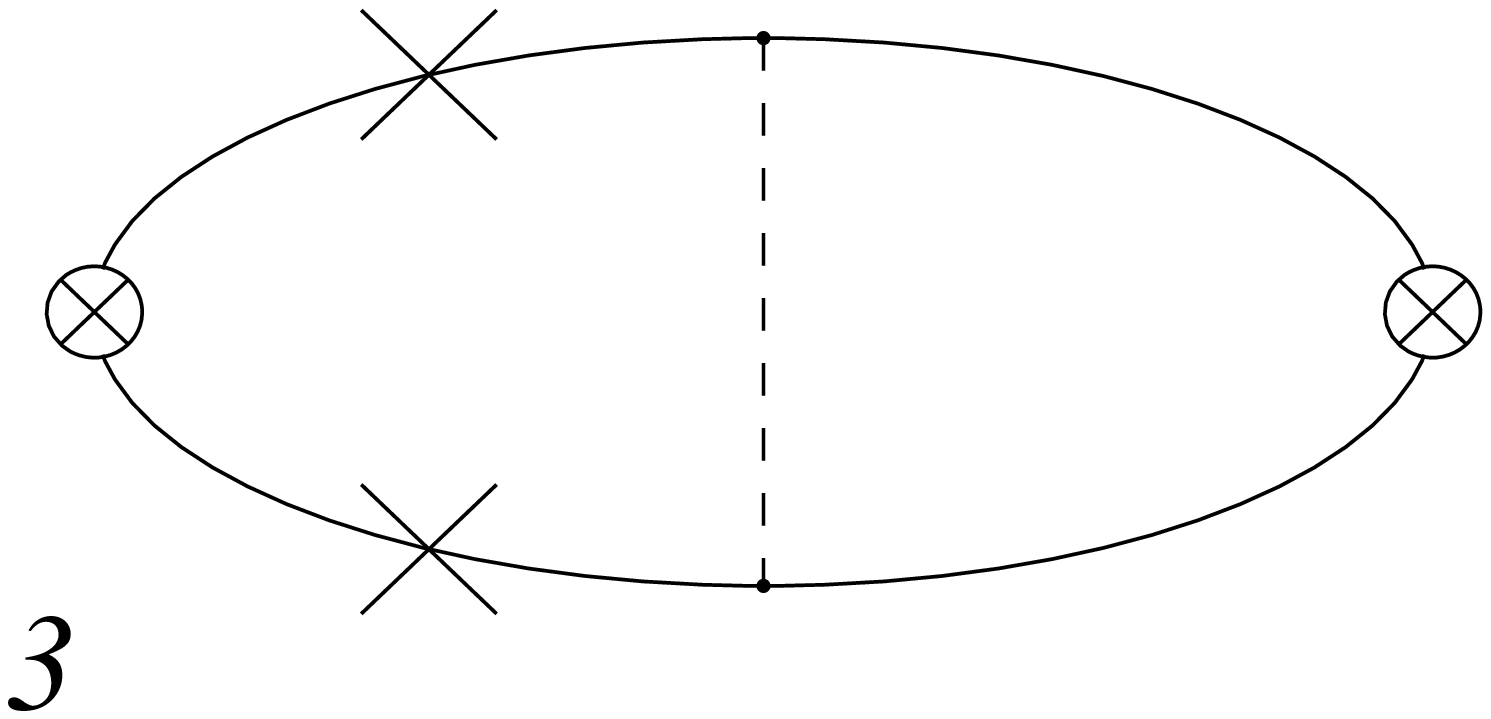}
\hspace{2.2cm}
\leavevmode
\epsfxsize=1.7cm
\epsffile[220 410 420 540]{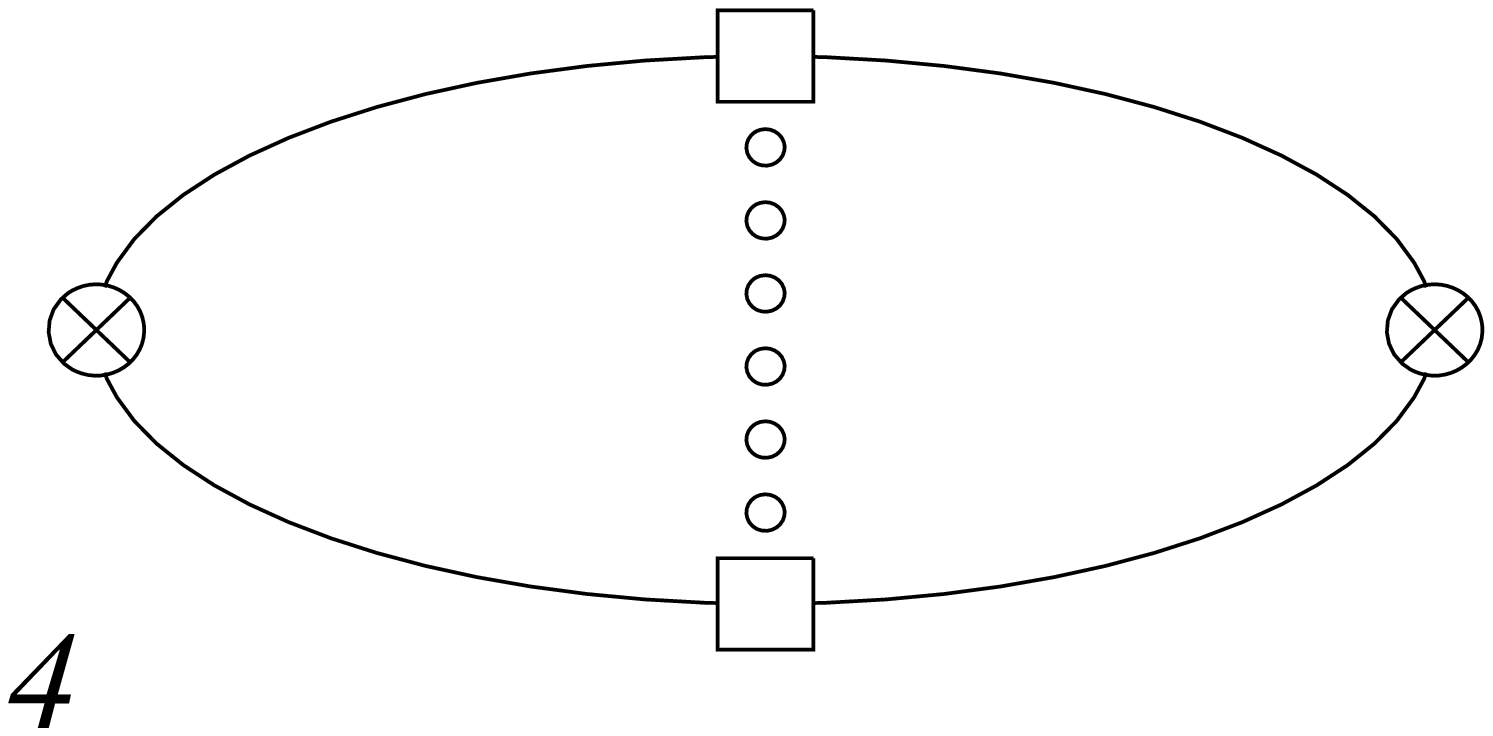}
\vskip  1.5cm
 \caption{\label{fignonrelcurrent1loop} 
Graphical representation of the  NRQCD vector-current correlators
diagrams needed to determine the non-relativistic
vector-current-induced cross section at ${\cal{O}}(\alpha_s)$ and NNLO
in the non-relativistic expansion.
}
 \end{center}
\end{figure}

\begin{figure}[htb] 
\begin{center}
\leavevmode
\epsfxsize=1.7cm
\epsffile[220 410 420 540]{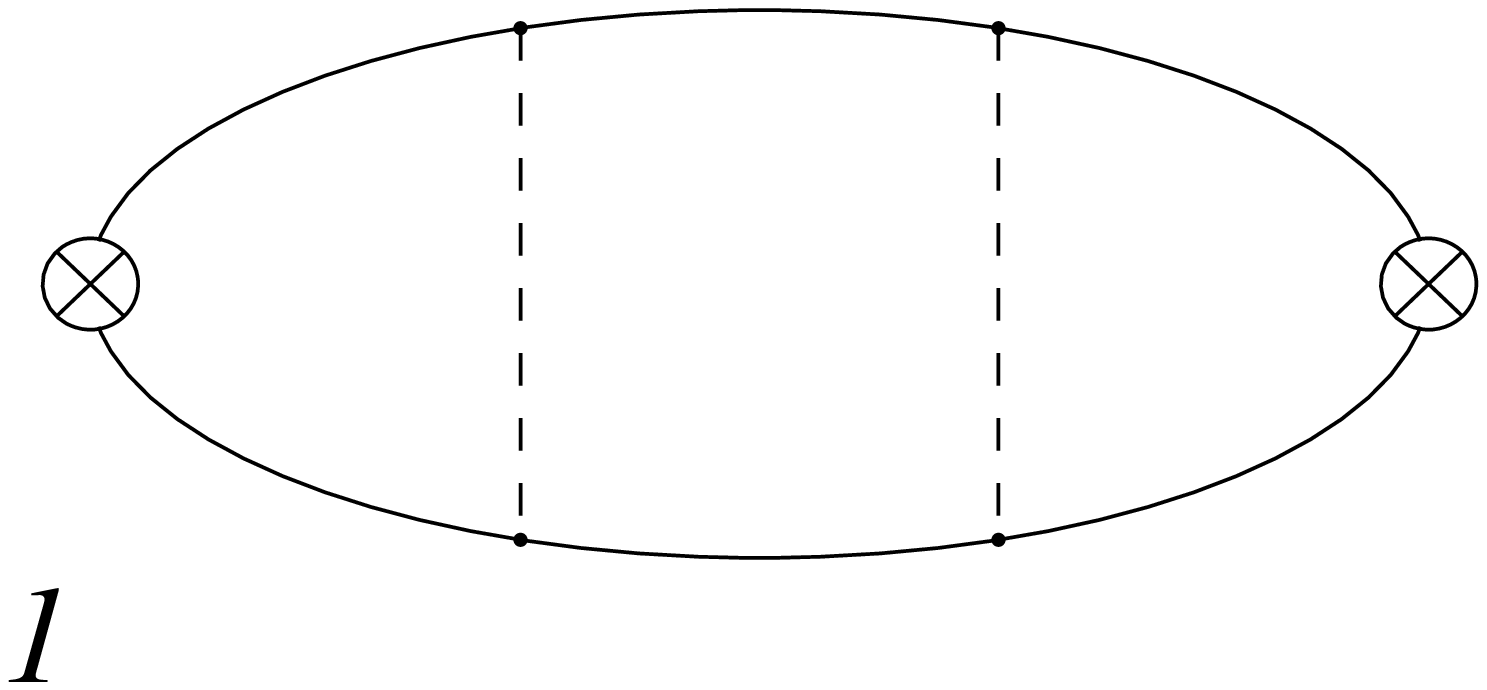}
\hspace{2.2cm}
\leavevmode
\epsfxsize=1.7cm
\epsffile[220 410 420 540]{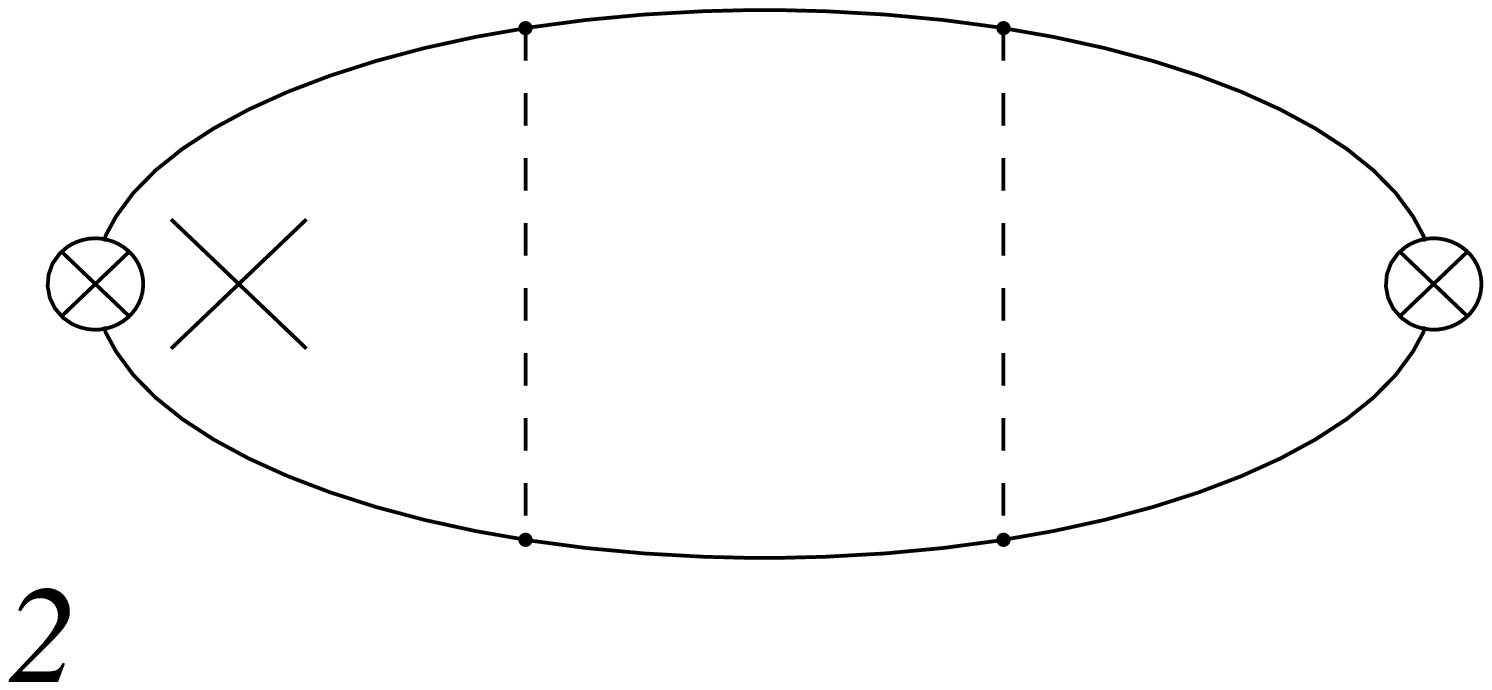}
\hspace{2.2cm}
\leavevmode
\epsfxsize=1.7cm
\epsffile[220 410 420 540]{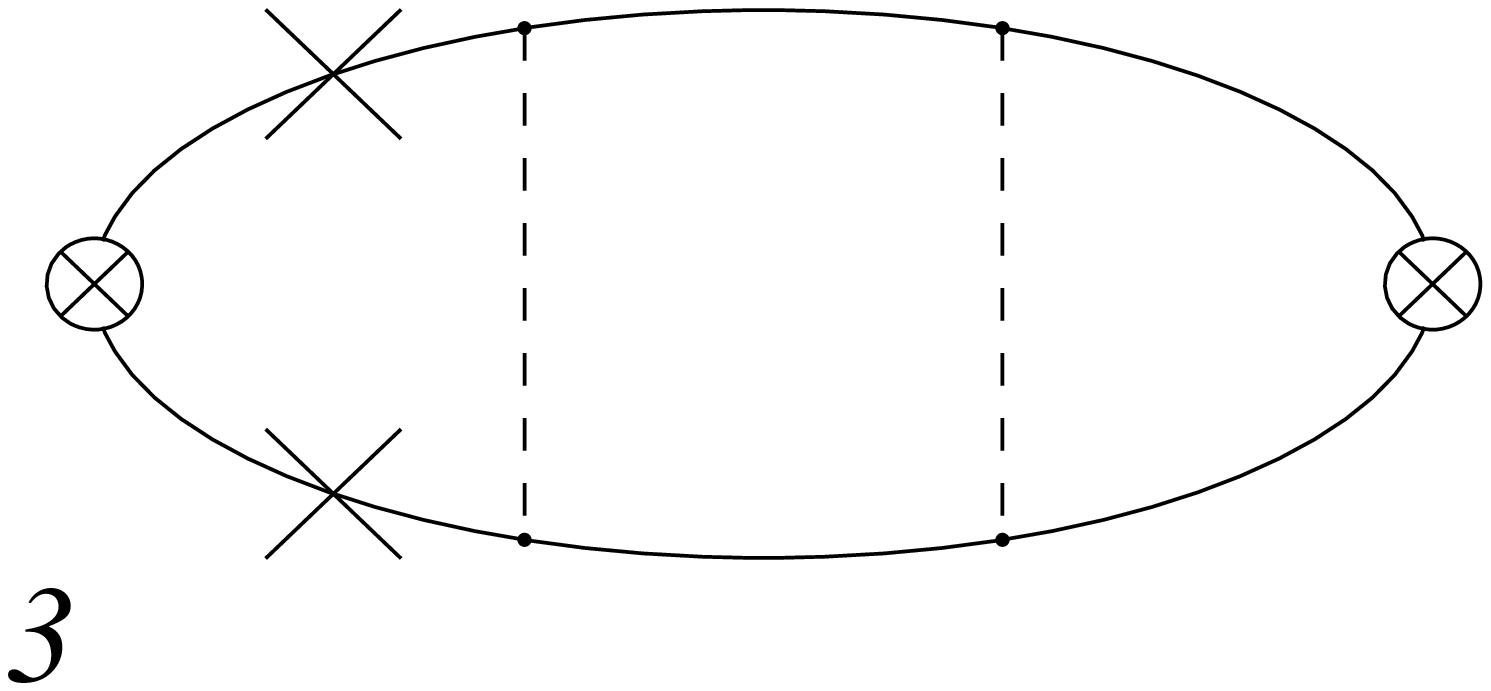}
\\[1.3cm]
\leavevmode
\epsfxsize=1.7cm
\epsffile[220 410 420 540]{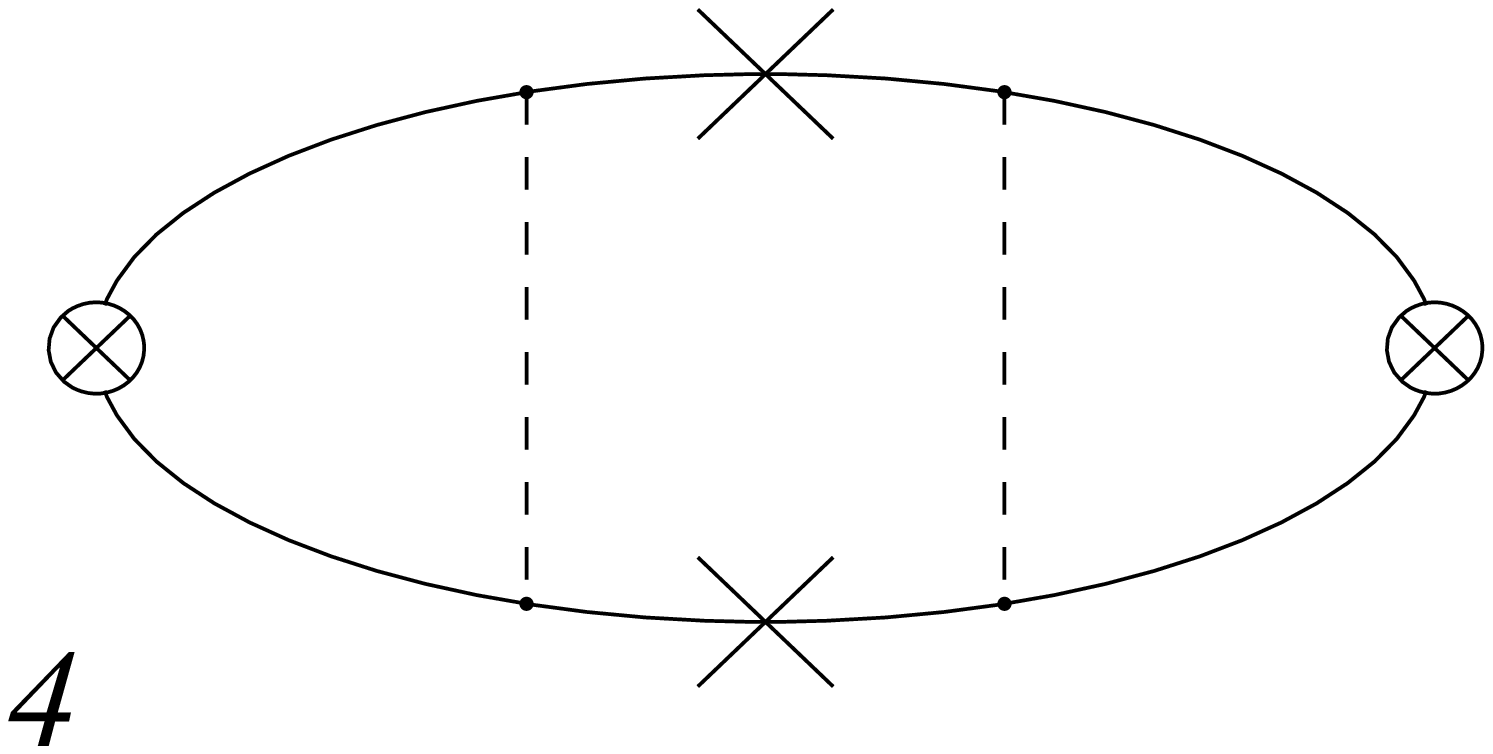}
\hspace{2.2cm}
\leavevmode
\epsfxsize=1.7cm
\epsffile[220 410 420 540]{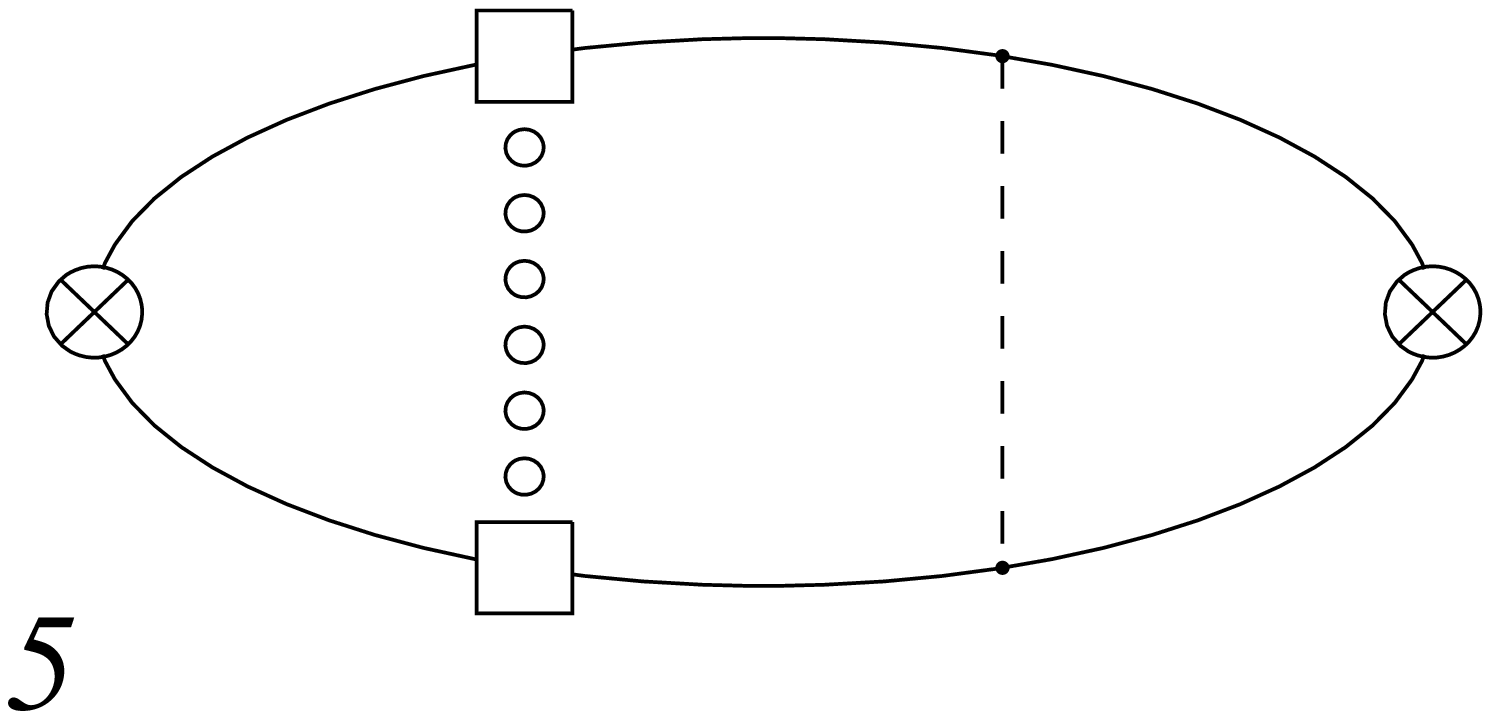}
\hspace{2.2cm}
\leavevmode
\epsfxsize=1.7cm
\epsffile[220 410 420 540]{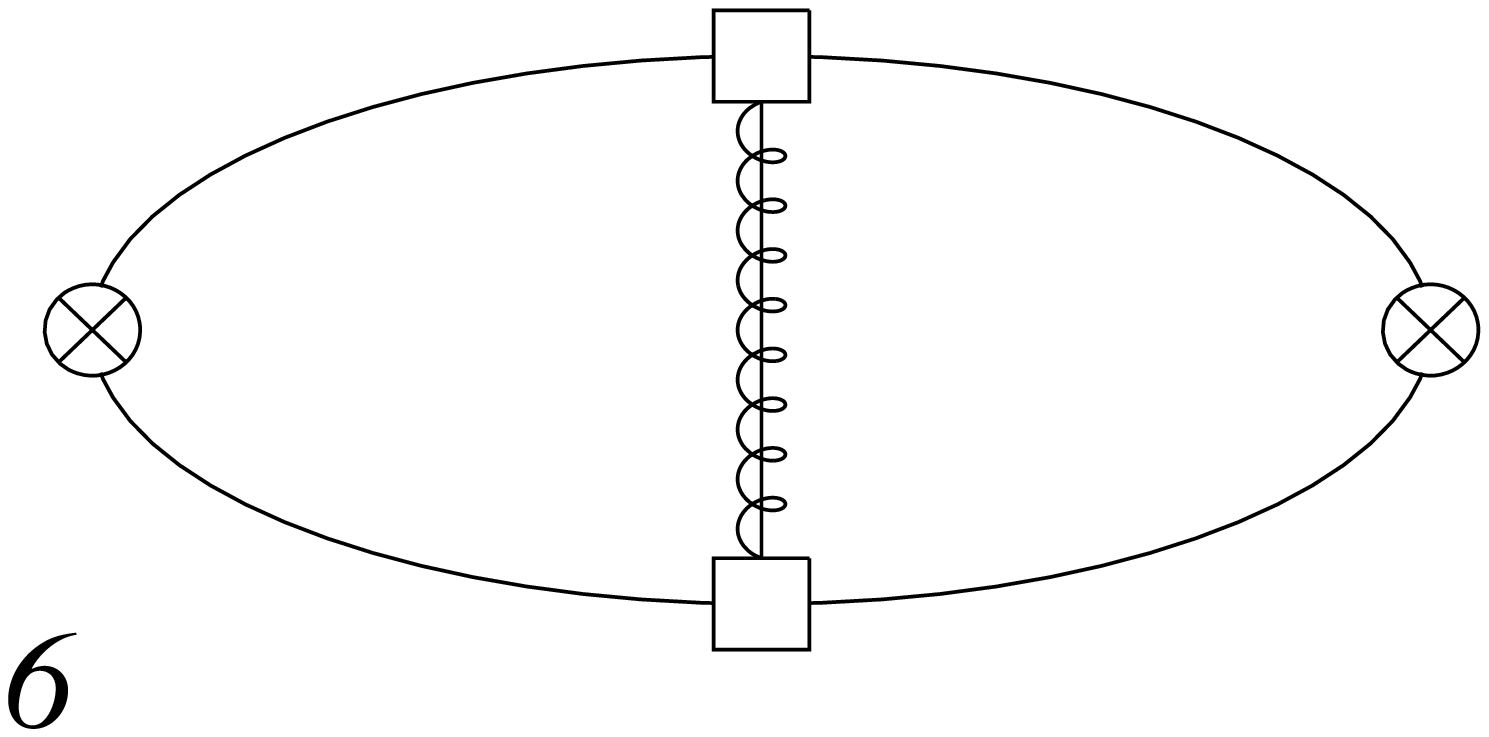}
\\[1.3cm]
\leavevmode
\epsfxsize=1.7cm
\epsffile[220 410 420 540]{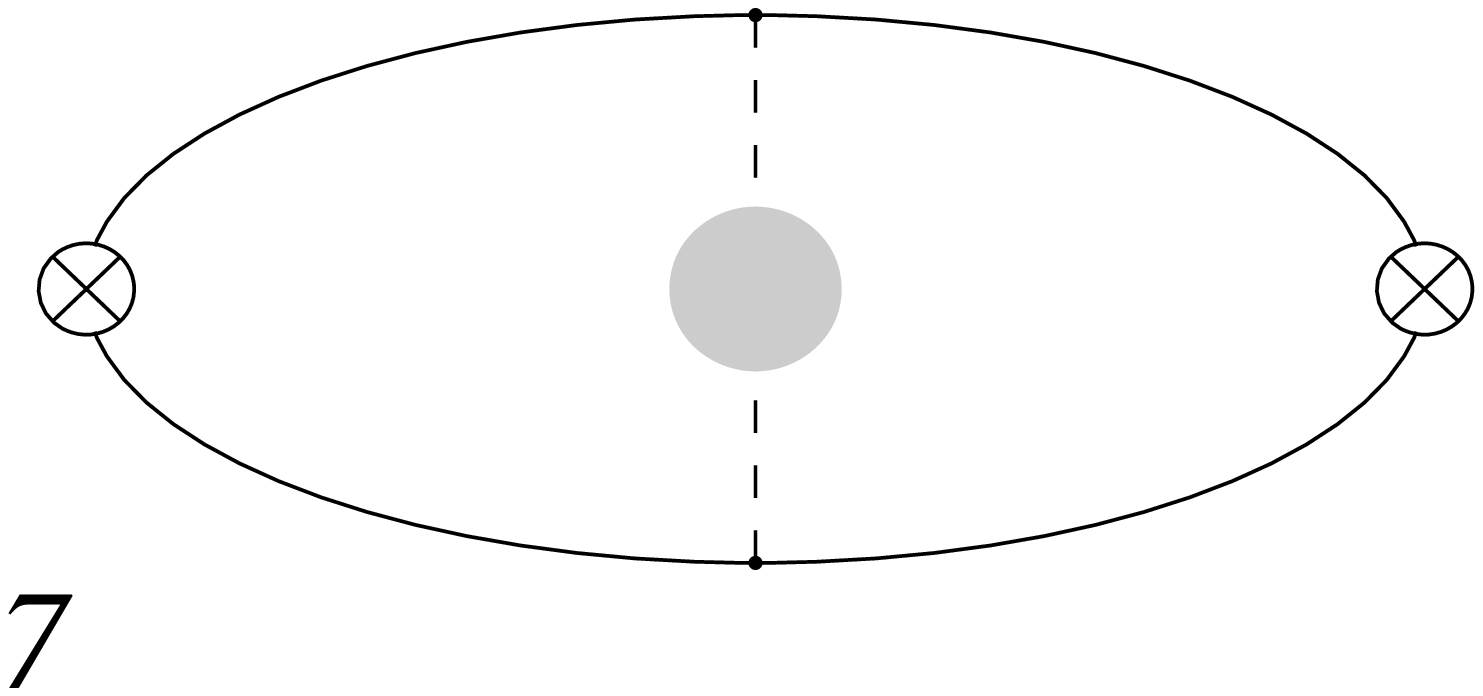}
\hspace{2.2cm}
\leavevmode
\epsfxsize=1.7cm
\epsffile[220 410 420 540]{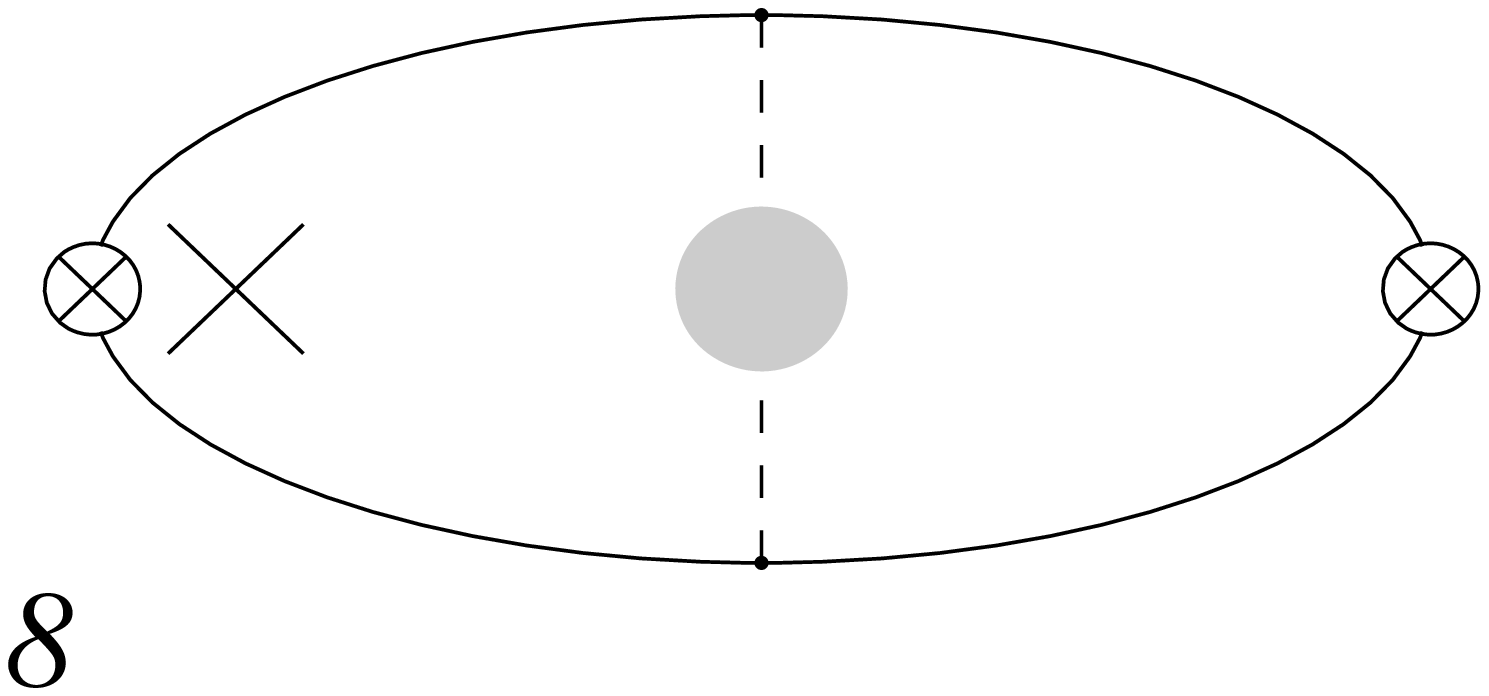}
\hspace{2.2cm}
\leavevmode
\epsfxsize=1.7cm
\epsffile[220 410 420 540]{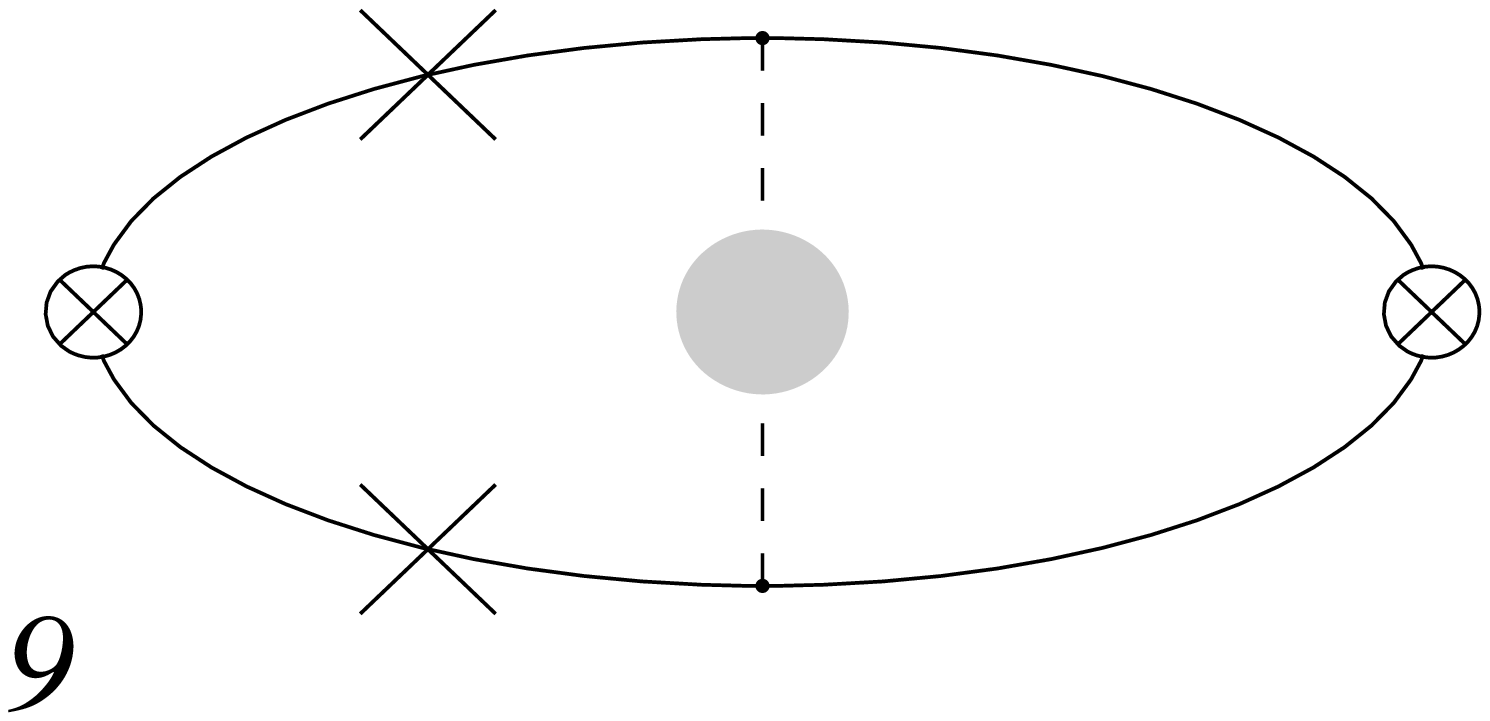}
\vskip  1.5cm
 \caption{\label{fignonrelcurrent2loop} 
Graphical representation of the NRQCD vector-current correlators
diagrams needed to determine the non-relativistic
vector-current-induced cross section at ${\cal{O}}(\alpha_s^2)$ and NNLO in
the non-relativistic expansion.
}
 \end{center}
\end{figure}

\begin{figure}[t] 
\begin{center}
\leavevmode
\epsfxsize=2.3cm
\epsffile[220 410 420 540]{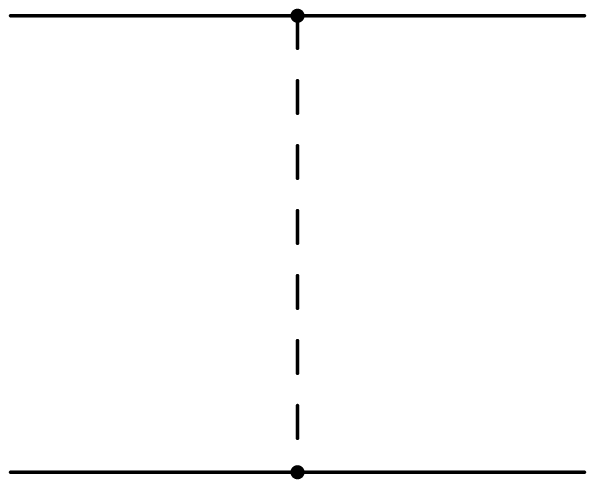}
\hspace{1cm}
\leavevmode
\epsfxsize=2.3cm
\epsffile[220 410 420 540]{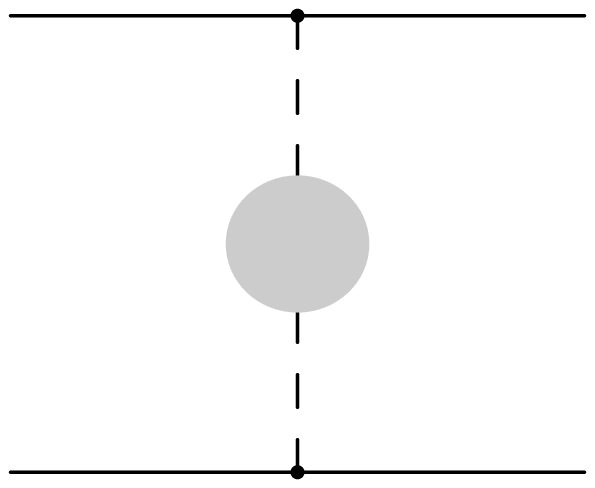}
\hspace{1cm}
\leavevmode
\epsfxsize=2.3cm
\epsffile[220 410 420 540]{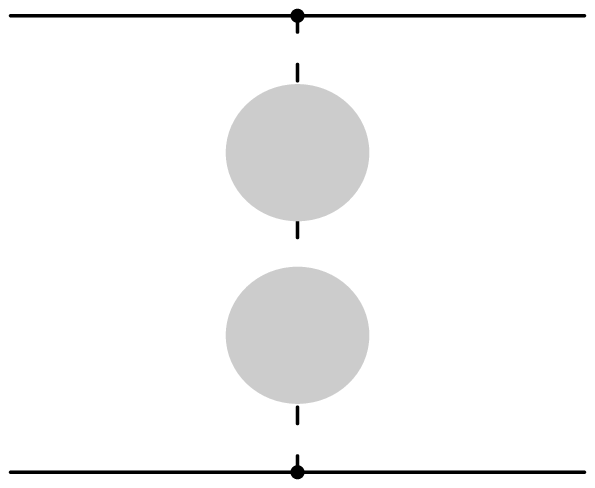}
\hspace{1cm}
\leavevmode
\epsfxsize=2.3cm
\epsffile[220 410 420 540]{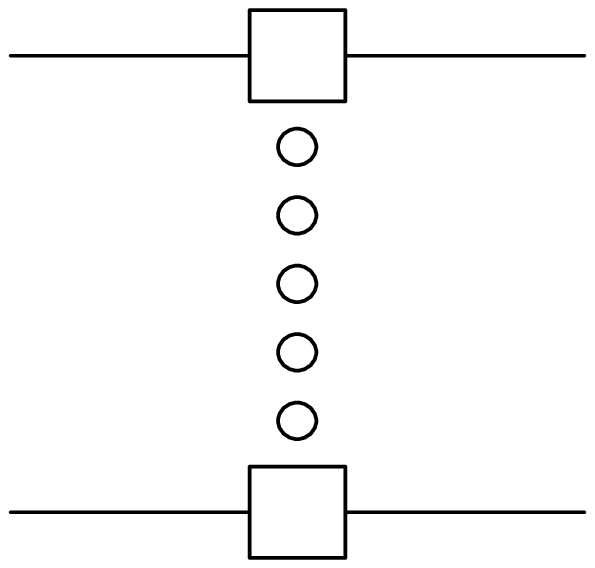}\\[1.5cm]
$V_c^{\tiny \mbox{LO}}$
\mbox{\hspace{2.65cm}}
$V_c^{\tiny \mbox{NLO}}$
\mbox{\hspace{2.55cm}}
$V_c^{\tiny \mbox{NNLO}}$ 
\mbox{\hspace{2.55cm}}
$V_{\mbox{\tiny BF}}$
\mbox{\hspace{1mm}}\\[.2cm]
\hspace{1cm}
\leavevmode
\epsfxsize=2.3cm
\epsffile[220 410 420 540]{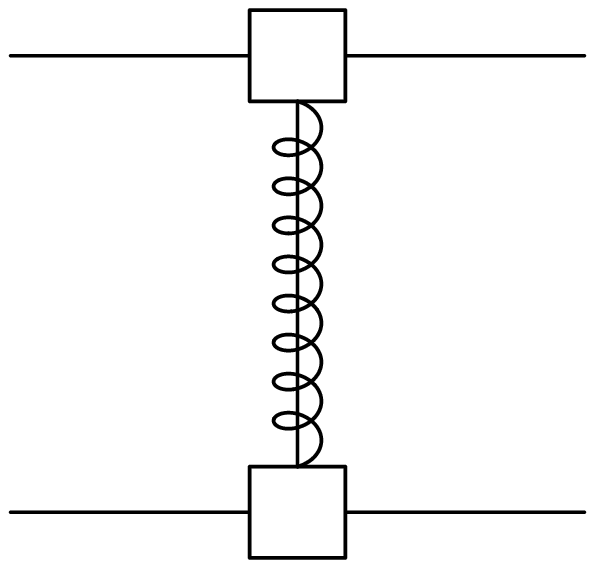}
\hspace{1cm}
\leavevmode
\epsfxsize=2.3cm
\epsffile[220 410 420 540]{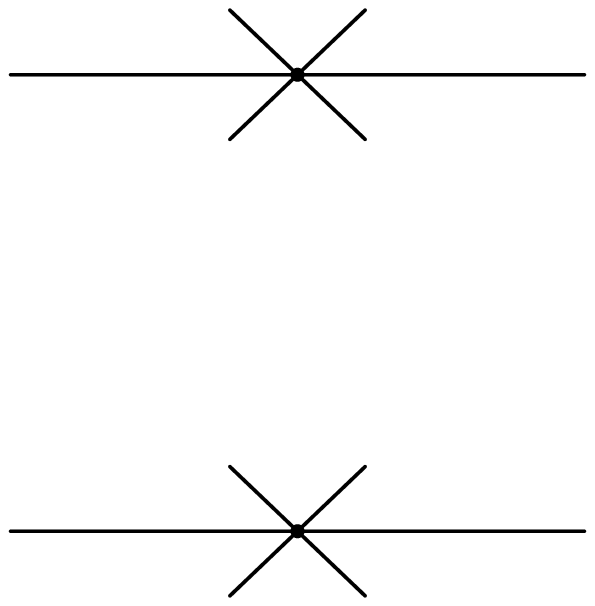}
\hspace{1cm}
\leavevmode
\epsfxsize=2.3cm
\epsffile[220 410 420 540]{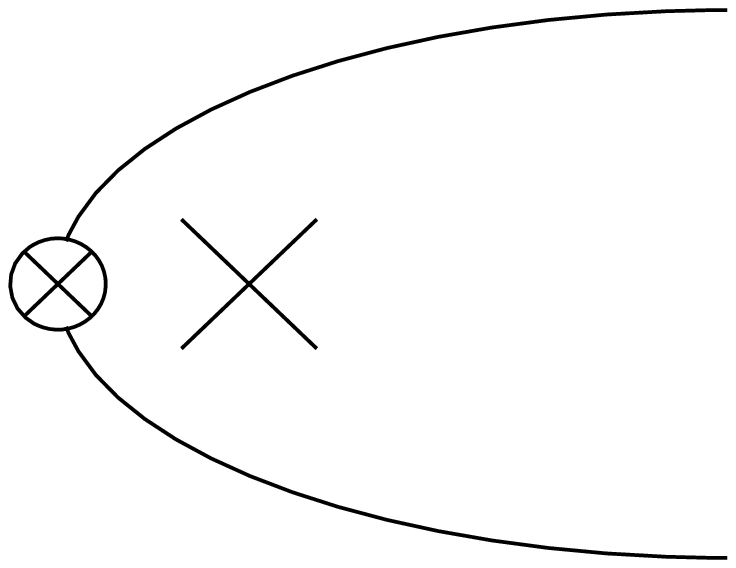}\\[1.5cm]
\mbox{\hspace{2.cm}}
$V_{\mbox{\tiny NA}}$
\mbox{\hspace{2.65cm}}
$\delta H_{\mbox{\tiny kin}}$
\mbox{\hspace{1.55cm}}
${\tilde \psi}^\dagger \sigma_i
(\mbox{$-\frac{i}{2}$} 
\stackrel{\leftrightarrow}{\mbox{\boldmath $D$}})^2
 \tilde \chi$ 
\mbox{\hspace{.25cm}}
\vskip  .2cm
 \caption{\label{fignonrealcurrentsymbols}
Symbols describing the interactions potentials $V_c^{\tiny \mbox{LO}}$,
$V_c^{\tiny \mbox{NLO}}$, 
$V_c^{\tiny \mbox{NNLO}}$, $V_{\mbox{\tiny BF}}$ and $V_{\mbox{\tiny
NA}}$ and the kinetic energy correction 
$\delta H_{\mbox{\tiny kin}} = (p_0^4-\vec k^4)/(4 M_t^3)$.
$V_c^{\tiny \mbox{LO}}$, $V_c^{\tiny \mbox{NLO}}$ and $V_c^{\tiny
  \mbox{NNLO}}$ refer to the Born, one-loop
and two-loop contributions to the Coulomb potential presented in
Eq.~(\ref{NNLOCoulomb}).}
 \end{center}
\end{figure}

To determine the corresponding total vector-current-induced cross
section in NRQCD, we have to calculate the absorptive part of the
correlator diagrams depicted in Figs.~\ref{fignonrelcurrentborn}, 
\ref{fignonrelcurrent1loop} and \ref{fignonrelcurrent2loop}.
The various symbols are defined in
Fig.~\ref{fignonrealcurrentsymbols}. We emphasize that we neglect
multiple insertions of NNLO contributions.
 
The results for the absorptive
parts of the individual diagrams read
($a\equiv C_F \alpha_s$, 
$D(\mbox{\boldmath $k$})\equiv 
M_t/(\mbox{\boldmath $k$}^2-p_0^2-i\epsilon)$):
\begin{figure}[t] 
\begin{center}
\leavevmode
\epsfxsize=1.7cm
\epsffile[220 410 420 540]{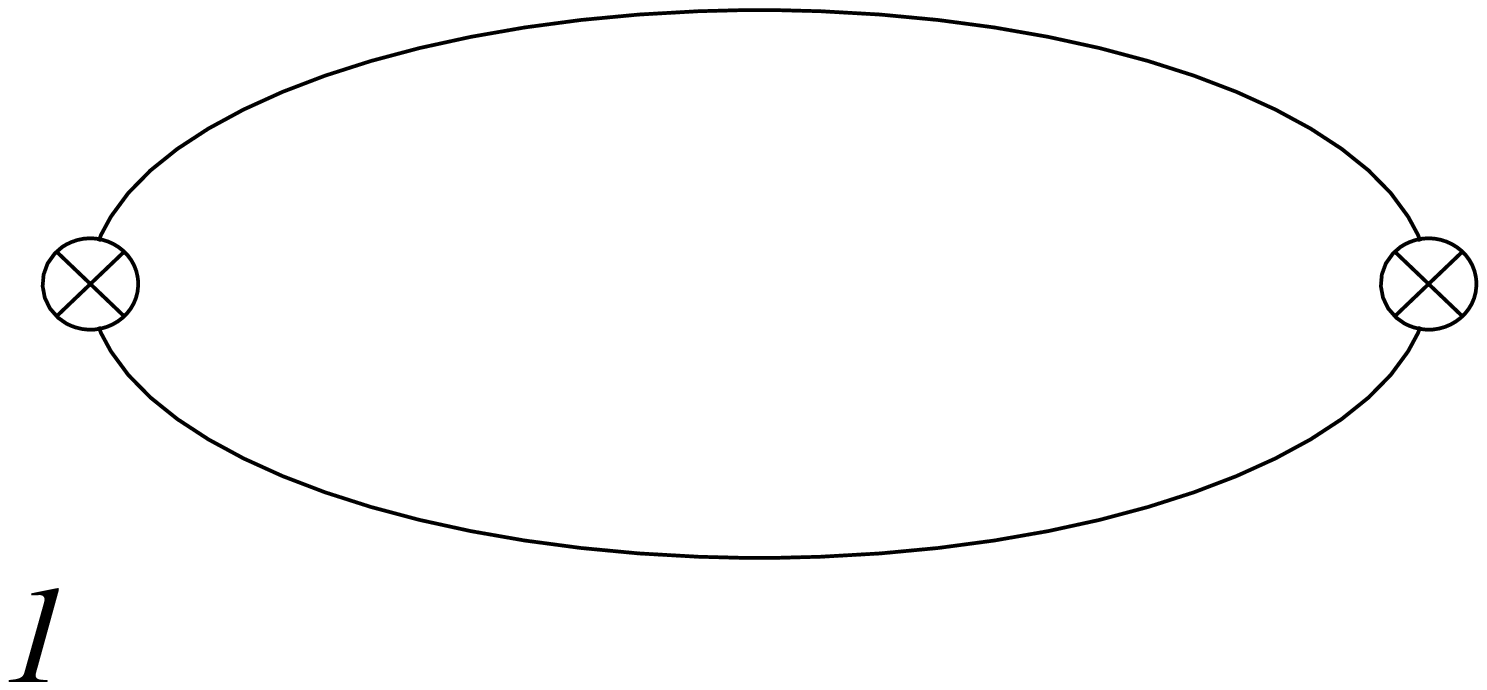}
\hspace{2.2cm}
\leavevmode
\epsfxsize=1.7cm
\epsffile[220 410 420 540]{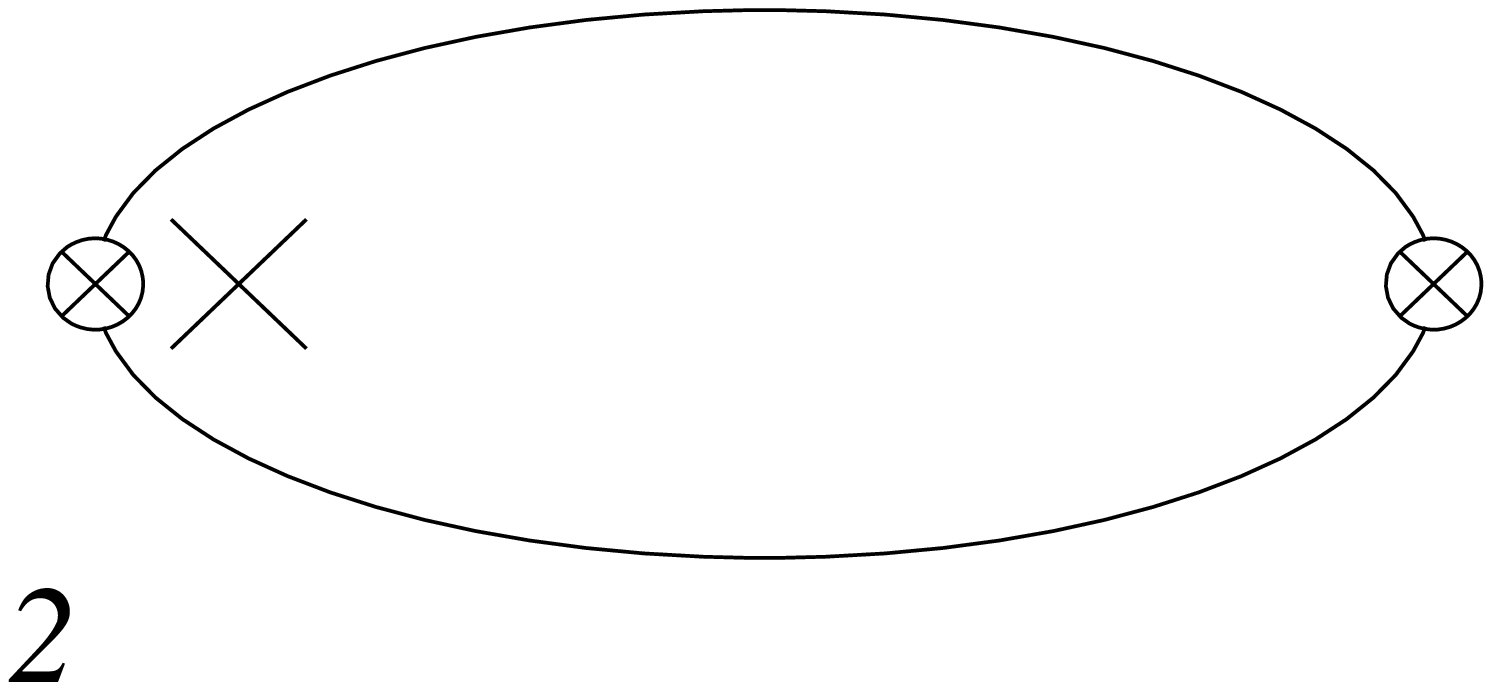}
\hspace{2.2cm}
\leavevmode
\epsfxsize=1.7cm
\epsffile[220 410 420 540]{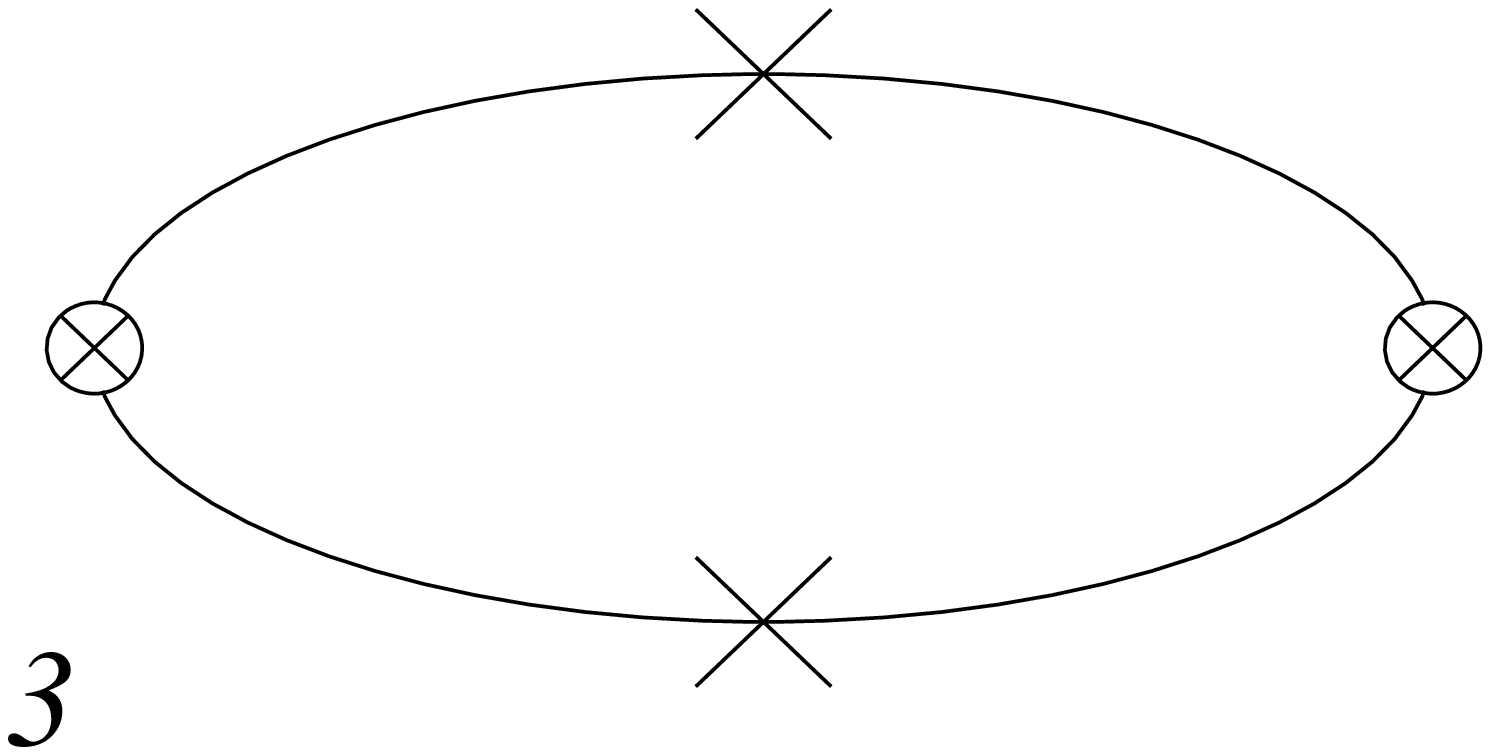}
\vskip  1.5cm
 \caption{\label{fignonrelcurrentborn} 
Graphical representation of the NRQCD vector-current correlators
diagrams needed to determine the non-relativistic
vector-current-induced cross section at the Born level and NNLO in the
non-relativistic expansion.
}
 \end{center}
\end{figure}
\begin{eqnarray}
I_1^{(0)} & = & \mbox{Im}\,\bigg[\,
\int\frac{d^3 \mbox{\boldmath $k$}}{(2\pi)^3}\,
D(\mbox{\boldmath $k$})
\,\bigg]
\, = \, 
\frac{M_t^2}{4 \pi}\,\frac{p_0}{M_t}
\,,
\\[2mm]
I_2^{(0)}  & = & \mbox{Im}\,\bigg[\,
2\,\int\frac{d^3 \mbox{\boldmath $k$}}{(2\pi)^3}\,
\bigg(-\frac{\mbox{\boldmath $k$}^2}{6\,M_t^2}\,\bigg)\,
D(\mbox{\boldmath $k$})
\,\bigg]
\, = \, -
\frac{M_t^2}{4 \pi}\,\frac{p_0^3}{3\,M_t^3}
\,,
\\[2mm]
I_3^{(0)}  & = & \mbox{Im}\,\bigg[\,
\int\frac{d^3 \mbox{\boldmath $k$}}{(2\pi)^3}\,
\bigg(\,\frac{\mbox{\boldmath $k$}^2+p_0^2}{4\,M_t^2}\,\bigg)\,
D(\mbox{\boldmath $k$})
\,\bigg]
\, = \,  
\frac{M_t^2}{4 \pi}\,\frac{p_0^3}{2\,M_t^3}
\,,
\\[5mm]
I_1^{(1)}  & = &  \mbox{Im}\,\bigg[\,
\int\frac{d^3 \mbox{\boldmath $k_1$}}{(2\pi)^3}\,
\int\frac{d^3 \mbox{\boldmath $k_2$}}{(2\pi)^3}\,
D(\mbox{\boldmath $k_1$})\,
  \frac{4\pi a}{(\mbox{\boldmath $k_1$}-\mbox{\boldmath $k_2$})^2}\,
D(\mbox{\boldmath $k_2$})
\,\bigg]
\, = \, 
\frac{a\,M_t^2}{4\, \pi^2}\,\bigg[\,
\frac{\pi^2}{2}-\frac{4\,p_0}{\Lambda} + 
{\cal{O}}\bigg( \frac{p_0^3}{M_t^3} \bigg)
\,\bigg]
\,,
\\[2mm]
I_2^{(1)}  & = &  \mbox{Im}\,\bigg[\,2\,
\int\frac{d^3 \mbox{\boldmath $k_1$}}{(2\pi)^3}\,
\int\frac{d^3 \mbox{\boldmath $k_2$}}{(2\pi)^3}\,
\bigg(-\frac{\mbox{\boldmath $k_1$}^2}{6\,M_t^2}\,\bigg)\,
D(\mbox{\boldmath $k_1$})\,
  \frac{4\pi a}{(\mbox{\boldmath $k_1$}-\mbox{\boldmath $k_2$})^2}\,
D(\mbox{\boldmath $k_2$})
\,\bigg]
\nonumber
\\[2mm]
& = & -\,
\frac{a\,M_t^2}{4\, \pi^2}\,\bigg[\,
\frac{2\,\Lambda\,p_0}{3\,M_t^2} + 
\frac{p_0^2\,\pi^2}{6\,M_t^2} +
{\cal{O}}\bigg( \frac{p_0^3}{M_t^3} \bigg)
\,\bigg]
\,,
\\[2mm]
I_3^{(1)}  & = &  \mbox{Im}\,\bigg[\,2\,
\int\frac{d^3 \mbox{\boldmath $k_1$}}{(2\pi)^3}\,
\int\frac{d^3 \mbox{\boldmath $k_2$}}{(2\pi)^3}\,
\bigg(\,\frac{\mbox{\boldmath $k_1$}^2+p_0^2}{4\,M_t^2}\,\bigg)\,
D(\mbox{\boldmath $k_1$})\,
  \frac{4\pi a}{(\mbox{\boldmath $k_1$}-\mbox{\boldmath $k_2$})^2}\,
D(\mbox{\boldmath $k_2$})
\,\bigg]
\nonumber
\\[2mm]
& = & 
\frac{a\,M_t^2}{4\, \pi^2}\,\bigg[\, 
\frac{\Lambda\,p_0}{M_t^2} +
\frac{p_0^2\,\pi^2}{2\,M_t^2} +  
{\cal{O}}\bigg( \frac{p_0^3}{M_t^3} \bigg)
\,\bigg]
\,,
\\[2mm]
I_4^{(1)}  & = &  \mbox{Im}\,\bigg[\,
\int\frac{d^3 \mbox{\boldmath $k_1$}}{(2\pi)^3}\,
\int\frac{d^3 \mbox{\boldmath $k_2$}}{(2\pi)^3}\,
D(\mbox{\boldmath $k_1$})\,
\bigg(\,
2\,\frac{\pi\,a}{M_t^2}\,
\frac{\mbox{\boldmath $k_1$}^2+{\mbox{\boldmath $k_2$}}^2}
   {(\mbox{\boldmath $k_1$}-\mbox{\boldmath $k_2$})^2} -
\frac{11}{3}\,\frac{\pi\,a}{M_t^2}\,
\bigg)\,
D(\mbox{\boldmath $k_2$})
\,\bigg]
\nonumber
\\[2mm]
& = & 
\frac{a\,M_t^2}{4\, \pi^2}\,\bigg[\,
- \frac{5\,\Lambda\,p_0}{3\,M_t^2}
+ \frac{p_0^2\,\pi^2}{2\,M_t^2} 
+ {\cal{O}}\bigg( \frac{p_0^3}{M_t^3} \bigg)
\,\bigg]
\,,
\\[5mm]
I_1^{(2)}  & = &  \mbox{Im}\,\bigg[\,
\int\frac{d^3 \mbox{\boldmath $k_1$}}{(2\pi)^3}\,
\int\frac{d^3 \mbox{\boldmath $k_2$}}{(2\pi)^3}\,
\int\frac{d^3 \mbox{\boldmath $k_3$}}{(2\pi)^3}\,
D(\mbox{\boldmath $k_1$})\,
  \frac{4\pi a}{(\mbox{\boldmath $k_1$}-\mbox{\boldmath $k_2$})^2}\,
D(\mbox{\boldmath $k_2$})\,
  \frac{4\pi a}{(\mbox{\boldmath $k_2$}-\mbox{\boldmath $k_3$})^2}\,
D(\mbox{\boldmath $k_3$})
\,\bigg]
\nonumber
\\[2mm]
& = & 
\frac{a^2\,M_t^2}{4\, \pi^3}\,\bigg[\,
\frac{M_t\,\pi^4}{12\,p_0}-\frac{2\,M_t\,\pi^2}{\Lambda} +
\frac{M_t\,p_0\,(12-\pi^2)}{2\,\Lambda^2} + 
{\cal{O}}\bigg( \frac{p_0^2}{M_t^2} \bigg)
\,\bigg]
\,,
\\[2mm]
I_2^{(2)}  & = &  \mbox{Im}\,\bigg[\,2\,
\int\frac{d^3 \mbox{\boldmath $k_1$}}{(2\pi)^3}\,
\int\frac{d^3 \mbox{\boldmath $k_2$}}{(2\pi)^3}\,
\int\frac{d^3 \mbox{\boldmath $k_3$}}{(2\pi)^3}\,
\bigg(-\frac{\mbox{\boldmath $k_1$}^2}{6\,M_t^2}\,\bigg)\,
D(\mbox{\boldmath $k_1$})\,
  \frac{4\pi a}{(\mbox{\boldmath $k_1$}-\mbox{\boldmath $k_2$})^2}\,
D(\mbox{\boldmath $k_2$})\,
  \frac{4\pi a}{(\mbox{\boldmath $k_2$}-\mbox{\boldmath $k_3$})^2}\,
D(\mbox{\boldmath $k_3$})
\,\bigg]
\nonumber
\\[2mm]
& = & \, - \,
\frac{a^2\,M_t^2}{4\, \pi^3}\,\bigg[\,
\frac{\Lambda\,\pi^2}{3\,M_t} - 
\frac{p_0\,(84+3\,\pi^2-\pi^4)}{36\,M_t} + 
{\cal{O}}\bigg( \frac{p_0^2}{M_t^2} \bigg)
\,\bigg]
\,,
\\[2mm]
I_3^{(2)}  & = &  \mbox{Im}\,\bigg[\,2\,
\int\frac{d^3 \mbox{\boldmath $k_1$}}{(2\pi)^3}\,
\int\frac{d^3 \mbox{\boldmath $k_2$}}{(2\pi)^3}\,
\int\frac{d^3 \mbox{\boldmath $k_3$}}{(2\pi)^3}\,
\bigg(\,\frac{\mbox{\boldmath $k_1$}^2+p_0^2}{4\,M_t^2}\,\bigg)\,
D(\mbox{\boldmath $k_1$})\,
  \frac{4\pi a}{(\mbox{\boldmath $k_1$}-\mbox{\boldmath $k_2$})^2}\,
D(\mbox{\boldmath $k_2$})\,
  \frac{4\pi a}{(\mbox{\boldmath $k_2$}-\mbox{\boldmath $k_3$})^2}\,
D(\mbox{\boldmath $k_3$})
\,\bigg]
\nonumber
\\[2mm]
& = & \, 
\frac{a^2\,M_t^2}{4\, \pi^3}\,\bigg[\,
\frac{\Lambda\,\pi^2}{2\,M_t} - 
\frac{p_0\,(84+3\,\pi^2-2\,\pi^4)}{24\,M_t} + 
{\cal{O}}\bigg( \frac{p_0^2}{M_t^2} \bigg)
\,\bigg]
\,,
\\[2mm]
I_4^{(2)}  & = &  \mbox{Im}\,\bigg[\,
\int\frac{d^3 \mbox{\boldmath $k_1$}}{(2\pi)^3}\,
\int\frac{d^3 \mbox{\boldmath $k_2$}}{(2\pi)^3}\,
\int\frac{d^3 \mbox{\boldmath $k_3$}}{(2\pi)^3}\,
D(\mbox{\boldmath $k_1$})\,
  \frac{4\pi a}{(\mbox{\boldmath $k_1$}-\mbox{\boldmath $k_2$})^2}\,
\bigg(\,\frac{\mbox{\boldmath $k_2$}^2+p_0^2}{4\,M_t^2}\,\bigg)\,
D(\mbox{\boldmath $k_2$})\,
  \frac{4\pi a}{(\mbox{\boldmath $k_2$}-\mbox{\boldmath $k_3$})^2}\,
D(\mbox{\boldmath $k_3$})
\,\bigg]
\nonumber
\\[2mm]
& = & \, 
\frac{a^2\,M_t^2}{4\, \pi^3}\,
\bigg[\,
\frac{p_0}{24\,M_t}\,\bigg(\,
12\,\pi^2 + \pi^4- 42\,\zeta_3 - 
12\,\pi^2\,\ln\Big(\frac{2\,p_0}{\Lambda}\Big) 
\,\bigg) + 
{\cal{O}}\bigg( \frac{p_0^2}{M_t^2} \bigg)
\,\bigg] 
\,,
\\[2mm]
I_5^{(2)}  & = &  \mbox{Im}\,\bigg[\,2\,
\int\frac{d^3 \mbox{\boldmath $k_1$}}{(2\pi)^3}\,
\int\frac{d^3 \mbox{\boldmath $k_2$}}{(2\pi)^3}\,
\int\frac{d^3 \mbox{\boldmath $k_3$}}{(2\pi)^3}\,
D(\mbox{\boldmath $k_1$})\,
\bigg(\,
2\,\frac{\pi\,a}{M_t^2}\,
\frac{\mbox{\boldmath $k_1$}^2+{\mbox{\boldmath $k_2$}}^2}
   {(\mbox{\boldmath $k_1$}-\mbox{\boldmath $k_2$})^2} -
\frac{11}{3}\,\frac{\pi\,a}{M_t^2}\,
\bigg)\,
\nonumber
\\[2mm] & &
\mbox{\hspace{3cm}}
\times\,
D(\mbox{\boldmath $k_2$})\,
  \frac{4\pi a}{(\mbox{\boldmath $k_2$}-\mbox{\boldmath $k_3$})^2}\,
D(\mbox{\boldmath $k_3$})
\,\bigg]
\nonumber
\\[2mm]
& = & 
\frac{a^2\,M_t^2}{4\, \pi^3}\,\bigg[\,
-\,\frac{5\,\Lambda\,\pi^2}{6\,M_t} + 
\frac{p_0}{12\,M_t}\,
  \bigg(92+21\,\pi^2 + 2\,\pi^4 - 7\,\zeta_3
- 2\,\pi^2\,\ln\Big(\frac{2\,p_0}{\Lambda}\Big)\,\bigg) + 
{\cal{O}}\bigg( \frac{p_0^2}{M_t^2} \bigg)
\,\bigg]
\,,
\\[2mm]
I_6^{(2)}  & = &  \mbox{Im}\,\bigg[\,
\int\frac{d^3 \mbox{\boldmath $k_1$}}{(2\pi)^3}\,
\int\frac{d^3 \mbox{\boldmath $k_2$}}{(2\pi)^3}\,
D(\mbox{\boldmath $k_1$})\,
\bigg(\,
\frac{C_A}{C_F}\,\frac{\pi^2\,a^2}
   {M_t\,|\mbox{\boldmath $k_1$}-\mbox{\boldmath $k_2$}|}
\bigg)\,
D(\mbox{\boldmath $k_2$})
\,\bigg]
\nonumber
\\[2mm]
& = & \,-\,
\frac{a^2\,M_t^2}{4\, \pi^3}\,\bigg[\,
\frac{C_A\,p_0\,\pi^2}{C_F\,M_t}\,
  \bigg(\,-1 + \ln\Big(\frac{2\,p_0}{\Lambda}\Big)\,\bigg)
+ {\cal{O}}\bigg( \frac{p_0^2}{M_t^2} \bigg)
\,\bigg]
\,,
\\[2mm]
I_7^{(2)}  & = &  \mbox{Im}\,\bigg[\,
\int\frac{d^3 \mbox{\boldmath $k_1$}}{(2\pi)^3}\,
\int\frac{d^3 \mbox{\boldmath $k_2$}}{(2\pi)^3}\,
D(\mbox{\boldmath $k_1$})\,
  \frac{4\pi a}{(\mbox{\boldmath $k_2$}-\mbox{\boldmath $k_3$})^2}\,
\frac{a}{4\,C_F\,\pi}
\bigg(\,  -\beta_0\,
\ln\Big(\frac{(\mbox{\boldmath $k_1$}-\mbox{\boldmath $k_2$})^2}{\mu^2}\Big)
+ a_1
\bigg)\,
D(\mbox{\boldmath $k_2$})
\,\bigg]
\nonumber
\\[2mm]
& = & 
\frac{a^2\,M_t^2}{4\,C_F\, \pi^3}\,\bigg[\,
\beta_0\,\bigg(\,
 -\frac{\pi^2}{4}\,\ln\Big(\frac{2\,p_0}{\mu}\Big)
 +\frac{2\,p_0}{\Lambda}\,
       \bigg(1 + \ln\Big(\frac{\Lambda}{\mu}\Big)\bigg)
\,\bigg)
+ a_1\,\bigg(\,
  \frac{\pi^2}{8} -\frac{p_0}{\Lambda} \,\bigg)
+ {\cal{O}}\bigg( \frac{p_0^2}{M_t^2} \bigg)
\,\bigg]
\,,
\\[2mm]
I_8^{(2)}  & = &  \mbox{Im}\,\bigg[\,2
\int\frac{d^3 \mbox{\boldmath $k_1$}}{(2\pi)^3}\,
\int\frac{d^3 \mbox{\boldmath $k_2$}}{(2\pi)^3}\,
\bigg(-\frac{\mbox{\boldmath $k_1$}^2}{6\,M_t^2}\,\bigg)\,
\nonumber\\ & &
\mbox{\hspace{3cm}}
\times\,
D(\mbox{\boldmath $k_1$})\,
  \frac{4\pi a}{(\mbox{\boldmath $k_2$}-\mbox{\boldmath $k_3$})^2}\,
\frac{a}{4\,C_F\,\pi}
\bigg(\,  -\beta_0\,
\ln\Big(\frac{(\mbox{\boldmath $k_1$}-\mbox{\boldmath $k_2$})^2}{\mu^2}\Big)
+ a_1
\bigg)\,
D(\mbox{\boldmath $k_2$})
\,\bigg]
\nonumber
\\[2mm]
& = & 
\frac{a^2\,M_t^2}{4\,C_F\, \pi^3}\,\bigg[\,
\beta_0\,\bigg(\,
\frac{\Lambda\,p_0}{3\,M_t^2}\,
\bigg(-1 + \ln\Big(\frac{\Lambda}{\mu}\Big)\,\bigg)
\,\bigg)
- a_1\,
 \frac{\Lambda\,p_0}{6\,M_t^2} 
+ {\cal{O}}\bigg( \frac{p_0^2}{M_t^2}
\,\bigg)
\,\bigg]
\,,
\\[2mm]
I_9^{(2)}  & = &  \mbox{Im}\,\bigg[\,2
\int\frac{d^3 \mbox{\boldmath $k_1$}}{(2\pi)^3}\,
\int\frac{d^3 \mbox{\boldmath $k_2$}}{(2\pi)^3}\,
\bigg(\,\frac{\mbox{\boldmath $k_1$}^2+p_0^2}{4\,M_t^2}\,\bigg)\,
\nonumber\\ & &
\mbox{\hspace{3cm}}
\times\,
D(\mbox{\boldmath $k_1$})\,
  \frac{4\pi a}{(\mbox{\boldmath $k_2$}-\mbox{\boldmath $k_3$})^2}\,
\frac{a}{4\,C_F\,\pi}
\bigg(\,  -\beta_0\,
\ln\Big(\frac{(\mbox{\boldmath $k_1$}-\mbox{\boldmath $k_2$})^2}{\mu^2}\Big)
+ a_1
\bigg)\,
D(\mbox{\boldmath $k_2$})
\,\bigg]
\nonumber
\\[2mm]
& = & 
\frac{a^2\,M_t^2}{4\,C_F\, \pi^3}\,\bigg[\,
\beta_0\,\bigg(\,
\frac{\Lambda\,p_0}{2\,M_t^2}\,
   \bigg(1 - \ln\Big(\frac{\Lambda}{\mu}\Big)\bigg)
\,\bigg)
+ a_1\,
\frac{\Lambda\,p_0}{4\,M_t^2}
+ {\cal{O}}\bigg( \frac{p_0^2}{M_t^2} \bigg)
\,\bigg]
\,,
\end{eqnarray}
where the upper index of the functions $I_j^{(i)}$ corresponds to the
power of the strong coupling of the diagrams and the lower index to
the numeration given in Figs.~\ref{fignonrelcurrentborn}, 
\ref{fignonrelcurrent1loop} and
\ref{fignonrelcurrent2loop}. Combinatorial factors are taken into
account. We note that the results above have been expanded in
$p_0/M_t, p_0/\Lambda \ll 1$; no condition has been assumed for the
ratio $\Lambda/M_t$. 

Summing
all terms leads to the total vector-current-induced
cross section in NRQCD:  
\begin{eqnarray}
R_{\mbox{\tiny NNLO}}^{v,\mbox{\tiny thr}} & = &
\frac{6\,\pi\,N_c}{M_t^2\,(1+\frac{p_0^2}{M_t^2})}\,C^v\,
\bigg[\,
\sum\limits_{i=1}^{3}\,I^{(0)}_i +
\sum\limits_{i=1}^{4}\,I^{(1)}_i +
\sum\limits_{i=1}^{9}\,I^{(2)}_i
\,\bigg]
\label{RphotonfullNRQCD}
\end{eqnarray}
The short-distance coefficient $C^v$ has to be chosen such that the
RHS of Eqs.~(\ref{RphotonfullQCD}) and (\ref{RphotonfullNRQCD}) are
equal for all terms up to order $\alpha_s^2$ and NNLO in the
non-relativistic expansion. The result
reads  
\begin{eqnarray}
C^v(\Lambda,\mu) \, = \, 1 \, + \, c^v_{\mbox{\tiny NLO}}(\Lambda,\mu) \, + \,
c^v_{\mbox{\tiny NNLO}}(\Lambda,\mu)
\,, 
\label{CVNNLOshortdistance}
\end{eqnarray}
where
\begin{eqnarray}
c^v_{\mbox{\tiny NLO}}(\Lambda,\mu) & = &
  \frac{4\,a}{\pi}\,
\bigg[\,
-1 + \frac{M_t}{\Lambda}
\,\bigg]
\,,
\label{cvNLOshortdistance}
\\
c^v_{\mbox{\tiny NNLO}}(\Lambda,\mu) & = &
\frac{4\,a\,\Lambda}{3\,\pi\,M_t}
 +
\frac{a^2}{\pi^2}\,\bigg[\,
\frac{\beta_0}{C_F}\,\bigg(\,
- \frac{\Lambda^2 + 12\,M_t^2}{6\,\Lambda\,M_t} 
+ \frac{\Lambda^2 - 12\,M_t^2}
     {6\,\Lambda\,M_t}\,\ln\Big(\frac{\Lambda}{\mu}\Big) 
+ 2\,\ln\Big(\frac{M_t}{\mu}\Big)
\,\bigg)
\nonumber
\\[2mm] & &
- \frac{a_1}{C_F}\,\frac{\Lambda^2 - 12\,M_t^2}{12\,\Lambda\,M_t} 
+ \pi^2\,\bigg(\,\frac{2}{3}+\frac{C_A}{C_F}\,\bigg)\,
    \ln\Big(\frac{2\,M_t}{\Lambda}\,\Big)
+ \frac{\pi^2\,\kappa}{C_F^2} 
\nonumber
\\[2mm] & &
+ \frac{16\,\Lambda^2}{9\,M_t^2} 
- \frac{16\,\Lambda}{3\,M_t} 
- \frac{16\,M_t}{\Lambda} 
+ \frac{M_t^2\,(20 + \pi^2)}{2\,\Lambda^2}
- \frac{53\,\pi^2}{24} - \frac{C_A\,\pi^2}{C_F} + \frac{25}{6} + 
\frac{7}{3}\zeta_3
\,\bigg]
\,.
\label{cvNNLOshortdistance}
\end{eqnarray}
In Eq.~(\ref{CVNNLOshortdistance}) we have displayed the NLO and NNLO
short-distance contributions separately. We note that the NLO
short-distance contributions in Eq.~(\ref{cvNLOshortdistance}) differ
from the commonly quoted one by the term
$4\frac{C_F\alpha_s}{\pi}\frac{M_t}{\Lambda}$. This is a consequence
of our
cutoff regularization scheme, which excludes loop momenta with spatial
components larger than $\Lambda$, even if the corresponding integration
is UV-convergent. We also point out that the NNLO short-distance
contribution in Eq.~(\ref{cvNNLOshortdistance}) contains the term
$\frac{4\alpha_s}{3\pi}\frac{\Lambda}{M_t}$, which is of order $\alpha_s$
only. This term is a manifestation of the power-counting breaking
effects discussed in Sec.~\ref{sectionregularization}. The term exists
because it subtracts the power-counting breaking terms originating
from the linear UV-divergent behaviour of the Breit-Fermi
potential, $V_{\mbox{\tiny BF}}$, the kinetic energy correction,
$(p_0^4-\mbox{\boldmath $k$}^4)/(4 M_t^3)$, and the dimension-5
NRQCD vector-current in the non-relativistic current correlators. Thus
it is important to consider this term as NNLO. It is a conspicuous
fact that the NLO short-distance coefficient $c^v_{\mbox{\tiny NLO}}$
vanishes for the choice $\Lambda=M_t$. We emphasize, however, that
this cancellation is purely accidental. Nevertheless, comparing the
short-distance constant $C^v$ calculated in our cutoff regularization
scheme with the corresponding coefficient obtained in the
$\overline{\mbox{MS}}$ scheme~\cite{Melnikov4,Beneke2}
\begin{eqnarray}
C^v_{\tiny \overline{\mbox{MS}}}(\mu) \, = \, 1 \, + \, 
c^v_{{\tiny \overline{\mbox{MS}}, \tiny NLO}}(\Lambda,\mu) \, + \,
c^v_{{\tiny \overline{\mbox{MS}}, \tiny NNLO}}(\Lambda,\mu)
\,, 
\label{CVNNLOshortdistanceMSbar}
\end{eqnarray}
where ($\alpha_s\equiv\alpha_s(\mu)$)
\begin{eqnarray}
c^v_{\tiny \overline{\mbox{MS}}, \mbox{\tiny NLO}}(\mu) & = &
-4\,C_F\,\frac{\alpha_s}{\pi}
\,,
\label{cvNLOshortdistanceMSbar}
\\
c^v_{\tiny \overline{\mbox{MS}}, \mbox{\tiny NNLO}}(\mu) & = &
\frac{\alpha_s^2}{\pi^2}\,\bigg[\,
C_F^2\,\bigg(\,\frac{39}{4} - \frac{79\,\pi^2}{18} 
       + 2\,\pi^2\,\ln 2 
       + \frac{\pi^2}{3}\,\ln\Big(\frac{M_t^2}{\mu^2}\Big) 
       - \zeta_3 \,\bigg)
\nonumber
\\[2mm] & &
+ C_A\,C_F\,\bigg( -\frac{151}{36} + \frac{89\,\pi^2}{72} 
       - \frac{5\,\pi^2}{3}\,\ln 2 
       + \frac{\pi^2}{2}\,\ln\Big(\frac{M_t^2}{\mu^2}\Big) 
       - \frac{13}{2}\,\zeta_3  \,\bigg)
\nonumber
\\[2mm] & & 
+ C_F\,T\,\bigg(\,\frac{44}{9} - \frac{4\,\pi^2}{9}\,\bigg) 
+ C_F\,T\,n_f\,\frac{11}{9}
\,\bigg]
\,,
\label{cvNNLOshortdistanceMSbar}
\end{eqnarray}
we find that the perturbative corrections are in general smaller in
the cutoff scheme. 
\begin{table}[t] 
\vskip 7mm
\begin{center}
\begin{tabular}{|c||c|c|c||c||c|c|c||c|} \hline
   $\mu [\mbox{GeV}]$ 
   & \multicolumn{3}{|c||}
        {$c^v_{\mbox{\tiny NLO}}(\Lambda [\mbox{GeV}])$}
   & $c^v_{\tiny \overline{\mbox{MS}}, \mbox{\tiny NLO}}$
   & \multicolumn{3}{|c||}
        {$c^v_{\mbox{\tiny NNLO}}(\Lambda [\mbox{GeV}])$}
   & $c^v_{\tiny \overline{\mbox{MS}}, \mbox{\tiny NNLO}}$
\\ \hline
 & $90$ &  $175$ & $350$ &    & $90$ & $175$ & $350$ &
\\ \hline\hline
$15$  & $0.261$ &  $0$ & $-0.138$ & $-0.276$ & $-0.018$ & $-0.101$ & $-0.040$ & $0.128$
\\\hline
$30$  & $0.228$ &  $0$ & $-0.120$ & $-0.241$ & $0.018$ & $-0.069$ & $-0.029$ & $0.025$
\\\hline
$60$  & $0.202$ &  $0$ & $-0.107$ & $-0.214$ & $0.039$ & $-0.048$ & $-0.022$ & $0.037$
\\\hline
$175$ & $0.172$ &  $0$ & $-0.091$ & $-0.182$ & $0.056$ & $-0.028$ & $-0.015$ & $-0.091$
\\ \hline\hline 
\end{tabular}
\caption{\label{tabshortdistance} 
The NLO and NNLO contributions to the short-distance coefficient $C^v$
in our cutoff scheme and in the $\overline{\mbox{MS}}$ scheme for
various choices of the cutoff $\Lambda$ and the renormalization scale
$\mu$. We have chosen $\alpha_s(M_z)=0.118$, and two-loop running for
the strong coupling has been employed.
}
\end{center}
\vskip 3mm
\end{table}
In Table~\ref{tabshortdistance} we have displayed the NLO and NNLO
short-distance corrections for exemplary choices of the
renormalization scale $\mu$ and the cutoff $\Lambda$ for
$\alpha_s(M_Z)=0.118$ and using two-loop running for the strong coupling.
We are not aware of any principle reason why the
short-distance corrections should be, in general, better convergent in our
cutoff scheme than when using the $\overline{\mbox{MS}}$ regularization. 
We finally
note that the potentially large logarithmic term 
$\alpha_s^2 C_A C_F \ln(M_t/\mu)$ 
in $C^v_{\tiny
\overline{\mbox{MS}}}$, which corresponds to an anomalous dimension of
the dimension-3 NRQCD vector-current 
${\tilde \psi}^\dagger \mbox{\boldmath $\sigma$} \tilde \chi$, does
not exist in $C^v$, since in our cutoff scheme the corresponding
logarithmic divergence
in the NRQCD diagrams is cut off at the scale $\Lambda\sim M_t$ rather
than $\mu\sim M_t v$, as in the $\overline{\mbox{MS}}$
scheme. However, we emphasize that the absence of this
logarithmic term in $C^v$ is traded for the existence of logarithms of
the ratio $2p_0/\Lambda$ in the non-relativistic correlator, which are
not present in the  $\overline{\mbox{MS}}$ scheme. The logarithms of
$\Lambda/\mu$ and $M_t/\mu$ in $c^v_{\mbox{\tiny NNLO}}$,
Eq.~(\ref{cvNNLOshortdistance}), originate from the running of the
strong coupling and are not related to an anomalous dimension. 
\end{appendix}

\vspace{1.5cm}
%
%
%
%
\sloppy
\raggedright
\def\app#1#2#3{{\it Act. Phys. Pol. }{\bf B #1} (#2) #3}
\def\apa#1#2#3{{\it Act. Phys. Austr.}{\bf #1} (#2) #3}
\def\lhc{Proc. LHC Workshop, CERN 90-10}
\def\npb#1#2#3{{\it Nucl. Phys. }{\bf B #1} (#2) #3}
\def\nP#1#2#3{{\it Nucl. Phys. }{\bf #1} (#2) #3}
\def\plb#1#2#3{{\it Phys. Lett. }{\bf B #1} (#2) #3}
\def\prd#1#2#3{{\it Phys. Rev. }{\bf D #1} (#2) #3}
\def\pra#1#2#3{{\it Phys. Rev. }{\bf A #1} (#2) #3}
\def\pR#1#2#3{{\it Phys. Rev. }{\bf #1} (#2) #3}
\def\prl#1#2#3{{\it Phys. Rev. Lett. }{\bf #1} (#2) #3}
\def\prc#1#2#3{{\it Phys. Reports }{\bf #1} (#2) #3}
\def\cpc#1#2#3{{\it Comp. Phys. Commun. }{\bf #1} (#2) #3}
\def\nim#1#2#3{{\it Nucl. Inst. Meth. }{\bf #1} (#2) #3}
\def\pr#1#2#3{{\it Phys. Reports }{\bf #1} (#2) #3}
\def\sovnp#1#2#3{{\it Sov. J. Nucl. Phys. }{\bf #1} (#2) #3}
\def\sovpJ#1#2#3{{\it Sov. Phys. LETP Lett. }{\bf #1} (#2) #3}
\def\jl#1#2#3{{\it JETP Lett. }{\bf #1} (#2) #3}
\def\jet#1#2#3{{\it JETP Lett. }{\bf #1} (#2) #3}
\def\zpc#1#2#3{{\it Z. Phys. }{\bf C #1} (#2) #3}
\def\epj#1#2#3{{\it Eur. Phys. J. }{\bf C #1} (#2) #3}
\def\ptp#1#2#3{{\it Prog.~Theor.~Phys.~}{\bf #1} (#2) #3}
\def\nca#1#2#3{{\it Nuovo~Cim.~}{\bf #1A} (#2) #3}
\def\ap#1#2#3{{\it Ann. Phys. }{\bf #1} (#2) #3}
\def\hpa#1#2#3{{\it Helv. Phys. Acta }{\bf #1} (#2) #3}
\def\ijmpA#1#2#3{{\it Int. J. Mod. Phys. }{\bf A #1} (#2) #3}
\def\ZETF#1#2#3{{\it Pis'ma Zh. Eksp. Teor. Fiz. }{\bf #1} (#2) #3}
\def\jmp#1#2#3{{\it J. Math. Phys. }{\bf #1} (#2) #3}
\def\yf#1#2#3{{\it Yad. Fiz. }{\bf #1} (#2) #3}
\def\aspn#1#2#3{{\it Arch. Sci. Phys. Nat. }{\bf #1} (#2) #3}

\end{document}